\documentclass[preprint,
amsmath,amssymb,aps,nofootinbib,superscriptaddress]{revtex4-1}
\usepackage[pdftex]{graphicx}
\usepackage{dcolumn}
\usepackage{subfigure}
\usepackage{bm}
\usepackage{natbib}
\usepackage[pdftex]{hyperref}
\hypersetup{
    colorlinks=true,
    linkcolor=blue,
    citecolor=blue,
    filecolor=blue,
    urlcolor=blue,
    }

\newcommand{\Slash}[1]{{\ooalign{\hfil/\hfil\crcr$#1$}}}%
\makeatletter \def\@eqnnum{{\normalsize \normalcolor (\theequation)}}  \makeatother %

\begin{document}
\preprint{KEK-TH-1798}

\title{Higgs sector extension of the neutrino minimal standard model with thermal freeze-in production mechanism}
\author{Hiroki Matsui}
\email{matshiro@post.kek.jp}
\affiliation{KEK Theory Center, Tsukuba, Ibaraki 305-0801, Japan\\SOKENDAI (The Graduate University for Advanced Studies)\\Department of Particle and Nuclear Physics, Tsukuba, Ibaraki 305-0801, Japan}
\author{Mihoko Nojiri}
\email{nojiri@post.kek.jp}
\affiliation{KEK Theory Center, Tsukuba, Ibaraki 305-0801, Japan\\SOKENDAI (The Graduate University for Advanced Studies)\\Department of Particle and Nuclear Physics, Tsukuba, Ibaraki 305-0801, Japan}
\affiliation{Kavli IPMU (WPI), Tokyo University,
\\Kashiwanoha 5-1-5, Kashiwa, Chiba 277-8583, Japan}

\begin{abstract}
The neutrino minimal Standard Model ($\nu$MSM) is the minimum extension
of the standard model. In this model, the Dodelson-Widrow mechanism (DW)
produces keV sterile neutrino dark matter (DM) and the degenerate GeV
heavy Majorana neutrinos lead to leptogenesis.  However, the DW
mechanism has been ruled out by Lyman-$\alpha$ bounds and X-ray
constraints.  An alternative scenario that evades these constraints has
been proposed, where the sterile neutrino DM is generated by the thermal
freeze-in mechanism via a singlet scalar.  In this paper, we consider
a Higgs sector extension of the $\nu$MSM to improve dark matter
sectors and leptogenesis scenarios, focusing on the thermal
freeze-in production mechanism. We discuss various thermal freeze-in
scenarios for the production of keV-MeV sterile neutrino DM with a singlet
scalar, and reinvestigate the Lyman-$\alpha$ bounds and the X-ray
constraints on the parameter regions. Furthermore, we propose thermal
freeze-in leptogenesis scenarios in the extended $\nu$MSM. The singlet
scalar needs to be TeV scale in order to generate the observed DM relic
density and baryon number density with the thermal freeze-in production
mechanism.

\end{abstract}
\maketitle

\section{Introduction}
The standard model of particle physics (SM) has demonstrated great success in high energy physics. However, it is not the complete theory because the Standard Model cannot treat gravity consistently nor explain several observed phenomena, such as neutrino oscillations, cosmological dark matter, the baryon asymmetry of the universe, the horizon and flatness problems, etc. In the dark matter sector, TeV-scale supersymmetry provides weakly interacting massive particles (WIMPs) as leading dark matter candidates~\cite{Vecchi:2013iza}. However, the first run of the LHC experiment excludes a significant region of parameter space for the weak super-partners, and the recent results from LUX~\cite{Faham:2014hza,Akerib:2013tjd} and  XENON100~\cite{Lavina:2013zxa,Aprile:2013doa,2012PhRvL.109r1301A} have severely restricted the WIMP cross section. This situation is the same in other beyond standard models. Feebly Interacting Massive Particles (FIMPs)~\cite{Hall:2009bx,Blennow:2013jba,Yaguna:2011ei,Dev:2013yza} are not constrained by such direct detection experiments due to their much smaller couplings. Therefore, FIMPs are an interesting candidate for dark matter. 

A sterile neutrino can be a FIMP. In particular, keV sterile neutrinos are a candidate for warm dark matter.
The neutrino minimal Standard Model ($\nu$MSM)~\cite{Asaka:2005an,Asaka:2005pn,Canetti:2012kh} has three right-handed neutrinos below the electroweak-scale and the lightest right-handed neutrino may have a mass around the keV scale. The lightest right-handed neutrino $N_{1}$ becomes the keV sterile neutrino dark matter and is produced via active-sterile neutrino oscillation, which is called the Dodelson-Widrow mechanism~\cite{Dodelson:1993je}.  The other heavy right-handed neutrinos $N_ {2}$ and $N_ {3}$ lead to leptogenesis via CP-violating oscillations~\cite{Akhmedov:1998qx,Canetti:2012vf},
where the heavy right-handed neutrinos should satisfy $1\ {\rm GeV} \le  m_{N_{2,3}}\le 20\ {\rm GeV}$. Furthermore, it is possible realise Higgs inflation by introducing a non-minimal coupling between the Higgs and gravity~\cite{Bezrukov:2007ep,Bezrukov:2010jz}. Thus, the $\nu$MSM can explain a large number of phenomena with a minimum number of parameters. 

However, the $\nu$MSM is severely constrained by recent observations.  In particular, there are severe constraints on sterile neutrino DM. The Dodelson-Widrow mechanism is known to be excluded by Lyman-$\alpha$ bounds and X-ray constraints~\cite{Abazajian:2006yn}. To evade these constraints, the production of sterile neutrino DM by the thermal freeze-in mechanism has been considered in Refs.\cite{Shaposhnikov:2006xi,Kusenko:2006rh,Petraki:2007gq,Merle:2013wta,Adulpravitchai:2014xna,Roland:2014vba,Merle:2014xpa,Merle:2015oja}~\footnote{Ref.\cite{Merle:2015oja} gives a comprehensive study of keV sterile neutrino DM via a singlet scalar, but our purpose is to estimate the scale in the extended $\nu$MSM to improve the dark matter sectors and the leptogenesis scenarios rather than such a generic study of the sterile neutrino DM.}. 
Ref.\cite{Shaposhnikov:2006xi} considers a scenario in which the inflaton decays into sterile neutrino DM. Refs.\cite{Kusenko:2006rh,Petraki:2007gq} show that a GeV-scale singlet scalar produces the sterile neutrino DM. Refs.\cite{Merle:2013wta,Adulpravitchai:2014xna} consider the non-thermal production via the decay of a singlet scalar. Furthermore, the singlet scalar can improve the electroweak vacuum stability or the Higgs inflation as well as the dark matter sectors.

In this paper,  we do not discuss the theoretical merits of the singlet scalar in the $\nu$MSM such as the electroweak vacuum stability~\cite{Chen:2012faa,Lebedev:2012zw}, Higgs inflation~\cite{Giudice:2010ka} and scale invariance~\cite{Kang:2014cia}, although we are motivated by these theoretical aspects.  Instead, 
we concentrate on estimating the scale of the singlet scalar to improve the dark matter sector and leptogenesis scenarios. We discuss various thermal freeze-in production scenarios for keV-MeV sterile neutrinos in the extended $\nu$MSM with a singlet scalar. In particular, we revisit the Lyman-$\alpha$ bounds and the X-ray constraints and show that the singlet scalar cannot be heavier than the TeV scale. We also discuss thermal freeze-in leptogenesis scenarios, which are able to produce a larger lepton asymmetry than is produced in thermal leptogenesis due to the contribution from the singlet scalar. In the leptogenesis scenarios, the singlet scalar needs to be lighter than 1 TeV in order to generate the observed baryon asymmetry. 

This paper is organized as follows. In section~\ref{sec:2}, we review the scalar singlet extension of the $\nu$MSM. In section~\ref{sec:3}, we consider two thermal freeze-in scenarios, one utilizing a thermal singlet scalar and the other a non-thermal singlet scalar. In section~\ref{sec:4}, we review the X-ray constraints and the lifetime bounds on the sterile neutrino dark matter. In section~\ref{sec:5}, we investigate the free streaming horizon and the Lyman-$\alpha$ constraints in our scenarios. In section~\ref{sec:6}, we discuss thermal freeze-in leptogenesis with the singlet scalar. Section~\ref{sec:7} is devoted to discussion and conclusions.

\section{The scalar singlet extensions of the $\nu$MSM}\label{sec:2}
In this section, we review the extended $\nu$MSM which contains three right-handed sterile neutrinos $N_{a}$ ($a=1,2,3$) and one real singlet scalar $S$~\footnote{Here we do not consider a complex singlet scalar because in that case a light Nambu-Goldstone boson appears with $U(1)_{L}$ breaking. The presence of such light bosons would make the sterile neutrinos unstable.}.
In this model, the vacuum expectation value (VEV) of $S$ generates a Majorana mass $M_{a}$ for the right-handed neutrino $N_{a}$. The Lagrangian is given as follows,
\begin{equation}
\mathcal{L} ={\mathcal{L}}_{ SM }+\frac { 1 }{ 2 } \left( { \partial  }_{ \mu  }S \right) \left( { \partial  }^{ \mu  }S \right) +i\overline{ { N }_{ a } }\Slash{\partial}{ N }_{ a }- { y }_{ \alpha a }H^{\dagger}\overline { { \ell  }_{ \alpha  } } N_{R_{a}}-\frac { { \kappa  }_{ a } }{ 2 } S\overline { { N }_{ a }^{ c } } { N }_{ a }-V\left( H,S \right) +h.c.,
\end{equation}
where ${ \ell  }_{ \alpha  }$ are the lepton doublets, $H$ is the Higgs doublet, ${ y }_{ \alpha a }$ and ${ \kappa  }_{ a }$ are the Yukawa couplings. $V\left( H,S \right)$ is the Higgs potential.
After spontaneous symmetry breaking, the Higgs doublet and the scalar singlet develop the VEVs $\left< H \right>=\frac { 1 }{ \sqrt { 2 }  }v$ and $\left< S \right> = f$, respectively, where $v$ = 247 GeV and
$S=s+\left< S \right>$.  The right-handed neutrinos acquire the Majorana masses $M_{a}={\kappa_{a}\left< S \right>}$. Without loss of generality, we can choose the mass basis where the Majorana mass term is diagonal. The Lagrangian is written as follows,
\begin{equation}
\mathcal{L} ={\mathcal{L}}_{ SM }+\frac { 1 }{ 2 } \left( { \partial  }_{ \mu  }s \right) \left( { \partial  }^{ \mu  }s \right) +i\overline{ N_{R_{i}} }\Slash{\partial}N_{R_{i}}-{ y }_{ \alpha i 
 }H^{\dagger}\overline { { \ell  }_{ \alpha  } } N_{R_{i}}-\frac { { M }_{ i } }{ 2 }\overline { N_{R_{i}}^{ c } }N_{R_{i}}-V\left( H,S \right) +h.c..
\end{equation} 
If the Dirac masses are much smaller than the Majorana masses, then, as a result of the type I seesaw mechanism, the left-handed neutrino masses can be expressed as follows,
\begin{gather}
{ m }_{ \nu  }\simeq{ m }_{ D }{ M }^{ -1 }{ m }_{ D }\simeq \frac { { \left( y\left< H \right>  \right)  }^{ 2 } }{ {\kappa \left< S \right>}}, \label{eq:3} \\ { m }_{ N }\simeq{ M\simeq  }{\kappa}\left< S \right>,  \ \ \theta \simeq \frac { { m }_{ D } }{ M }. \label{eq:4}
\end{gather}
For the scalar potential $V\left( H,S \right)$, we impose the softly broken discrete symmetry $\mathbb{Z}_{2}$, where the scalar singlet is $\mathbb{Z}_{2}$-odd ($S\rightarrow-S$) and all the other fields are $\mathbb{Z}_{2}$-even.  We can then construct the following scalar potential with even powers,
\begin{equation}
V\left( H,S \right) =-{ \mu  }_{ H }^{ 2 }{ H }^{ \dagger  }H-\frac { 1 }{ 2 } { \mu  }_{ S }^{ 2 }{ S }^{ 2 }+{ \lambda  }_{ H }{ \left( { H }^{ \dagger  }H \right)  }^{ 2 }+\frac { 1 }{ 4 } { \lambda  }_{ S }{ S }^{ 4 }+2\lambda \left( { H }^{ \dagger  }H \right) { S }^{ 2 }+\omega S,
\label{eq:5}
\end{equation}
where $\omega S$ is a soft $\mathbb{Z}_{2}$ breaking term. The spontaneous breaking of the discrete symmetries $\mathbb{Z}_{N}$ could produce domain walls~\cite{Zeldovich:1974uw}. The soft $\mathbb{Z}_{2}$ breaking term $\omega S$ makes the vacua of the singlet scalar degenerate so that the domain wall problem is evaded~\cite{Zeldovich:1974uw,Vilenkin:1984ib,Kibble:1976sj}. 
The minima of the scalar potential are given by the following equations, 
\begin{equation}
\begin{cases} { \mu  }_{ H }^{ 2 }={ \lambda  }_{ H }{ v }^{ 2 }+2\lambda { f }^{ 2 }, \\ { \mu  }_{ S }^{ 2 }={ \lambda  }_{ S }{ f }^{ 2 }+2\lambda { v }^{ 2 }+\frac { \omega  }{ f }.  \end{cases}
\end{equation}
The mass eigenstates of the Higgs and singlet scalar are $h$ and $s$, where $h$ approximately corresponds to the SM higgs boson. The physical masses of $h$ and $s$ are given by, 
\begin{gather}
{ m }_{ h }^{ 2 }\simeq{ \lambda  }_{ H }{ v }^{ 2 }-\frac { { \left( 2\lambda fv \right)  }^{ 2 } }{ { \lambda  }_{ S }f^{ 2 }-{ \lambda  }_{ H }{ v }^{ 2 } },\\ 
{ m }_{ s }^{ 2 }\simeq{ \lambda  }_{ S }{ f }^{ 2 }+\frac { { \left( 2\lambda fv \right)  }^{ 2 } }{ { \lambda  }_{ S }f^{ 2 }-{ \lambda  }_{ H }{ v }^{ 2 } }.     
\end{gather}
The Higgs portal coupling $\lambda$ induces doublet-singlet mixing.
In this paper, we consider a TeV-scale singlet scalar which decays into sterile neutrino DM and heavy Majorana neutrinos. Therefore, there is essentially no constraint on the coupling $\lambda$. However, the size of the coupling $\lambda$ still affects the thermal history of $s$. The references~\cite{Petraki:2007gq,McDonald:1993ex} show that $s$ is out of thermal equilibrium if the Higgs portal coupling satisfies $\lambda \ll 10^{-6}$. 
In this paper, we assume that the singlet scalar is out of thermal equilibrium for $\lambda \ll 10^{-6}$ and concentrate on the thermal freeze-in production mechanism~\footnote{If the singlet scalar is not directly produced from inflatons and the reheating temperature is low enough, the singlet scalar can be out of thermal equilibrium even if  $\lambda > 10^{-6}$.}.

\section{Sterile neutrino dark matter from the thermal freeze-in production mechanism}\label{sec:3}
In this section, we consider various scenarios for the production of sterile neutrino DM in the extended $\nu$MSM with a singlet scalar. Sterile neutrinos could be produced by thermal freeze-out, thermal freeze-in or non-thermal decay. These scenarios also depend on whether the singlet scalar is generated by freeze-out or freeze-in. In addition, the Dodelson-Widrow mechanism can produce sterile neutrinos via active-sterile oscillations. 

It is possible to constrain these scenarios using the mass relation of the seesaw mechanism. For simplicity, we assume that the lightest right-handed neutrino $N_{1}$ is sterile neutrino DM, with a mass of about $10$ keV. We will see later that a sterile neutrino with mass above 1 MeV is not favored by X-ray constraints and lifetime bounds. If the Yukawa coupling of the singlet scalar and the right-handed neutrino is $\kappa_{1}\approx 10 ^ {- 8} $ and the vacuum expectation value of the singlet scalar is $\left< S \right> \approx 1$ TeV, then the following relations can be derived from the seesaw mechanism,
\begin{gather}
{ m }_{ \nu  }\simeq { m }_{ D }{ M }^{ -1 }{ m }_{ D }\simeq \frac { { \left( y\left< H \right>  \right)  }^{ 2 } }{ {\kappa_{1} \left< S \right>}}\simeq y ^{ 2 }\left( { 10 }^{ 18 }\ {\rm eV} \right), \\ { m }_{ N_{1} }\simeq{ M_{1}\simeq  }{\kappa_{1}} \left< S \right>\simeq 10\ {\rm keV}.
\end{gather}
The Yukawa couplings $y$, $\kappa_{1}$ are very small $y\approx 10 ^ {-10}$ and $\kappa_{1}\approx 10 ^ {- 8}$. If the reheating temperature $T_{RE}$ satisfies $T_{RE} \lesssim 10^{16}\  \rm{GeV}$, the sterile neutrino DM does not come into thermal equilibrium for $\kappa_{1} \ll 10^{-6}$~\cite{Vilja:1993uw,Enqvist:1992va}. Therefore, we may regard the sterile neutrino DM as non-thermal particles in the early universe. In such a case, we find only two realistic dark matter scenarios to realize keV-MeV-scale sterile neutrinos.  We will now proceed to describe these two scenarios.

\subsection{The singlet scalar is in thermal equilibrium}\label{sec:3a}
If the Higgs portal coupling is relatively large $\lambda > 10^{-6}$, the singlet scalar $s$ enters into thermal equilibrium and the sterile neutrino DM can be produced via the thermal freeze-in of $s$. In addition, $h$ couples to $N_{1}\nu$ with suppressed coupling, so after the EW symmetry breaking there is a small mixing between $\nu_{L}$ and $s$. To check the effect of this mixing we consider the $h\rightarrow \nu_{e}N_{1}$ contribution to sterile neutrino production as well.  
The production by the singlet scalar has been considered in Ref.\cite{Kusenko:2006rh,Petraki:2007gq}. The thermal freeze-in production is caused by the Yukawa interaction of $s$ and $N_{1}$ or $h$ and $N_{1}$. Under the assumption $m_{s}\gg m_{h}$, as the universe is expanding the temperature becomes low and $s$ disappears first. The Higgs boson $h$, however, is still in thermal equilibrium and thermal freeze-in production by $h$ is effective until $T\sim m_{h}$.  

The dark matter yield can be calculated by solving the Boltzmann equations. In this scenario, the relevant Boltzmann equations for ${ Y }_{ { N }_{ 1 } }=n_{N_{1}}/s$ are given as follows,
\begin{equation}
\frac { d{ Y }_{ { N }_{ 1 } } }{ dT } =\frac { d{ Y }_{ { N }_{ 1 } }^{ Ds }}{ dT } +\frac { d{ Y }_{ { N }_{ 1 } }^{ Dh  }}{ dT }, 
\end{equation}
where ${ Y }_{ { N }_{ 1 } }^{ Ds } ({ Y }_{ { N }_{ 1 } }^{ Dh })$ is ${ Y }_{ { N }_{ 1 } }$ from $s$ ($h$) decays respectively.

\begin{figure*}[t]
	\begin{tabular}{ccc}
	\begin{minipage}{.33\textwidth}
		\centering
		\subfigure[$\ h\rightarrow \nu_{e}N_{1}$]{
		\includegraphics[width=57mm]{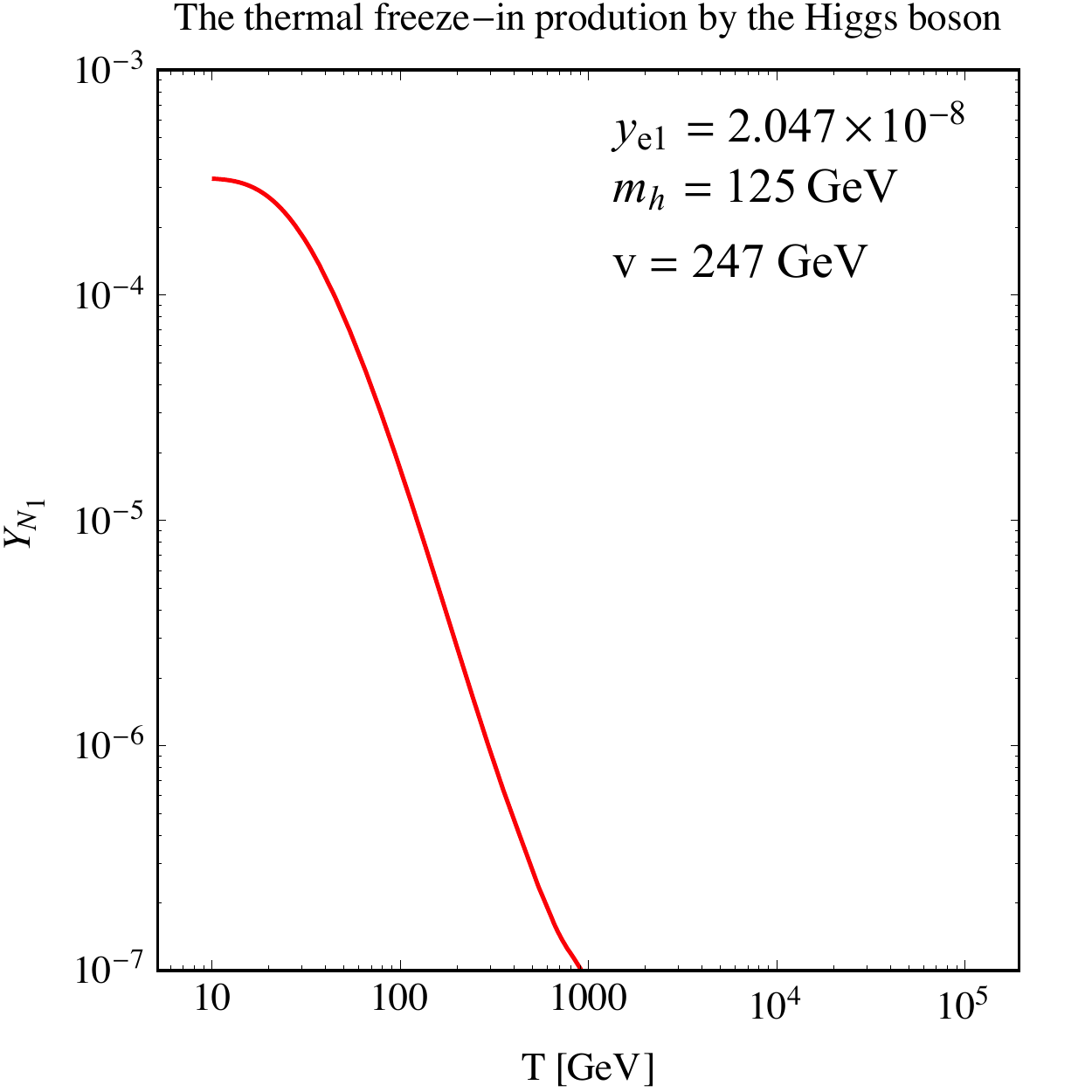}
		\label{Fig:1a}}
	\end{minipage}
		\begin{minipage}{.33\textwidth}
		\centering
		\subfigure[$\ s\rightarrow N_{1}N_{1}$]{
		\includegraphics[width=57mm]{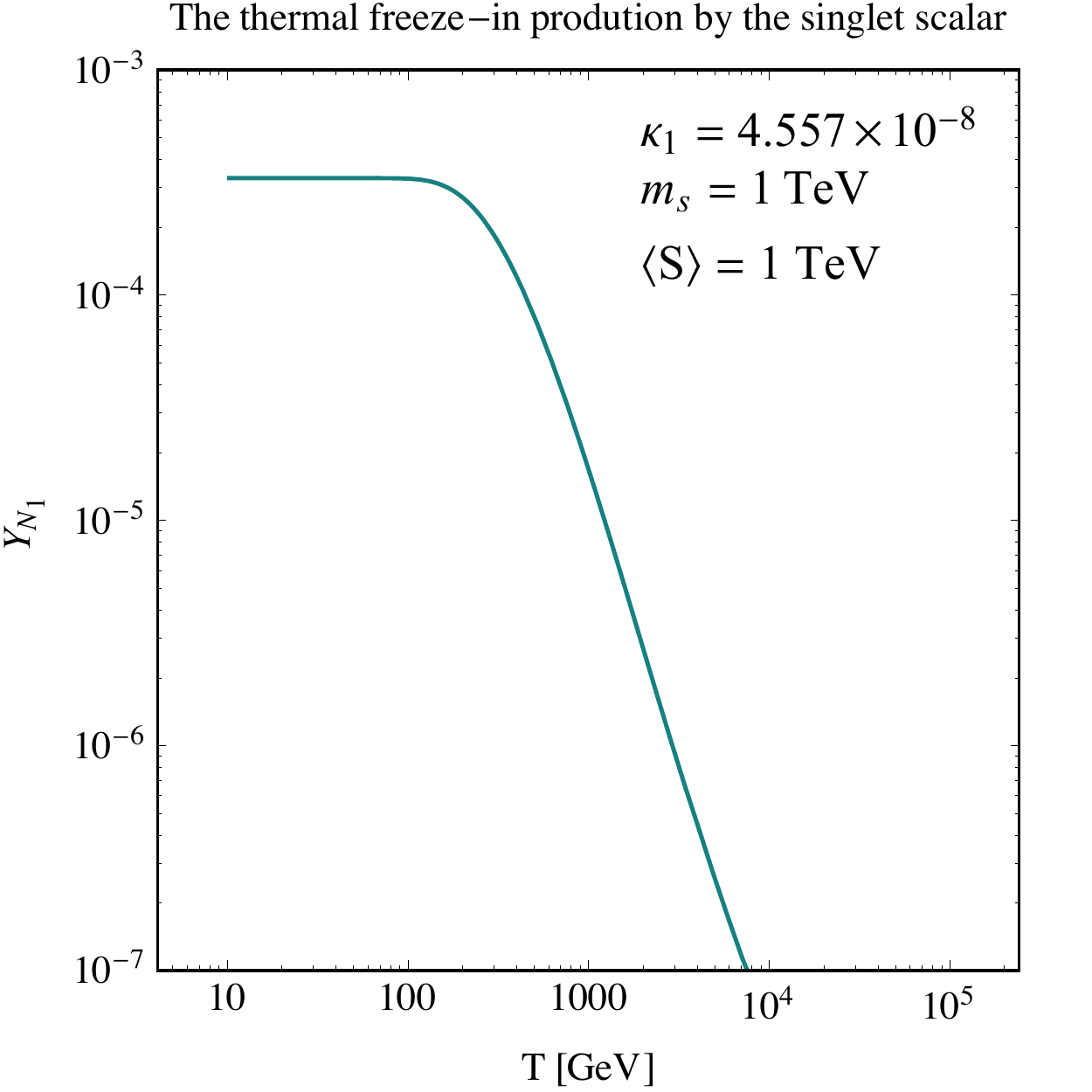}
		\label{Fig:1b}}
	\end{minipage}
		\begin{minipage}{.33\textwidth}
		\centering
		\subfigure[$\ h\rightarrow s\rightarrow N_{1}N_{1}$]{
	        \includegraphics[width=57mm]{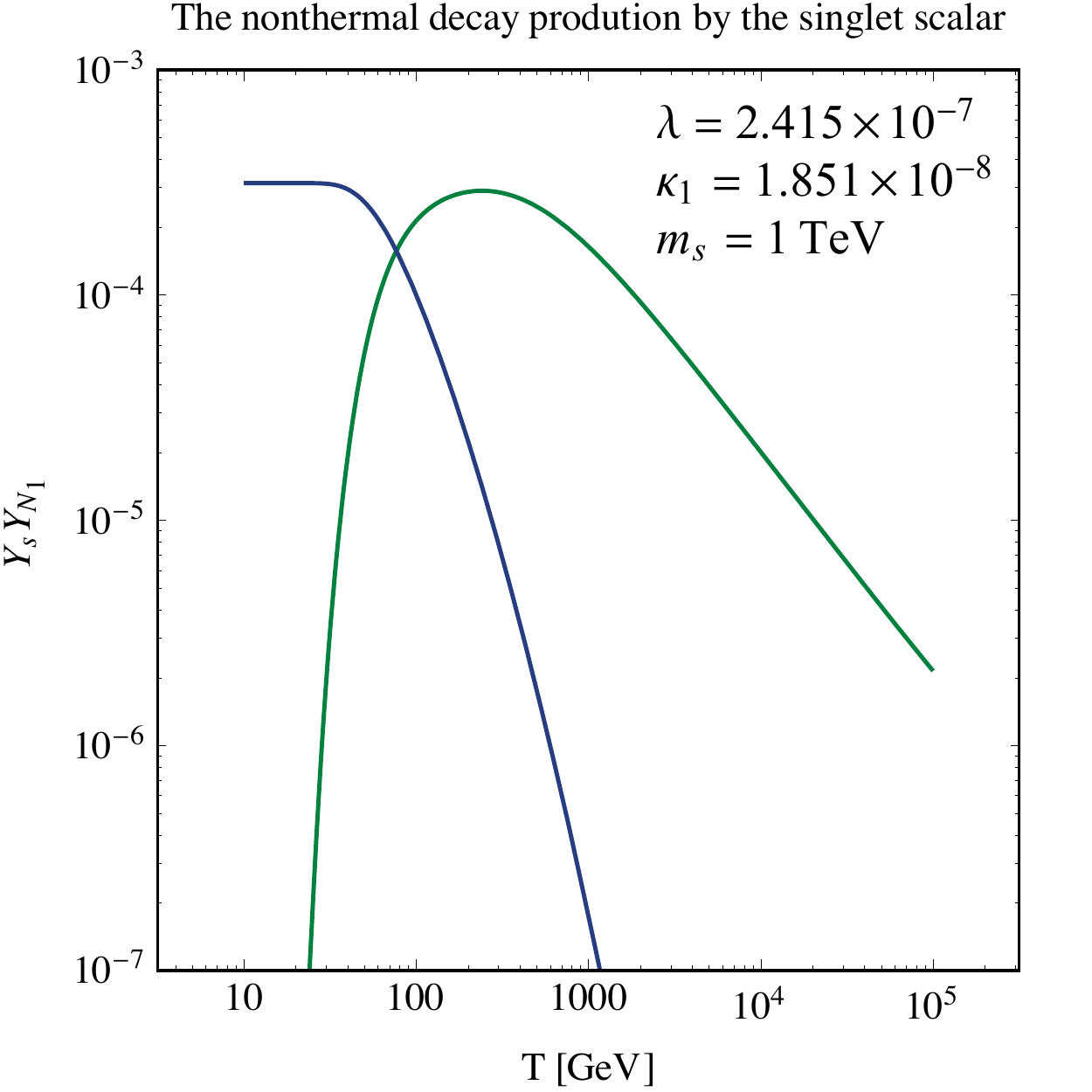}
		\label{Fig:1c}}
	\end{minipage}
        \end{tabular}
\caption{We describe the evolution of the yields ${ Y }_{ { N }_{ 1 } }$ and  ${ Y }_{ s }$ as the temperature $T$ decreases. The three thermal freeze-in production processes are shown in FIG.\ref{Fig:1a}, \ref{Fig:1b} and \ref{Fig:1c}. The sterile neutrino is generated by the thermal freeze-in mechanism of the Higgs boson (FIG.\ref{Fig:1a}), by the thermal freeze-in production of the singlet scalar (FIG.\ref{Fig:1b}), by non-thermal decay production of the singlet scalar (FIG.\ref{Fig:1c}). }\label{Fig:1}
\end{figure*}
In FIG.\ref{Fig:1a}, FIG.\ref{Fig:1b} and FIG.\ref{Fig:1c}, we show numerical results for the evolution of the sterile neutrino yield ${ Y }_{ { N }_{ 1 } }$ and the singlet scalar yield ${ Y }_{ s }$ for various thermal freeze-in mechanisms.  Sterile neutrino DM is generated by the thermal freeze-in production of $h$ in FIG.\ref{Fig:1a}, the thermal freeze-in production of $s$ in FIG.\ref{Fig:1b}, and the non-thermal decay production of $s$ in FIG.\ref{Fig:1c}.

Now, $Y_{N_{1}}$ is obtained from the following calculation. The Boltzmann equation 
for the sterile neutrino number density $n_{N_{1}}$ involving $s$ is written as,
\begin{eqnarray}  \frac { d }{ dt } { n }_{ { N }_{ 1 } }+3H{ n }_{ { N }_{ 1 } }&&=\sum _{ spin }{ \int { d{ \pi  }_{ s } } d{ \pi  }_{ { N }_{ 1 } } } d{ \pi  }_{ { N' }_{ 1 } }{ \left( 2\pi  \right)  }^{ 4 }{ \delta  }^{ \left( 4 \right)  }\left( { p }_{ s }-{ p }_{ { N }_{ 1 } }-{ p }_{ { N' }_{ 1 } } \right) \nonumber \\ && \left\{ {\left| M \right| }_{ s\rightarrow{ N }_{ 1 }{N' }_{ 1 } }^{ 2 }{ f }_{ s }\left( 1-{ f }_{ { N }_{ 1 } } \right) \left( 1-{ f }_{ { N' }_{ 1 } } \right)   -{ \left| M \right| }_{ { N }_{ 1 }{N' }_{ 1 }\rightarrow s }^{ 2 }{ f }_{ { N }_{ 1 } }{ f }_{ { N' }_{ 1 } }\left( 1-{ f }_{ s } \right) \right\}.  \end{eqnarray} 
We assume that the initial abundance of sterile neutrinos can be neglected and the singlet scalar enters into thermal equilibrium, such that, 
\begin{eqnarray}
\frac { d }{ dt } { n }_{ { N }_{ 1 } }+3H{ n }_{ { N }_{ 1 } }&=&\sum _{ spin }{ \int { d{ \pi  }_{ s } } d{ \pi  }_{ { N }_{ 1 } } } d{ \pi  }_{ { N' }_{ 1 } }{ \left( 2\pi  \right)  }^{ 4 }{ \delta  }^{ \left( 4 \right)  }\left( { p }_{ s }-{ p }_{ { N }_{ 1 } }-{ p }_{ { N' }_{ 1 } } \right) {\left| M \right| }_{ s\rightarrow{ N }_{ 1 }{N' }_{ 1 } }^{ 2 }{ f }_{ s }^{eq} \nonumber \\
&=&2\frac{K_{1}\left(m_{s}/T\right)}{K_{2}\left(m_{s}/T\right)}\Gamma \left( s\rightarrow { N }_{ 1 }{ N }_{ 1 } \right). 
\end{eqnarray} 
The sterile neutrino yield $Y_{N_{1}}=\frac{n_{N_{1}}}{s}$ can be obtained from the entropy density $s=\frac{2\pi^{2}}{45}h_{\rm eff}T^{3}$, and satisfies the following equation,
\begin{eqnarray}
\frac { d{ Y }_{ { N }_{ 1 } }^{  Ds } }{ dT } &=&-\sqrt { \frac { 45 }{ { \pi  }^{ 3 }{ G }_{ N } }  } \frac { 1 }{ \sqrt { { g }_{\rm eff} }  } \frac { 1 }{ { T }^{ 3 } } \frac { { K }_{ 1 }\left( { { m }_{ s } }/{ T } \right)  }{ { K }_{ 2 }\left( { { m }_{ s } }/{ T } \right)  } \Gamma \left( s\rightarrow { N }_{ 1 }{ N }_{ 1 } \right) { Y }_{ s }^{ eq } \nonumber  \\
&=& -\frac { 135\sqrt { 5 }  }{ 4{ \pi  }^{ { 11 }/{ 2 }  } } \frac { { m }_{\rm pl } }{ { h }_{ \rm eff }\sqrt { { g }_{\rm eff } }  } \frac { { m }_{ s }^{ 2 }{ K }_{ 1 }\left( { { m }_{ s } }/{ T } \right)  }{ { T }^{ 5 } } \Gamma \left( s\rightarrow { N }_{ 1 }{ N }_{ 1 } \right).
\end{eqnarray}
Similarly, $h$ also produces ${ N }_{ 1 }$ as the following,
\begin{eqnarray}
\frac { d{ Y }_{ { N }_{ 1 } }^{ Dh  }}{ dT } &=&-\sqrt { \frac { 45 }{ { 4\pi  }^{ 3 }{ G }_{ N } }  } \frac { 1 }{ \sqrt { { g }_{ \rm eff } }  } \frac { 1 }{ { T }^{ 3 } } \frac { { K }_{ 1 }\left( { { m }_{ h } }/{ T } \right)  }{ { K }_{ 2 }\left( { { m }_{ h } }/{ T } \right)  } \Gamma \left( h\rightarrow { N }_{ 1 }\nu _{ e } \right) { Y }_{ h }^{ eq } \nonumber  \\ 
&=& -\frac { 135\sqrt { 5 }  }{ 8{ \pi  }^{ { 11 }/{ 2 } } } \frac { { m }_{ \rm pl } }{ { h }_{ \rm eff }\sqrt { { g }_{ \rm eff } }  } \frac { { m }_{ h }^{ 2 }{ K }_{ 1 }\left( { { m }_{ h } }/{ T } \right)  }{ { T }^{ 5 } } \Gamma \left( h \rightarrow { N }_{ 1 }{ \nu  }_{ e } \right),
\end{eqnarray}
where $m_{\rm pl}= 1.22 \times 10^{19}\ \rm{GeV} $ is the planck mass, ${ g }_{ \rm eff }$ and ${ h }_{\rm  eff }$ are the effective number of degrees of freedom for energy and entropy and ${ K }_{ n }\left( x \right)$ is the modified Bessel function of the second kind.  The equilibrium yields ${ Y }_{ s, h }^{ eq }$ are expressed as,
\begin{gather}
{ Y }_{ s }^{ eq }=\frac { 45{ g }_{ s }{ m }_{ s }^{ 2 } }{ 4{ \pi  }^{ 4 }{ T }^{ 2 } } \frac { { K }_{ 2 }\left( { { m }_{ s} }/{ T } \right)  }{ { h }_{ \rm eff } },\\
{ Y }_{ h }^{ eq }=\frac { 45{ g }_{ h }{ m }_{ h }^{ 2 } }{ 4{ \pi  }^{ 4 }{ T }^{ 2 } } \frac { { K }_{ 2 }\left( { { m }_{ h } }/{ T } \right)  }{ { h }_{ \rm eff } }. 
\end{gather}
The partial decay width of $s$ into $N_{1}N_{1}$ is obtained as,
\begin{equation}
\Gamma \left( s\rightarrow { N }_{ 1 }{ N }_{ 1 }\right) =\frac { { \kappa  }_{ 1 }^{ 2 }{ m }_{ s } }{ 16\pi  } \left( 1-\frac { 4{ { M }_{ { N }_{ 1 } }^{ 2 } } }{ { { m }_{ s }^{ 2 } } }  \right)^{{ 3 }/{ 2 }}\approx \frac { { \kappa  }_{ 1 }^{ 2 }{ m }_{ s } }{ 16\pi  }, 
\end{equation}
and the partial decay width of $h$ into $N_{1}{ \nu  }_{ e }$ is given as,
\begin{equation}
\Gamma \left( h\rightarrow { N }_{ 1 }{ \nu  }_{ e } \right) =\frac { y_{ e1 }^{ 2 }{ m }_{ h } }{ 8\pi  } \left( 1-\frac { { { M }_{ { N }_{ 1 } }^{ 2 } } }{ { { m }_{ h }^{ 2 } } }  \right)^{{ 3 }/{ 2 }}\approx \frac { y_{ e1 }^{ 2 }{ m }_{ h } }{ 8\pi  }.
\end{equation} 

\begin{figure*}[t]
	\begin{tabular}{cc}
	\begin{minipage}{0.5\hsize}
		\centering
		\subfigure[$\ s\rightarrow N_{1}N_{1}\quad \left< S \right>=10\ { \rm{TeV}}$]{
		\includegraphics[width=78mm]{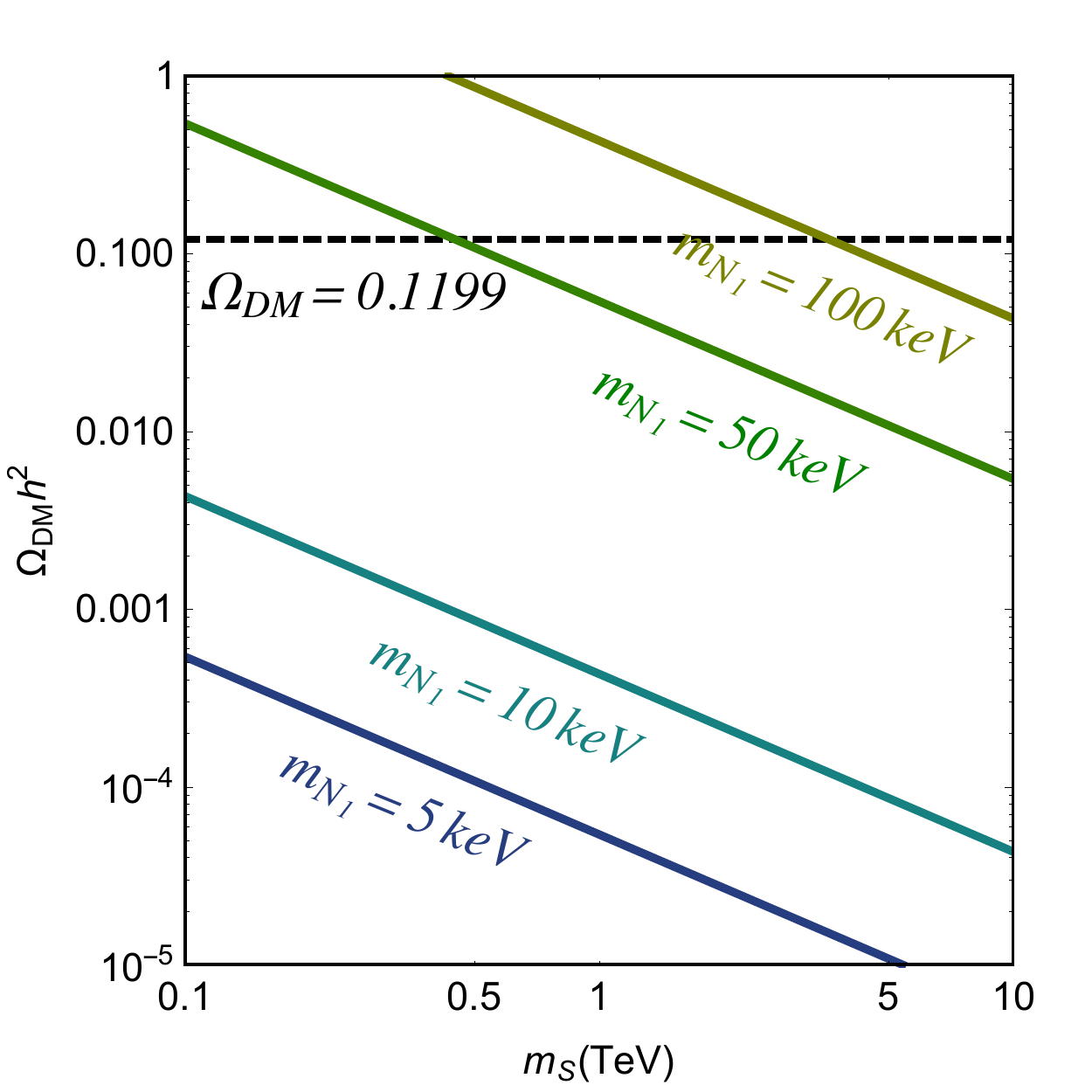}
		\label{Fig:2a}}
	\end{minipage}
   \begin{minipage}{0.5\hsize}
		\centering
		\subfigure[$\ s\rightarrow N_{1}N_{1}\quad \left< S \right>=1000\ { \rm{TeV}}$]{
		\includegraphics[width=78mm]{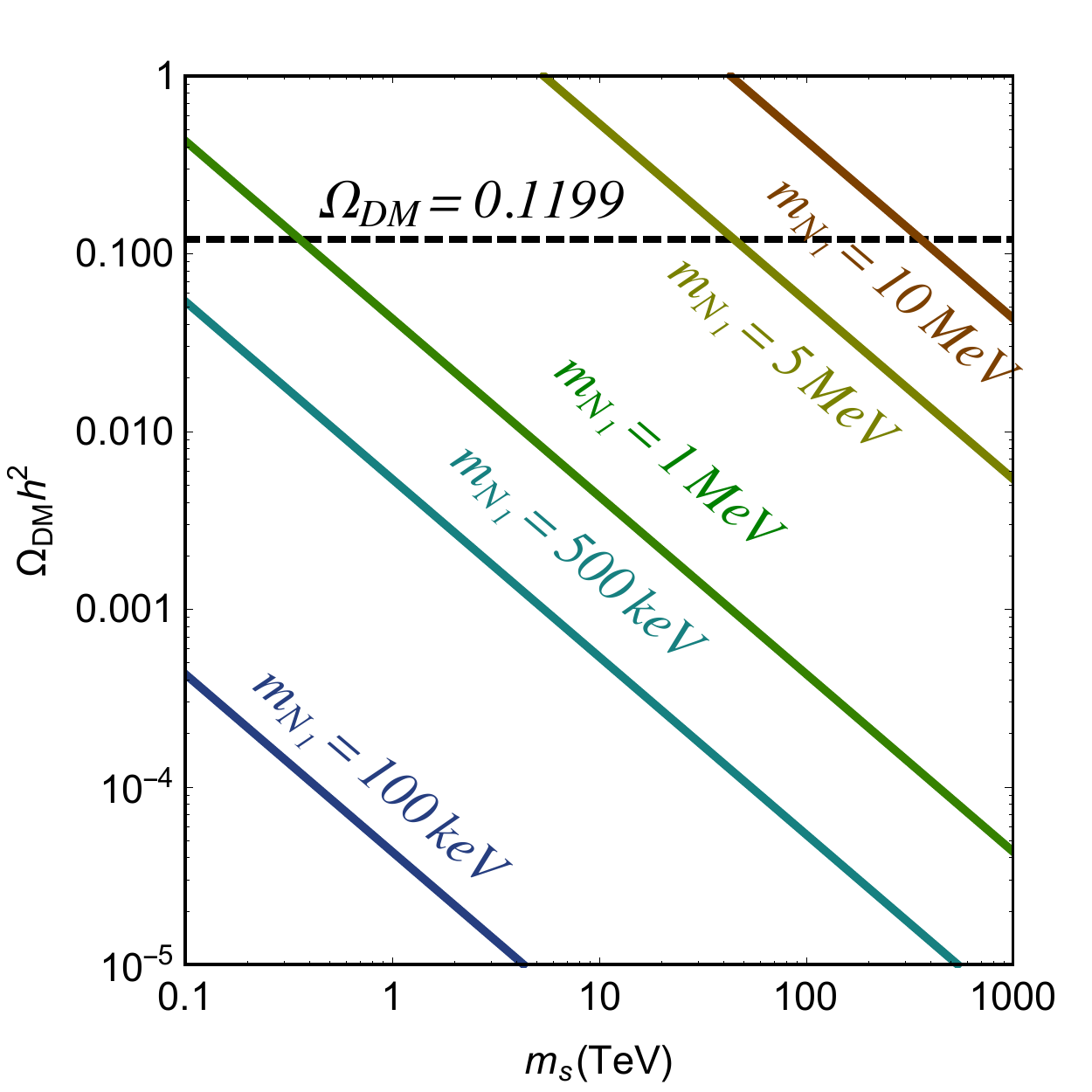}
		\label{Fig:2b}}
	\end{minipage}	
\end{tabular}
\caption{The relic density of sterile neutrinos
as a function of $m_{s}$ for different values of $m_{N_{1}}$ and $\left< S \right>$ in the case of thermal freeze-in production via $s$.}\label{Fig:2}
\end{figure*}

In order to estimate the yield, we analytically integrate the relevant Boltzmann equations. The yield at the temperature of the universe today $Y_{N_{1}}\left( T_{0}\right)$ is given for $s\rightarrow { N }_{ 1 }{ N }_{ 1 }$ as,
\begin{eqnarray}
{ Y }_{ { N }_{ 1 } }^{ { D }_{ s } }\left( { T }_{ 0 } \right) &=&-\frac { 135\sqrt { 5 }  }{ 4{ \pi  }^{ { 11 }/{ 2 } } } \frac { { m }_{ \rm pl } }{ { h }_{ \rm eff }\sqrt { g_{ \rm eff } }  } \int _{ { T }_{ RE } }^{ { T }_{ 0 } }{ \frac { { m }_{ s }^{ 2 }{ K }_{ 1 }\left( { { m }_{ s } }/{ T } \right)  }{ { T }^{ 5 } }  }\Gamma \left( s\rightarrow { N }_{ 1 }{ N }_{ 1 } \right) dT \nonumber \\
&\approx&-\frac { 135\sqrt { 5 }  }{ 4{ \pi  }^{ { 11 }/{ 2 } } } \frac { { m }_{ \rm pl } }{ { h }_{ \rm eff }\sqrt { g_{ \rm eff } }  } \int _{ \infty }^{ 0  }{ \frac { { m }_{ s }^{ 2 }{ K }_{ 1 }\left( { { m }_{ s } }/{ T } \right)  }{ { T }^{ 5 } }  }\Gamma \left( s\rightarrow { N }_{ 1 }{ N }_{ 1 } \right) dT \nonumber \\ 
&\approx&1.58\times{ 10 }^{ 14 }\left(\frac { { m }_{ { N }_{ 1 } }^{ 2 } }{ { m }_{ s } }\right)\left(\frac {1}{\left< S \right> }\right)^{2},
\end{eqnarray}
where we assume $h_{\rm eff}\approx g_{\rm eff}\approx 100$. The relic density of  the sterile neutrino DM can be obtained as,
\begin{eqnarray}
{ \Omega  }_{ { N }_{ 1 } }^{ { D }_{ s } }{ h }^{ 2 }&=&2.733\times { 10 }^{ 8 }\cdot { Y }_{ 0 }\cdot \left( \frac { { m }_{ DM } }{ \rm{GeV} }  \right) \nonumber 
\\ &=&4.32\times { 10 }^{ -5 }{ \left( \frac { { m }_{ { N }_{ 1 } } }{ \rm{keV} }  \right)  }^{ 3 }{ \left( \frac { \rm{TeV} }{ { m }_{ s } }  \right)  }\left(\frac { \ \rm{TeV}}{ \left< S \right> }\right)^{2}.\label{eq:21}
\end{eqnarray}
The DM relic density observed by Planck+WP~\cite{Ade:2013zuv} is estimated as,
\begin{equation}
{ \Omega  }_{ DM }{ h }^{ 2 }=0.1199\pm0.0027.
\end{equation}
The sterile neutrino mass required to explain the observed DM relic density is thus ${ m }_{{ N }_{ 1 }}\approx10\ \rm{keV}$ for $\left< S \right> \approx { m }_{ s }\approx 1\ \rm{TeV}$ and ${ m }_{{ N }_{ 1 }}\approx1 \ \rm{MeV}$ for $\left< S \right> \approx { m }_{ s }\approx 100\ \rm{TeV}$. FIG.\ref{Fig:2} shows the relic density of sterile neutrinos as a function of ${ m }_{ s }$ for different values of $m_{N_{1}}$. 

Similarly, we can integrate the Boltzmann equation of thermal freeze-in production via $h$, obtaining,
\begin{eqnarray}
{ Y }_{ { N }_{ 1 } }^{ { D }_{ h } }\left( { T }_{ 0 } \right) &=&-\frac { 135\sqrt { 5 }  }{ 8{ \pi  }^{ { 11 }/{ 2 } } } \frac { { m }_{ \rm pl } }{ { h }_{ \rm eff }\sqrt { g_{ \rm eff } }  } \int _{ { T }_{ RE } }^{ { T }_{ 0 } }{ \frac { { m }_{ h }^{ 2 }{ K }_{ 1 }\left( { { m }_{ h } }/{ T } \right)  }{ { T }^{ 5 } }  }\Gamma \left( h\rightarrow { N }_{ 1 }{ \nu  }_{ e } \right) dT \nonumber \\ 
&\approx&-\frac { 135\sqrt { 5 }  }{ 8{ \pi  }^{ { 11 }/{ 2 } } } \frac { { m }_{ \rm pl } }{ { h }_{ \rm eff }\sqrt { g_{\rm eff } }  } \int _{ \infty }^{ 0  }{ \frac { { m }_{ h }^{ 2 }{ K }_{ 1 }\left( { { m }_{ h } }/{ T } \right)  }{ { T }^{ 5 } }  }\Gamma \left( h\rightarrow { N }_{ 1 }{ \nu  }_{ e } \right) dT \nonumber \\
&\approx&1.04\times{ 10 }^{ 7 }\cdot \sin ^{ 2 }{ \theta  } \cdot { m }_{ { N }_{ 1 } }^{ 2 },
\end{eqnarray}
with the relic density of $N_{1}$ given as,
\begin{equation}
{ \Omega  }_{ { N }_{ 1 } }^{ { D }_{ h } }{ h }^{ 2 }=2.84\times { 10 }^{ -3 }\sin ^{ 2 }{ \theta  } { \left( \frac { { m }_{ { N }_{ 1 } } }{ \rm{keV} }  \right)  }^{ 3 }.\label{eq:24}
\end{equation}   
Finally, sterile neutrino DM is also produced by the thermal background of active neutrinos via coherent scattering (Dodelson-Widrow mechanism). The dark matter relic density is found to be given as~\cite{Abazajian:2005gj},
\begin{equation}
{ \Omega  }_{ { N }_{ 1 } }^{ DW }{ h }^{ 2 }=5.47\times { 10 }^{ 7 }\sin ^{ 2 }{ \theta  } { \left( \frac { { m }_{ { N }_{ 1 } } }{ \rm{keV} }  \right)  }^{1.63}.\label{eq:25}\end{equation}

In the case of keV-MeV-scale sterile neutrino DM, the contribution to ${ \Omega  }_{ { N }_{ 1 } }^{ DW }$ given in Eq.(\ref{eq:25}) is larger than that from the thermal freeze-in production via $h$ given in Eq.(\ref{eq:24}). In fact, there are additional contributions of the same order as those given in Eq.(\ref{eq:24}) that come from the decay of Z-bosons or W-bosons due to neutrino mixing, but we can safely ignore these contributions in the mass region under consideration. Altogether, the total sterile neutrino DM relic density is given as,
\begin{eqnarray}
{ \Omega  }_{ { N }_{ 1 } }{ h }^{ 2 }&=&4.32\times { 10 }^{ -5 }{ \left( \frac { { m }_{ { N }_{ 1 } } }{ \rm{keV} }  \right)  }^{ 3 }{ \left( \frac { \rm{TeV} }{ { m }_{ s } }  \right)  }\left(\frac { \rm{TeV}}{ \left< S \right> }\right)^{2}\nonumber \\ &&+5.47\times { 10 }^{ 7 }\sin ^{ 2 }{ \theta  } { \left( \frac { { m }_{ { N }_{ 1 } } }{ \rm{keV} }  \right)  }^{1.63}.
\end{eqnarray}
When the mass and the VEV of the singlet scalar are of order 1 TeV, the dominant mechanism for production of keV-MeV-scale sterile neutrinos is via the thermal freeze-in of the singlet scalar.

\subsection{The singlet scalar is out of thermal equilibrium}\label{sec:3b}
In this case $\lambda \ll 10^{-6}$ and both $s$ and $N_{1}$ never enter into thermal equilibrium in the early universe. Sterile neutrino DM is generated via the thermal freeze-in of $h$. The singlet scalar $s$ is also generated by thermal freeze-in production, and then proceeds to decay efficiently into $N_{1}$. 
In this section, we assume $m_{s} \ll m_{N_{2,3}}$, so that $s$ cannot decay into $N_{2}$ and $N_{3}$.

To calculate the yield of sterile neutrinos, we have to solve the Boltzmann equations given by the following two interrelated equations, 
\begin{eqnarray}
\frac { d{ Y }_{ s } }{ dT } &=&\frac { d{ Y }_{ s }^{ A } }{ dT } +\frac { d{ Y }_{ s }^{ Ds } }{ dT }\label{eq:27},\\ \nonumber \\
 \frac { d{ Y }_{ { N }_{ 1 } } }{ dT } &=&\frac { d{ Y }_{ { N }_{ 1 } }^{ Ds } }{ dT } +\frac { d{ Y }_{ { N }_{ 1 } }^{ Dh } }{ dT },
 \ \   \frac { d{ Y }_{ s }^{ Ds } }{ dT } =-\frac { 1 }{ 2 } \frac { d{ Y }_{ { N }_{ 1 } }^{ Ds } }{ dT }\label{eq:28}. 
\end{eqnarray}

\begin{figure*}[t]
	\begin{tabular}{cc}
	\begin{minipage}{0.5\hsize}
		\centering
		\subfigure[$\ h\rightarrow s\rightarrow N_{1}N_{1}\quad \lambda = 10^{-7}$]{
		\includegraphics[width=78mm]{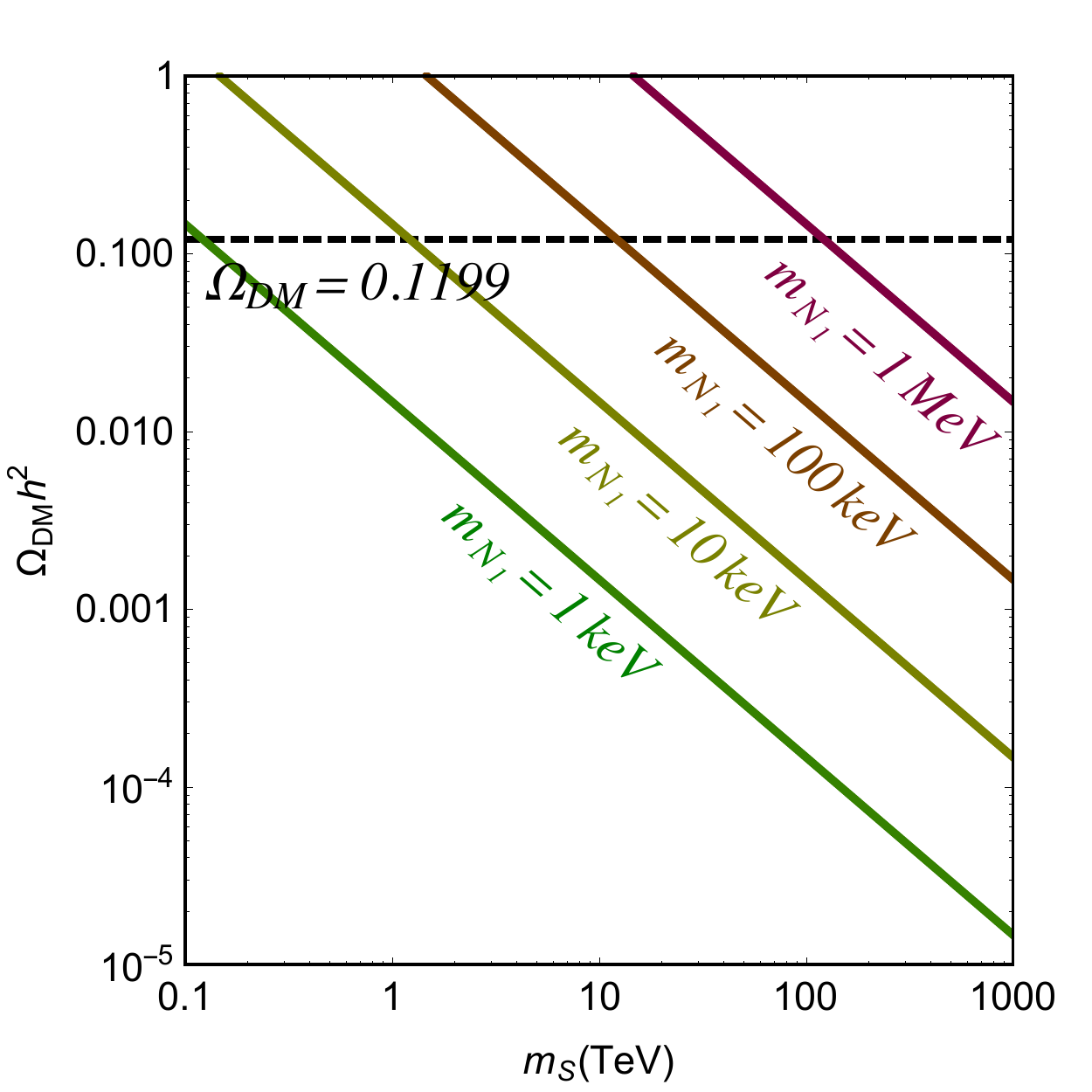}
		\label{Fig:3a}}
	\end{minipage}
   \begin{minipage}{0.5\hsize}
		\centering
		\subfigure[$\ h\rightarrow s\rightarrow N_{1}N_{1}\quad { m }_{ { N }_{ 1 } }=10\ { \rm{keV}}$]{
		\includegraphics[width=78mm]{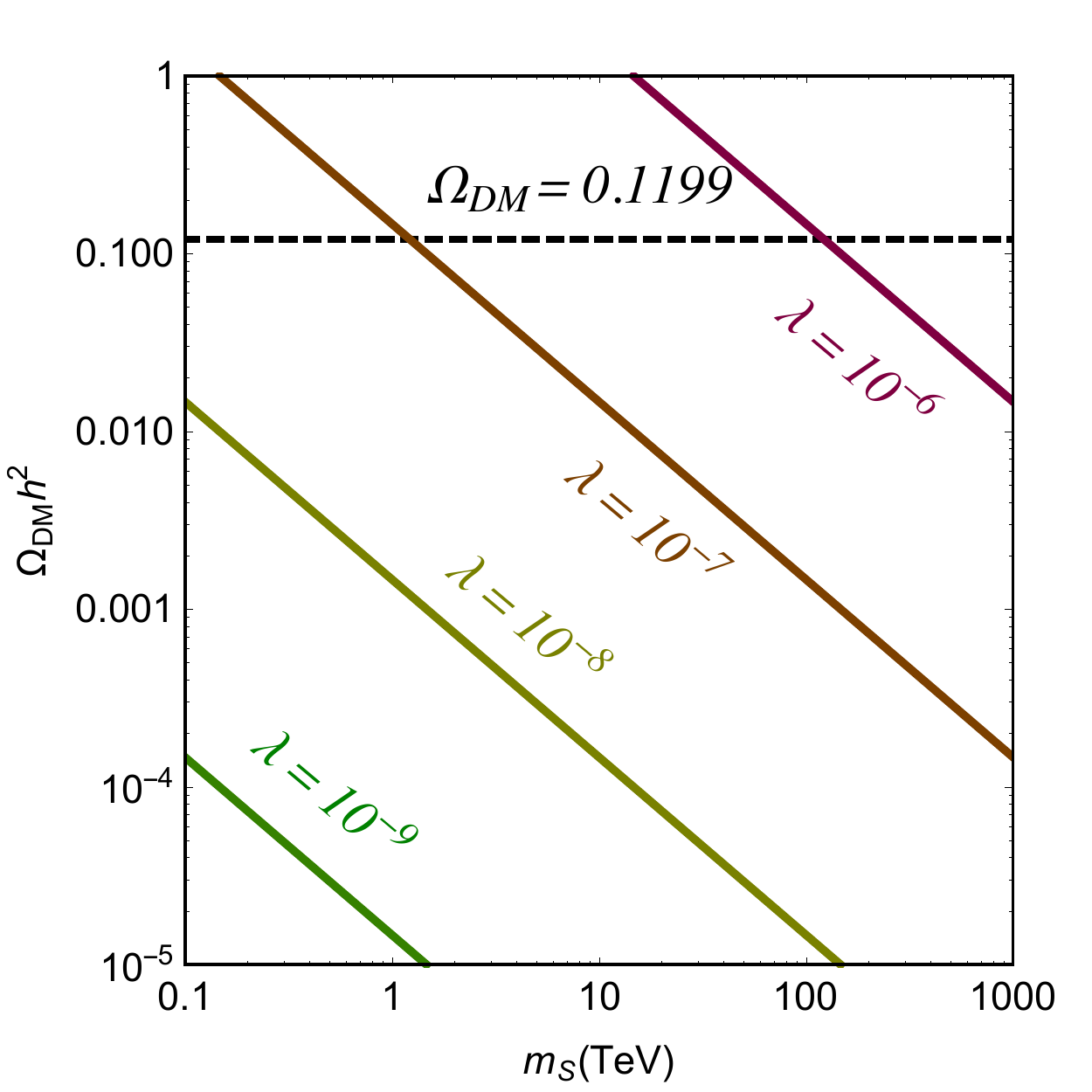}
		\label{Fig:3b}}
	\end{minipage}	
\end{tabular}
\caption{The relic density of sterile neutrinos as a function of $m_{s}$ for (a) different sterile neutrino masses $m_{N_{1}}$ and (b) different values of the Higgs portal coupling $\lambda$. }\label{Fig:3}
\end{figure*}  

Standard model particles in thermal equilibrium can annihilate into a singlet scalar. For simplicity, we concentrate on thermal Higgs annihilation as the dominant production mechanism of singlet scalars and ignore the other standard model effects. The Boltzmann equation for the annihilation process can be expressed as follows,
\begin{eqnarray}
 \frac { d{ Y }_{ s }^{ A } }{ dT }= -\frac { 135\sqrt { 5 }  }{ 64{ \pi  }^{ { 17 }/{ 2 } } } \frac { { m }_{ \rm pl } }{ { h }_{ \rm eff }\sqrt { { g }_{ \rm eff } }  } \frac { { \lambda  }^{ 2 }{ m }_{ s } }{ { T }^{ 3 } } { K }_{ 1 }\left( { 2{ m }_{ s } }/{ T } \right).\label{eq:29}
\end{eqnarray} 
We integrate Eq.(\ref{eq:27}) to estimate the yield of sterile neutrinos $Y_{N_{1}}$. There is no initial yield ($Y_{s}\left( { T }_{ RE } \right) = 0$) and no final yield ($Y_{s}\left( { T }_{ 0 } \right) = 0$), and therefore, the following equation can be obtained,
\begin{equation}
\int _{ { T }_{ RE } }^{ { T }_{ 0 } }{ \frac { d{ Y }_{ s }^{ Ds } }{ dT } dT} =-\int _{ { T }_{ RE } }^{ { T }_{ 0 } }{ \frac { d{ Y }_{ s }^{ A } }{ dT } dT }.\label{eq:31}
\end{equation}
The yield of sterile neutrinos at today's temperature can be obtained using Eq.(\ref{eq:29}) and Eq.(\ref{eq:31}), 
\begin{eqnarray}
{ Y }_{ { N }_{ 1 } }^{ { D }_{ s } }\left( { T }_{ 0 } \right) &=&-2\int _{ { T }_{ RE } }^{ { T }_{ 0 } }{ \frac { d{ Y }_{ s }^{ Ds } }{ dT } dT}=2\int _{ { T }_{ RE } }^{ { T }_{ 0 } }{ \frac { d{ Y }_{ s }^{ A } }{ dT } dT }\nonumber\\  &=&-\frac { 135\sqrt { 5 }  }{ 32{ \pi  }^{ { 17 }/{ 2 } } } \frac { { m }_{ \rm pl } }{ { h }_{ \rm eff }\sqrt { { g }_{ \rm eff } }  } \int _{ { T }_{ RE } }^{ { T }_{ 0 } }\frac { { \lambda  }^{ 2 }{ m }_{ s } }{ { T }^{ 3 } } { K }_{ 1 }\left( { 2{ m }_{ s } }/{ T } \right)dT.
\end{eqnarray}
Therefore, the yield at today's temperature $Y_{N_{1}}\left( T_{0}\right)$ is given as,
\begin{eqnarray}
{ Y }_{ { N }_{ 1 } }^{ { D }_{ s } }\left( { T }_{ 0 } \right) &=& -\frac { 135\sqrt { 5 }  }{ 32{ \pi  }^{ { 17 }/{ 2 } } } \frac { { m }_{ \rm pl } }{ { h }_{ \rm eff }\sqrt { { g }_{ \rm eff } }  } \int _{ { T }_{ RE } }^{ { T }_{ 0 } }\frac { { \lambda  }^{ 2 }{ m }_{ s } }{ { T }^{ 3 } } { K }_{ 1 }\left( { 2{ m }_{ s } }/{ T } \right)dT \nonumber \\
&\approx& -\frac { 135\sqrt { 5 }  }{ 32{ \pi  }^{ { 17 }/{ 2 } } } \frac { { m }_{ \rm pl } }{ { h }_{ \rm eff }\sqrt { { g }_{ \rm eff } }  }\int _{ \infty }^{ 0  }\frac { { \lambda  }^{ 2 }{ m }_{ s } }{ { T }^{ 3 } } { K }_{ 1 }\left( { 2{ m }_{ s } }/{ T } \right)dT \nonumber \\
&\approx&1.07\times{ 10 }^{ 13 }\left(\frac { \lambda^{ 2 } }{ { m }_{ s } }\right).
\end{eqnarray}
The sterile neutrino DM relic density resulting from the non-thermal decay mechanism is obtained as,
\begin{equation}
{ \Omega  }_{ { N }_{ 1 } }^{ D_{s}}{ h }^{ 2 }=2.93\times { 10^{-2} }{ \left( \frac { { m }_{ { N }_{ 1 } } }{ \rm{keV} }  \right)  }{ \left( \frac {\lambda }{ 10^{-7} }  \right)  }^{2} { \left( \frac { \rm{TeV} }{ { m }_{ s } }  \right)  }.\label{eq:33}
\end{equation}
FIG.\ref{Fig:3} shows the relic density of sterile neutrinos as a function of $m_{s}$ for different values of the Higgs portal coupling $\lambda$ and the sterile neutrino mass $m_{N_{1}}$. 

In this scenario, the Dodelson-Widrow mechanism can also produce sterile neutrino DM. Therefore, the total relic density is obtained as, 
\begin{eqnarray}
{ \Omega  }_{ { N }_{ 1 } }{ h }^{ 2 }&=&2.93\times { 10^{-2} }{ \left( \frac { { m }_{ { N }_{ 1 } } }{ \rm{keV} }  \right)  }{ \left( \frac {\lambda }{ 10^{-7} }  \right)  }^{2} { \left( \frac { \rm{TeV} }{ { m }_{ s } }  \right)  } \nonumber \\
& &+5.47\times { 10 }^{ 7 }\sin ^{ 2 }{ \theta  } { \left( \frac { { m }_{ { N }_{ 1 } } }{ \rm{keV} }  \right)  }^{1.63}.
\end{eqnarray}
The relic density formula depends on the Higgs portal coupling $\lambda$. The coupling $\lambda$ is bounded as $\lambda <10^{-6}$ so that $s$ does not come into thermal equilibrium. Therefore, $s$ can not be heavier than the TeV scale in order to produce keV-MeV sterile neutrino dark matter via the non-thermal decay production mechanism.

\section{X-ray constraints and lifetime bounds on sterile neutrino dark matter}\label{sec:4}
In this section, we will review the X-ray bounds and the lifetime bounds on sterile neutrino DM. Sterile neutrinos can decay into standard model particles through active-sterile neutrino mixings.  In the keV-MeV mass range, sterile neutrinos decay mainly into the three active neutrinos~\cite{2011arXiv1110.6479F,Abazajian:2001nj,Abazajian:2001vt}.
For the three-neutrino decay channel, the decay lifetime is expressed as,
\begin{equation}
{ \tau  }_{ 3\nu}\simeq2.88\times { 10 }^{ 19 }\ {\rm sec}\ { \left( \frac {  {\rm keV} }{{ m }_{ N_{1} } }  \right)  }^{ 5 }\frac { 1 }{ \sin ^{ 2 }{ \theta  }}.\end{equation}
Their lifetime must be longer than the age of the universe ($10^{17}\ {\rm sec}$) if sterile neutrinos are to constitute dark matter, which constrains the mixing angle and the sterile neutrino mass as follows,
\begin{equation}
\sin ^{ 2 }{ 2\theta  } < 2.88\times { 10 }^{ 2 }\ { \left( \frac { { m }_{ { N }_{ 1 } } }{ \rm{keV} }  \right)  }^{ -5 }.
\end{equation} 

If sterile neutrinos constitute dark matter, their radiative decay ($N_{1}\rightarrow \gamma\ \nu$) would lead to a cosmic X-ray background. We have not seen such an X-ray excess except for the recent observation of a 3.5 keV signal~\cite{Bulbul:2014sua,Boyarsky:2014jta} in galactic clusters. This puts an upper limit on the neutrino mixing angle for a given sterile neutrino mass. From the diffuse X-ray background observations XMM-Newton~\cite{Lumb:2002sw,Read:2003hw} and HEAO-1~\cite{Gruber:1999yr}, the authors of Ref.\cite{Abazajian:2006yn,Boyarsky:2005us} obtain the simple empirical formula,
\begin{equation}
\sin ^{ 2 }{ 2\theta  } < 1.15\times { 10 }^{ -4 }{ \left( \frac { { m }_{ { N }_{ 1 } } }{ \rm{keV} }  \right)  }^{ -5 }{ \left( \frac { { \Omega  }_{ { N }_{ 1 } } }{ 0.26 }  \right)^{-1}  }.
\end{equation}
The XMM-Newton observations of the Virgo and Coma galaxy clusters present the more stringent constraints~\cite{Abazajian:2006yn,Boyarsky:2006zi},
\begin{equation}
\sin ^{ 2 }{ 2\theta  } < 8\times { 10 }^{ -5 }{ \left( \frac { { m }_{ { N }_{ 1 } } }{ \rm{keV} }  \right)  }^{ -5.43 }{ \left( \frac { { \Omega  }_{ { N }_{ 1 } } }{ 0.26 }  \right)^{-1}  }.
\end{equation}
More precise X-ray constraints have been reported in Ref.\cite{Abazajian:2006yn}. Note that these bounds are given for sterile neutrino DM which explains the current dark matter density. If sterile neutrino DM only constitutes part of the total dark matter, the X-rays bounds become weaker.

\section{The free streaming horizon and Lyman-$\alpha$ constraints}\label{sec:5}
Recent observations such as the WMAP and Planck missions have proven that the $\Lambda$CDM model, which contains cold dark matter, is an extremely successful cosmological model~\cite{Ade:2013zuv}. However the $\Lambda$CDM can not solve the small-scale crises~\cite{Weinberg:2013aya}, including the missing satellite problem and the cuspy halo problem. Warm dark matter (WDM), which has an adequate free streaming horizon and suppresses the structure of dwarf galaxies size, may solve the problem. The upper bound on the free streaming scale of WDM is obtained from the observed Lyman-$\alpha$ forest, which refers to the absorption lines of intergalactic neutral hydrogen in the spectra of distant quasars and galaxies.

The free-streaming horizon corresponds to the average distance travelled by DM particles and is a good measure to classify CDM, WDM and HDM. The free streaming horizon is given as,
\begin{equation}
{ \lambda  }_{ FS }=\int _{ { t }_{ in } }^{ { t }_{ 0 } }{ \frac { \left< v\left( t \right)  \right>  }{ a\left( t \right)  }  } dt,
\end{equation} 
where $t_{in}$ is the DM production time, $t_{0}$ is the current time, $\left< v\left( t \right)  \right> $ is the average thermal velocity of the DM particles, and $a(t)$ is the scale factor. In this paper we assume that the free-streaming scale of CDM, WDM and HDM satisfy ${ \lambda  }_{ FS } < 0.01\ \rm Mpc$, $0.01\ \rm Mpc<{ \lambda  }_{ FS } < 0.1\ \rm Mpc$ and $0.1\ \rm Mpc<{ \lambda  }_{ FS }$, respectively.  
This is not an accurate definition, but gives a useful criteria to classify the thermal property of DM. Note that HDM is excluded by observations of the Lyman-$\alpha$ forest.

In order to determine the free-streaming horizon, we now consider the average thermal velocity $\left< v\left( t \right)  \right> $ of the sterile neutrino DM $N_{1}$.  We define $t_{nr}$ as the time when $N_{1}$ becomes non-relativistic, which we take to be when the equality $\left< p\left( t_{nr} \right)  \right> = { m }_{ { N }_{ 1 }}$ is satisfied.  The approximate average thermal velocity $\left< v\left( t \right)  \right> $ is then given as follows,
\begin{equation}
\left< v\left( t \right)  \right> \simeq \begin{cases} \begin{matrix}1 &\quad\  t<{ t }_{ nr }, \end{matrix} \\ \begin{matrix} \frac { \left< p\left( t \right)  \right>  }{ { m }_{  N _ 1  }} &   t\ge { t }_{ nr }. \end{matrix} \end{cases}
\end{equation}
The non-relativistic thermal velocity is expressed in terms of the average thermal momentum, which can be extracted from the distribution function $f\left( p\right)$ and depends on the DM production mechanism. In this section we will consider the average thermal momentum and the free streaming horizon when production is via thermal freeze-in of the singlet scalar, via the Dodelson-Widrow mechanism and via the non-thermal singlet scalar. Finally, we determine the Lyman-$\alpha$ constraints and the allowed parameter region for each production mechanisms.

\subsection{Production via thermal freeze-in of the singlet scalar}\label{sec:5a}
For production via the thermal freeze-in of the singlet scalar boson, the momentum distribution of sterile neutrino DM~\cite{Kamada:2013sh,Boyanovsky:2008nc} 
is given by,
\begin{equation}
f\left( p \right) =\frac { \beta  }{ { \left( { p }/{ T } \right)  }^{ { 1 }/{ 2 } } }\ { g }_{ { 5 }/{ 2 } }\left( { p }/{ T } \right), 
\end{equation}
where
\begin{equation}
{ g }_{ \nu  }\left( x \right) =\sum _{ n=1 }^{ \infty  }{ \frac { { e }^{ -n x } }{ { n }^{ \nu  } } }. 
\end{equation}
The normalization factor $\beta$ is determined by the Yukawa coupling $\kappa$ and the singlet scalar mass $m_{s}$, with $\beta \propto \kappa^{2}m_{s}^{-1}$. 
The average thermal momentum $\left< p\left( t\right)  \right>$ can be calculated as,
\begin{equation}
\left< p\left( t \right)  \right> =\frac { \int _{ 0 }^{ \infty  }{ dp\sqrt { T{ p }^{ 5 } } \sum _{ n=1 }^{ \infty  }{ \frac { { e }^{ -n\left( { { p } }/{ T } \right)  } }{ { n }^{ { 5 }/{ 2 } } }  }  }  }{ \int _{ 0 }^{ \infty  }{ dp\sqrt { T{ p }^{ 3 } } \sum _{ n=1 }^{ \infty  }{ \frac { { e }^{ -n\left( { { p } }/{ T } \right)  } }{ { n }^{ { 5 }/{ 2 } } }  }  }  } \approx 2.4527\ T.
\end{equation}
This average thermal momentum $\left< p\left( t\right)  \right>$ leads to the average thermal velocity $\left< v\left( t \right)  \right> $,
\begin{equation}
\left< v\left( t \right)  \right> \simeq \begin{cases} \begin{matrix}\quad 1 &\quad\quad\quad\quad\ \  
 t<{ t }_{ nr }, \end{matrix} \\ \begin{matrix} \frac { 2.45T }{  { m }_{  N _ 1  } }= \frac{a(t_{nr})}{a(t)}
 &   \ \  t\ge { t }_{ nr }. \end{matrix} \end{cases}
\end{equation}
\begin{figure*}[t]
	\begin{tabular}{cc}
	\begin{minipage}{0.5\hsize}
		\centering
		\subfigure[$\ s\rightarrow N_{1}N_{1}\quad\left< S \right>=1\ {\rm TeV}$]{
		\includegraphics[width=78mm]{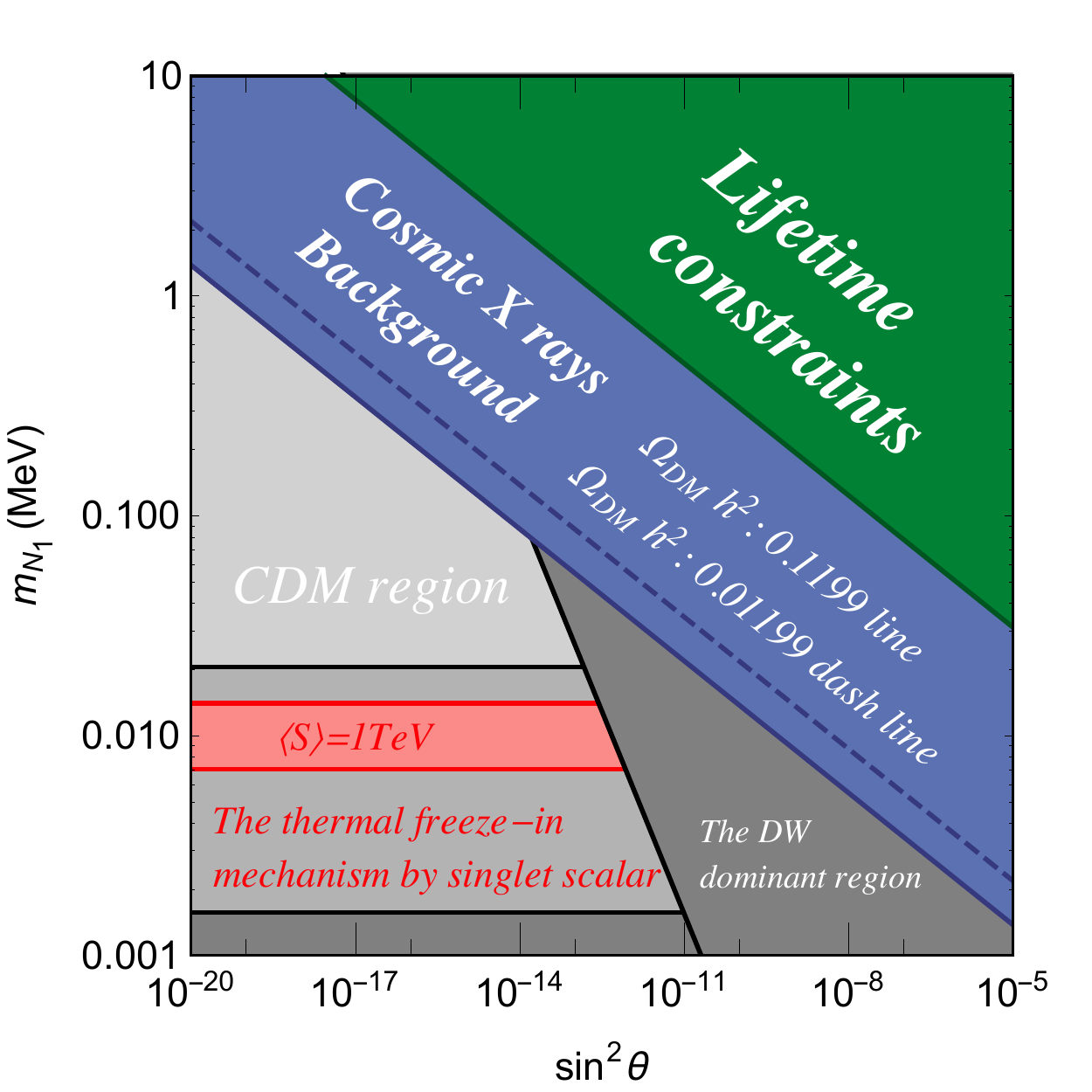}
		\label{Fig:4a}}
	\end{minipage}
		\begin{minipage}{0.5\hsize}
		\centering
		\subfigure[$\ s\rightarrow N_{1}N_{1}\quad\left< S \right>=100\ {\rm TeV}$]{
		\includegraphics[width=78mm]{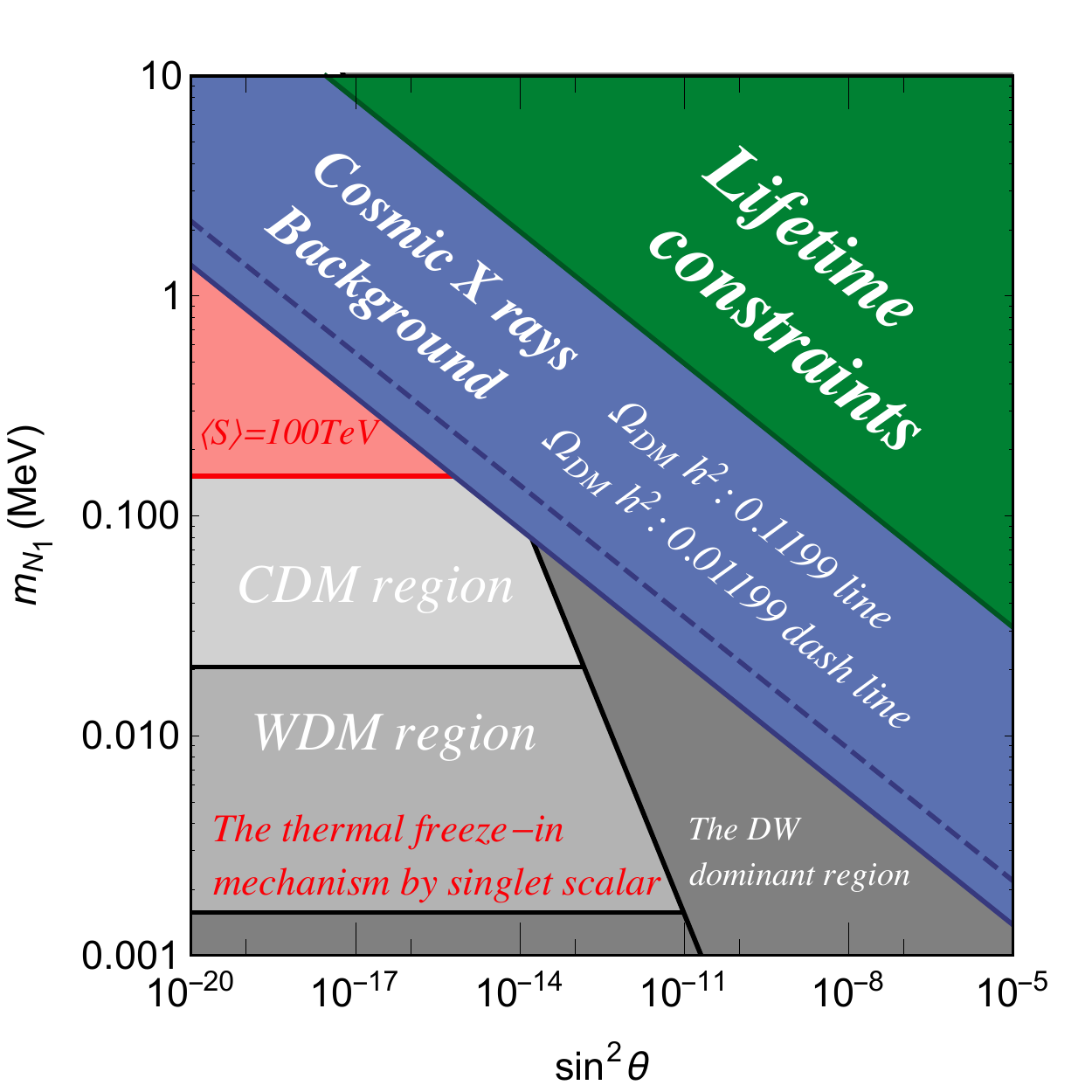}
		\label{Fig:4b}}
	\end{minipage}
        \end{tabular}
\caption{X-ray bounds, the free streaming horizon, HDM, WDM and CDM regions and other constraints for sterile neutrino production via thermal freeze-in of the singlet scalar for (a) $\left< S \right>=1\ {\rm TeV}$ and for (b) $\left< S \right>=100\ {\rm TeV}$.}\label{Fig:4}
\end{figure*}
The time when DM particles become non-relativistic is $t_{nr}^{1/2}\approx 2.45\left( \frac{\rm{MeV}}{ { m }_{  N _ 1  }} \right)$ sec. The free streaming horizon is calculated as,
\begin{eqnarray}
{ \lambda  }_{ FS }&=&\int _{ { t }_{ in } }^{ { t }_{ 0 } }{ \frac { \left< v\left( t \right)  \right>  }{ a\left( t \right)  }  } dt \nonumber \\\ &=& \int _{ { t }_{ in } }^{ { t }_{ nr } }{ \frac { dt }{ a\left( t \right)  }  } +\int _{ { t }_{ nr } }^{ { t }_{ eq } }{ \frac { \left< v\left( t \right)  \right>  }{ a\left( t \right)  }  } dt+\int _{ { t }_{ eq } }^{ { t }_{ 0 } }{ \frac { \left< v\left( t \right)  \right>  }{ a\left( t \right)  }  } dt \nonumber \\ &=&\frac { 5\sqrt { { t }_{ eq }{ t }_{ nr } }  }{ a\left( { t }_{ eq } \right)  } +\frac { \sqrt { { t }_{ eq }{ t }_{ nr } }  }{ a\left( { t }_{ eq } \right)  } \ln { \left( \frac { { t }_{ eq } }{ { t }_{ nr } }  \right)  } -\frac { 2\sqrt { { t }_{ eq }{ t }_{ in } }  }{ a\left( { t }_{ eq } \right)  } -\frac { 3\sqrt { { t }_{ eq }{ t }_{ nr } }  }{ { a\left( { t }_{ eq } \right)  }^{ { 1 }/{ 2 } } } \nonumber \\ &\simeq& \frac { \sqrt { { t }_{ eq }{ t }_{ nr } }  }{ a\left( { t }_{ eq } \right)  } \left[ 5+\ln { \left( \frac { { t }_{ eq } }{ { t }_{ nr } }  \right)  }  \right].\label{eq:44}
\end{eqnarray}
To obtain the last line, we neglect the third and the last terms of the third line. 

In this production mechanism, the DM is produced at high temperatures, $T\gtrsim1\ {\rm TeV}$, and entropy dilution affects the free streaming horizon. The effect of entropy dilution can be estimated by the factor $\xi^{-1/3}$ which is given by,
\begin{equation}
\xi =\frac { { g }_{ \rm eff }\left( {\rm high}\ T \right)  }{ { g }_{ \rm eff }\left( {\rm current}\ { T }_{ 0 } \right)  } \approx \frac { 109.5 }{ 3.36 }. 
\end{equation}
Now we assume that both $s$ and $N_{1}$ contribute to the effective number of degrees of freedom and ignore the tiny effect of the other heavy right-handed neutrinos $N_{2}$ and $N_{3}$. Taking entropy dilution into account and using the conversion factor $c=10^{-14}({\rm Mpc}/{\rm sec})$, the final expression is given as,
\begin{equation}
{ \lambda  }_{ FS }=\frac { c\sqrt { { t }_{ eq }{ t }_{ nr } }  }{ a\left( { t }_{ eq } \right)  } \left[ 5+\ln { \left( \frac { { t }_{ eq } }{ { t }_{ nr } }  \right)  }  \right] \frac { 1 }{ { \xi  }^{ { 1 }/{ 3 } } }.
\end{equation}
The Lyman-$\alpha$ bound on ${ m }_{  N _ 1  }$ is given by, 
\begin{equation}{ m }_{  N _ 1  }>1.57\ {\rm keV}.\end{equation}
The range of sterile neutrino mass corrsponding to WDM is obtained as,
\begin{equation}1.57\ {\rm keV}< { m }_{  N _ 1  }<20.5\ {\rm keV}.\end{equation}

In FIG.\ref{Fig:4}, we show the X-ray bounds and the HDM, WDM and CDM regions for sterile neutrino production via the thermal freeze-in of the singlet scalar. In this figure, we assume that $m_{s}$ is larger than $m_{h}= 125\ {\rm GeV}$ but smaller than (a) $\left< S \right> = 1 \ {\rm TeV}$ and (b) $\left< S \right> = 100 \ {\rm TeV}$. 
We also show the parameter region where more than 1 $\%$ of the DM is produced by the DW mechanism.
When $\left< S \right>$ is $1\ {\rm TeV}$ ($100 \ {\rm TeV}$), the sterile neutrino DM is warm (cold).
However, the scenario with $\left< S \right> >100 \ {\rm TeV}$ suffers from the X-ray constraints.

\subsection{Production via non-thermal decay of the singlet scalar}\label{sec:5b}
If the Higgs portal coupling is small and the singlet scalar is out of thermal equilibrium, it decays into the sterile neutrino. The free streaming horizon was considered in~\cite{Merle:2013wta,Adulpravitchai:2014xna}. The momentum distribution of the sterile neutrino DM is given as~\cite{Kaplinghat:2005sy,Hisano:2000dz,Strigari:2006jf,Aoyama:2011ba},
\begin{equation}
f\left( p \right) =\frac { \beta  }{ { p }/{ T_{DM} } }{\rm exp}\left( -\frac { { p }^{ 2 } }{ T_{DM}^{2} }  \right), 
\end{equation}
where $\beta$ is a normalization factor and the DM temperature is $T_{DM}=\frac{m_{s}a(t_{in})}{2a(t)}$. The average thermal momentum is given as,
\begin{equation}
\left< p\left( t \right)  \right> =\frac { \int { { d }^{ 3 }ppf\left( p \right)  }  }{ \int { { d }^{ 3 }pf\left( p \right)  }  } =\frac { \int _{ 0 }^{ \infty  }{ dp } { p }^{ 2 }{ e }^{ { -{ p }^{ 2 } }/{ T_{DM}^{ 2 } } } }{ \int _{ 0 }^{ \infty  }{ dp } { p }{ e }^{ { -{ p }^{ 2 } }/{  T_{DM}^{ 2 } } } } =\frac { \sqrt { \pi  }  }{ 2 } { T_{DM} }\label{eq:51}.
\end{equation}
From Eq.(\ref{eq:51}), the average thermal velocity is expressed as,
\begin{equation}
\left< v\left( t \right)  \right> \simeq \begin{cases} \begin{matrix}\quad1 &\ \ \quad\quad\quad\quad\quad\quad\   t<{ t }_{ nr }, \end{matrix} \\ \begin{matrix} \frac { \sqrt { \pi  } { m }_{ s }a\left( { t }_{ in } \right)  }{ 4 { m }_{  N _ 1  }a\left( t \right)  } =\frac { a\left( { t }_{ nr } \right)  }{ a\left( t \right)  } 
 &   \quad t\ge { t }_{ nr }. \end{matrix} \end{cases}
\end{equation}
\begin{figure*}[t]
	\begin{tabular}{cc}
	\begin{minipage}{0.5\hsize}
		\centering
		\subfigure[$\ h\rightarrow s\rightarrow N_{1}N_{1}\quad m_{s}=1\ {\rm TeV}$]{
		\includegraphics[width=78mm]{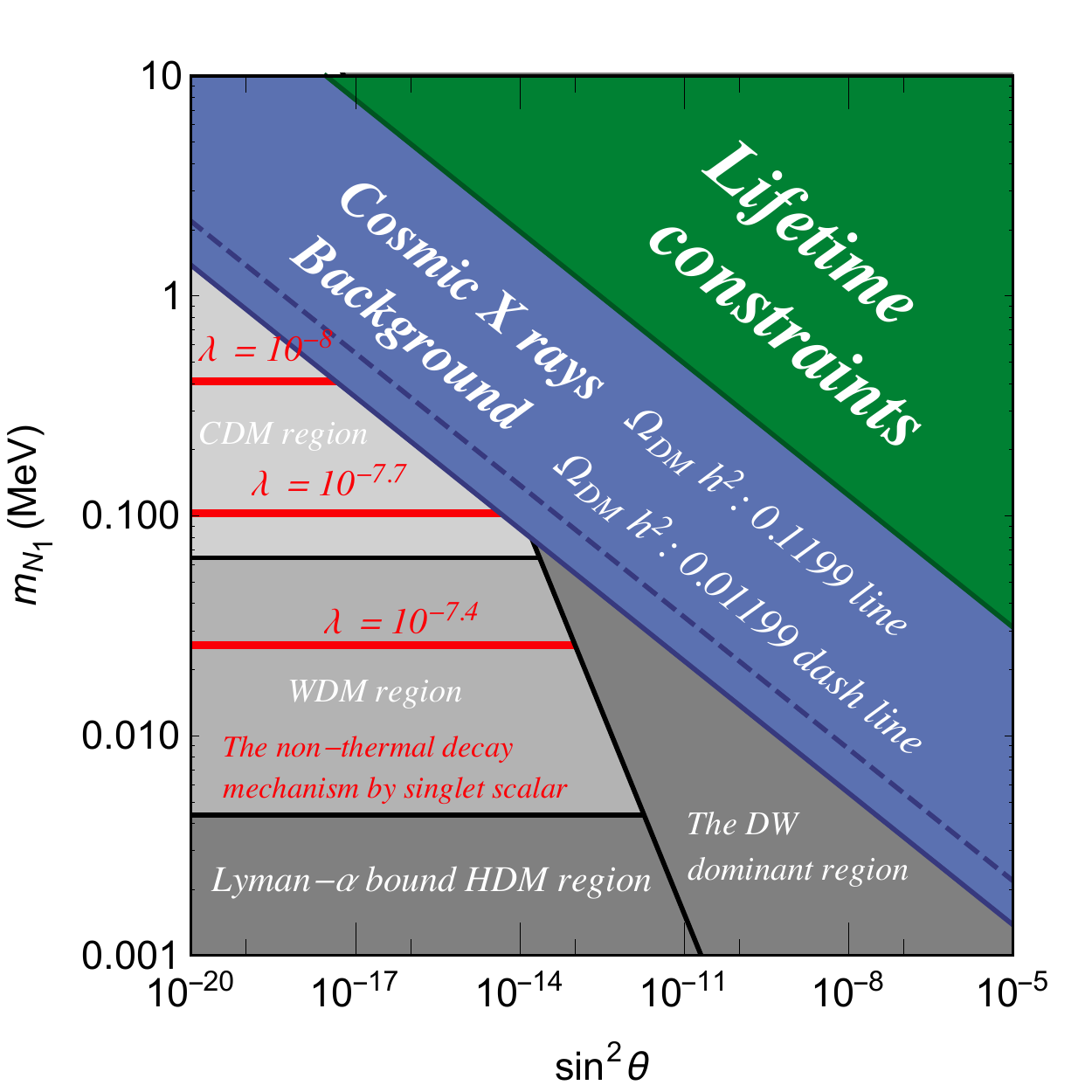}
		\label{Fig:7FSD}}
	\end{minipage}
		\begin{minipage}{0.5\hsize}
		\centering
		\subfigure[$\ h\rightarrow s\rightarrow N_{1}N_{1}\quad m_{s}=100\ {\rm TeV}$]{
		\includegraphics[width=78mm]{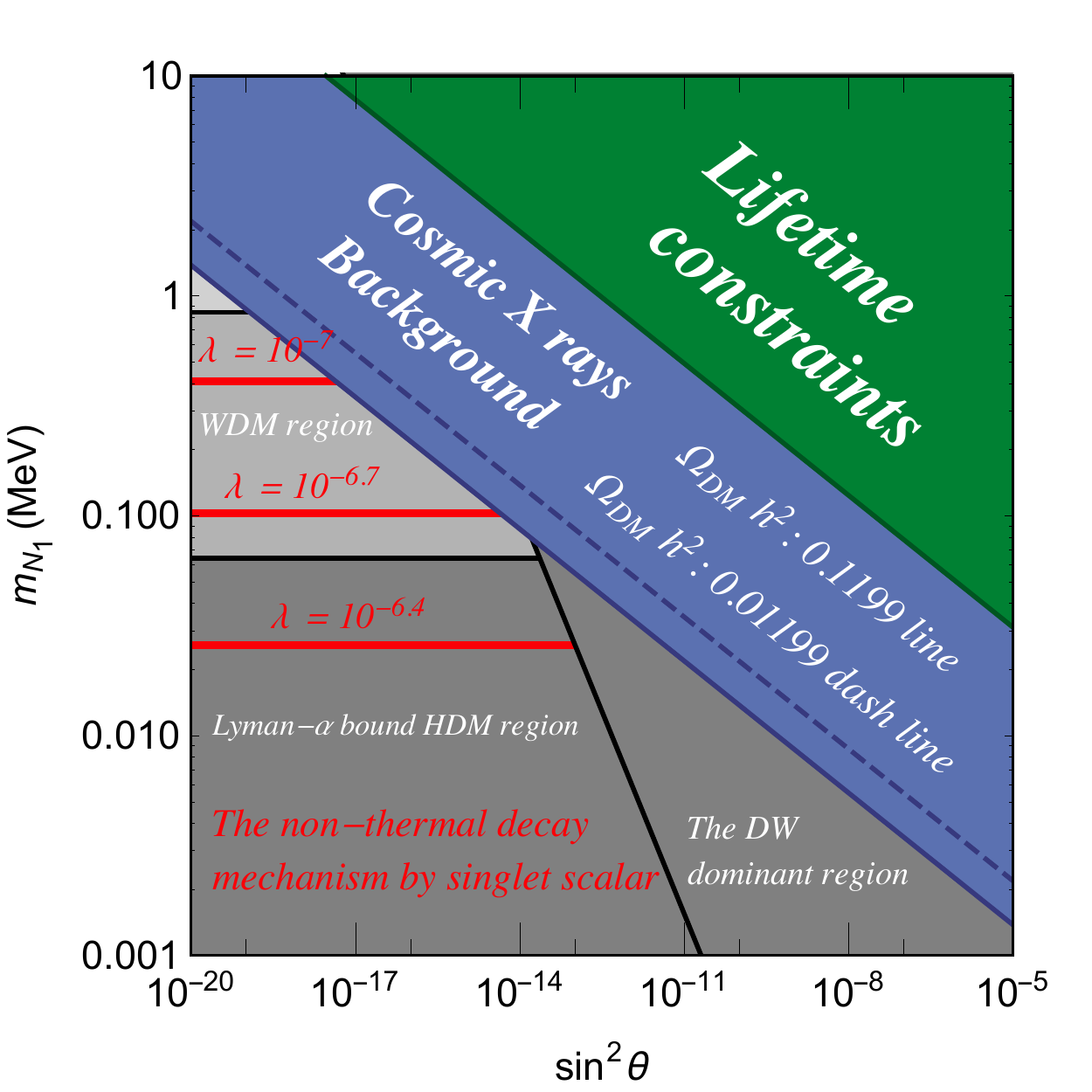}
		\label{Fig:8FSD}}
	\end{minipage}
        \end{tabular}
\caption{X-ray bounds, HDM, WDM and CDM regions and other constraints on the sterile neutrino mass $m_{N_1}$ and mixing angle $\theta$ in the case of production via the non-thermal decay of the singlet scalar for (a) $m_{s} = 1\ {\rm TeV}$ and for (b) $m_{s} = 100\ {\rm TeV}$.}\label{Fig:5}
\end{figure*}
Now, we assume that the production time is $t_{in}=t_{fe}+\tau$, where $t_{fe}$ is the freeze-in time of $s$ and is given as $t_{fe}\simeq{\left(\frac{{\rm MeV}}{T_{fe}}\right)}^{2}$ sec, where $T_{fe}\simeq m_{s}\  {\rm GeV}$ is the freeze-in temperature, and the lifetime of $s$ is $\tau=\hbar /\Gamma \left( s\rightarrow { N }_{ 1 }{ N }_{ 1 }\right)$. The time at which the sterile neutrinos become non-relativistic is given by $t_{nr}=\frac { \pi  }{ 16 } \frac { { m }_{ s }^{ 2 } }{  { m }_{  N _ 1  }^{ 2 } } { t }_{ in }$ sec and the time of matter-radiation equality is $t_{eq}=1.9\times10^{11}\ {\rm sec}$. We estimate the free-streaming horizon of the DM sterile neutrinos using the formula,
\begin{equation}
{ \lambda  }_{ FS }=\frac { c\sqrt { { t }_{ eq }{ t }_{ nr } }  }{ a\left( { t }_{ eq } \right)  } \left[ 5+\ln { \left( \frac { { t }_{ eq } }{ { t }_{ nr } }  \right)  }  \right] \frac { 1 }{ { \xi  }^{ { 1 }/{ 3 } } }.
\end{equation}
For $m_{s}=1\ {\rm TeV}$ and $\kappa_{1}=10^{-8}$, the Lyman-$\alpha$ bound on $m_{N_1}$ is obtained as,
\begin{equation}{ m }_{  N _ 1  }>4.36\ {\rm keV}.\end{equation}
The WDM sterile neutrino mass can be constrained as \begin{equation}4.36\ {\rm keV}< { m }_{  N _ 1  }<64.3\ {\rm keV}.\end{equation}
For $m_{s}=100\ {\rm TeV}$ and $\kappa_{1}=10^{-8}$, the Lyman-$\alpha$ bound is given as,
\begin{equation}{ m }_{  N _ 1  }>64.2\ {\rm keV}.\end{equation}
The WDM sterile neutrino mass range is obtained as  
\begin{equation}64.2\ {\rm keV}< { m }_{  N _ 1  }<840\ {\rm keV}.\end{equation}
Therefore, in this scenario, the singlet scalar can not be heavier than the TeV scale~\footnote{Ref.\cite{Merle:2015oja} presents more detailed calculations in this scenario, solving numerically the system of Boltzmann equations.  The singlet scalar mass could be more tightly restricted.}
This constraint is tighter than that in the DW mechanism $ { m }_{  N _ 1  }>10\ {\rm keV}$~\cite{Viel:2006kd,Viel:2005qj}. This is because the DM sterile neutrino is produced by the decay of non-thermal heavy particles. When the lifetime of the singlet scalar is not small, the Lyman-$\alpha$ constraint becomes tight. 

In FIG.\ref{Fig:5}, we show the X-ray bounds and the HDM, WDM and CDM regions in the $m_{N_1}$--$\theta$ plane for sterile neutrino DM production via non-thermal decay of the singlet scalar. In FIG.\ref{Fig:7FSD} we take $m_{s}= 1\ {\rm TeV}$ and plot the sterile neutrino constraints for the three different values of Higgs portal coupling $\lambda=10^{-7.4},10^{-7.7},10^{-8}$. The sterile neutrino DM is not constrained by the Lyman-$\alpha$ bounds. FIG.\ref{Fig:8FSD} shows the sterile neutrino constraints for the portal couplings $\lambda=10^{-6.4},10^{-6.7},10^{-7}$ and singlet scalar mass $m_{s}= 100\ {\rm TeV}$. When the mass of the singlet scalar is large, the produced sterile neutrinos are warmer and the scenario is constrained by the Lyman-$\alpha$ bounds.

\subsection{Sterile neutrino DM production mechanisms in the $\nu$MSM}\label{sec:5c}
\begin{figure*}[t]
	\begin{tabular}{cc}
	\begin{minipage}{0.5\hsize}
		\centering
		\subfigure[$\ h\rightarrow \nu_{e}N_{1}$]{
		\includegraphics[width=78mm]{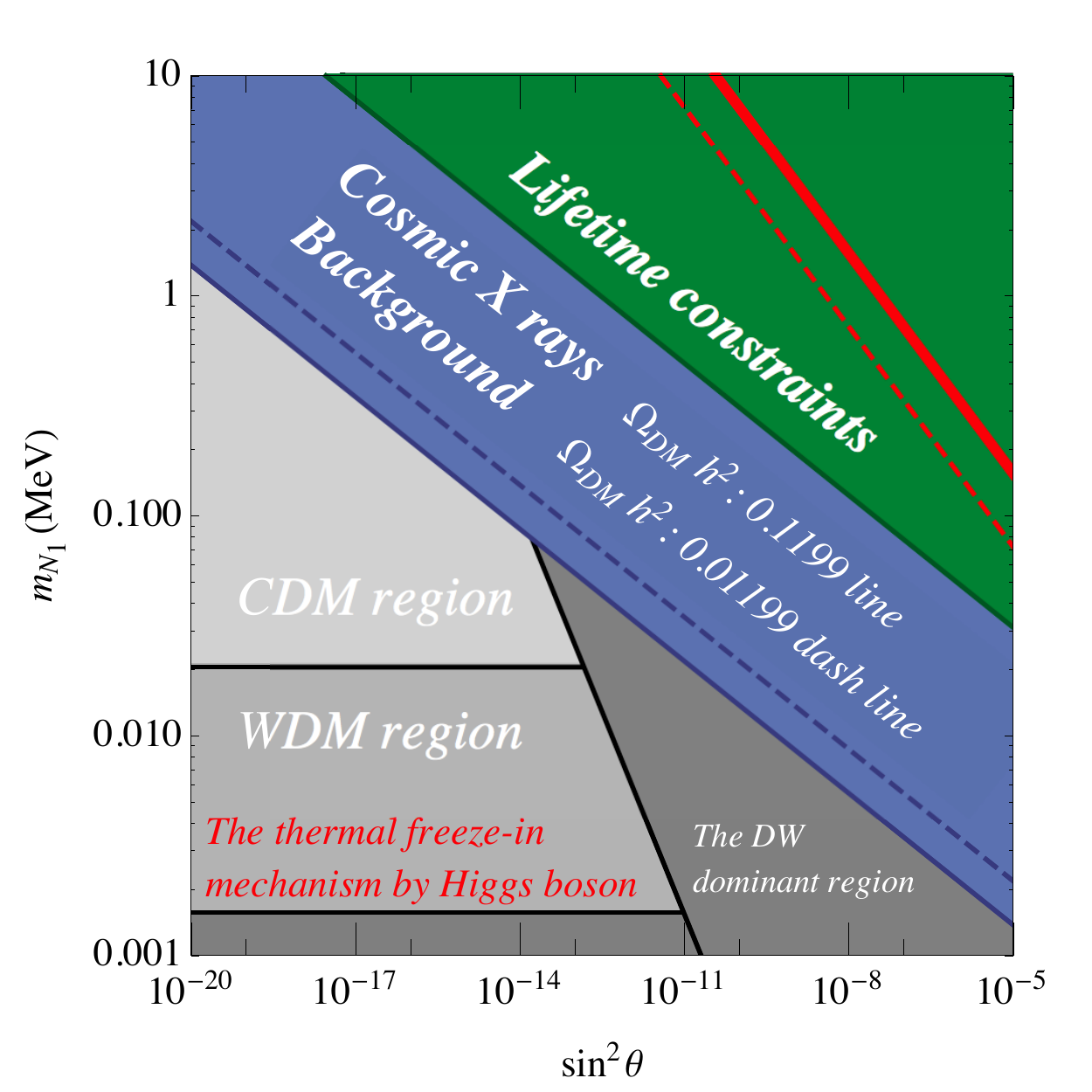}
		\label{Fig:6a}}
	\end{minipage}
\begin{minipage}{0.5\hsize}
		\centering
		\subfigure[$\ \nu_{e}\nu_{\mu}\nu_{\tau}\rightarrow N_{1}$]{
		\includegraphics[width=78mm]{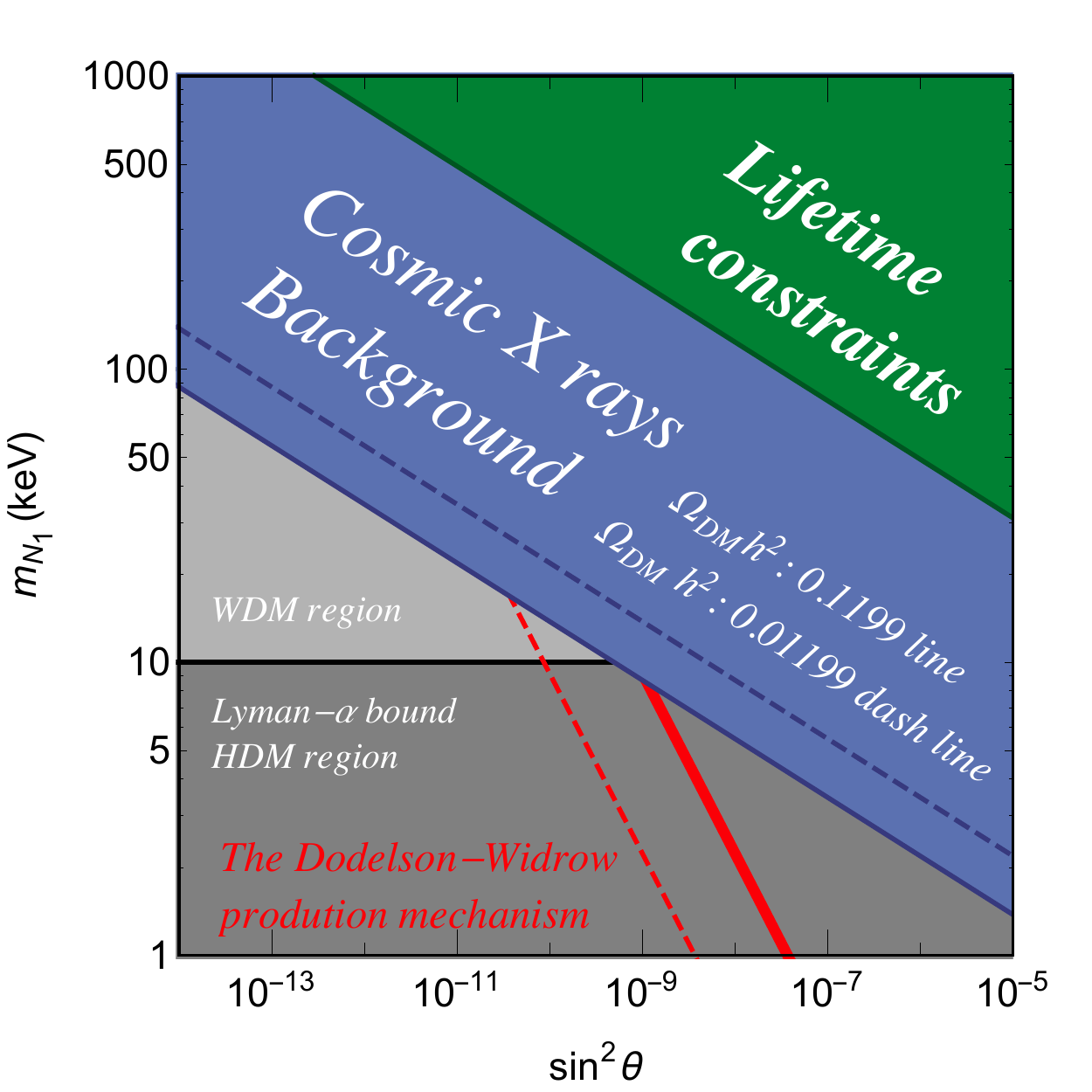}
		\label{Fig:6b}}
	\end{minipage}
        \end{tabular}
\caption{X-ray bounds, HDM, WDM and CDM regions and other constraints for (a) thermal freeze-in via Higgs boson interactions and (b) the Dodelson-Widrow mechanism. The solid red lines show the parameters for which the sterile neutrino DM density ${ \Omega  }_{ DM }h^{2}=0.1199$ and the red dashed lines the parameters for which the sterile neutrino DM density satisfies ${ \Omega  }_{ DM }h^{2}=0.01199$. }\label{Fig:6}
\end{figure*}

In the $\nu$MSM, sterile neutrino dark matter can be generated by the thermal freeze-in mechanism -- through interactions with the Higgs boson -- or the Dodelson-Widrow mechanism.  The free-streaming horizon of sterile neutrino DM produced by thermal freeze-in via the Higgs boson is as small as that of sterile neutrino DM that is produced by thermal freeze-in via $s$.  However, this production mechanism is in conflict with lifetime bounds and X-ray bounds (see FIG.\ref{Fig:6a}). 

In the Dodelson-Widrow mechanism, sterile neutrinos are produced from the thermal background of active neutrinos via coherent scattering. Therefore, the momentum distribution is thermal and of the Fermi-Dirac type~\cite{Dodelson:1993je}, i.e,
\begin {equation}
f\left( p \right) =\frac { \beta  }{ { e }^{ p/T}+1 }, 
\end{equation}
where $p$ denotes the comoving momentum of $N_{1}$ and $\beta \propto \theta^{2}M_{1}$. For the thermalized sterile neutrinos, the average thermal momentum $\left< p\left( t\right)  \right>$ is given as,
\begin{equation}
\left< p\left( t \right)  \right> =\frac { \int _{ 0 }^{ \infty  }{ dp\frac { { p }^{ 3 } }{ { e }^{ { { p } }/{ { T }} }+1 }  }  }{ \int _{ 0 }^{ \infty  }{ dp\frac { { p }^{ 2 } }{ { e }^{ { { p } }/{ { T } } }+1 }  }  } =\frac { 7{ \pi  }^{ 4 }T }{ 180\zeta \left( 3 \right)  } \approx 3.1513\ T.\label{eq:50}
\end{equation}
The average thermal momentum of the thermally produced sterile neutrinos thus satisfies the relation $\left< p \right>/3.15T\approx 1$, but the DW mechanism produces a colder distribution of sterile neutrinos, with $\left< p \right>/3.15T\approx0.9$. 
The free-streaming horizon of the sterile neutrino DM is given in terms of the average thermal momentum as~\cite{Abazajian:2001nj},
\begin{equation}
{ \lambda  }_{ FS }\approx 0.84\ {\rm Mpc}\left( \frac { \rm{keV} } { { m }_{ { N }_{ 1 } } } \right) \left( \frac { \left< p \right>  }{ 3.15\ T }  \right).
\end{equation}
In the case of the DW mechanism, observations of the Lyman-$\alpha$ forest lead to the severe constraint $ { m }_{  N _ 1  }>10\ {\rm keV}$~\cite{Viel:2006kd,Viel:2005qj}. This limit is in conflict with X-ray bounds (see FIG.\ref{Fig:6b}), meaning that the DW mechanism scenario is excluded. 

If the lepton asymmetry is relatively large in the early universe, the thermal production of sterile neutrinos can be enhanced by the MSW effect (Shi-Fuller mechanism). The Shi-Fuller mechanism leads to a colder thermal distribution with $\left< p \right>/3.15T\approx 0.6$. Therefore, the Shi-Fuller production mechanism can evade the Lyman-$\alpha$ bounds and X-ray constraints, although a relatively large lepton asymmetry is needed. In conclusion, the thermal background production of sterile neutrino DM, through mechanisms such as the DW mechanism, is severely constrained due to the large free-streaming scale and X-ray bounds. It should be emphasized that the $\nu$MSM fails to explain the dark matter sector, because the sterile neutrino dark matter cannot be generated by neither the DW mechanism nor thermal freeze-in production via the Higgs boson. This is why we have to extend the $\nu$MSM to explore new DM production scenarios.

\section{Thermal freeze-in leptogenesis via the singlet scalar}\label{sec:6}
In this section, we discuss leptogenesis scenarios that rely on the thermal freeze-in production mechanism. In this scenario, we assume that the heavy Majorana neutrinos are generated by either thermal freeze-in via the singlet scalar or non-thermal decay of the frozen-in singlet scalar. The produced Majorana neutrinos generate a lepton asymmetry, which is transferred into a baryon asymmetry via non-perturbative electroweak effects (Sphalerons). We assume that $N_{1}$ constitutes sterile neutrino DM and does not affect the leptogenesis scenarios, while $N_{2}$ and $N_{3}$ satisfy $T_{EW}< M_{2}<M_{3}$.

In line with the original motivation for the $\nu$MSM to explain BSM phenomena by introducing TeV-scale particles, we set the Majorana masses to around the TeV scale.  However, TeV-scale leptogenesis is in conflict with the Davidson-Ibarra bound~\cite{Davidson:2002qv}, which constrains the Majorana mass as $M_{2}>10^{9}\ \rm{GeV}$. It is possible to evade this lower bound when the mass difference between $N_{2}$ and $N_{3}$ is of the order of their decay width. In this case, resonant leptogenesis can occur~\cite{Pilaftsis:1997jf,Pilaftsis:2003gt}. The resonant CP asymmetry is obtained as,
\begin{equation}
{ \epsilon  }_{ i }=\frac { \Gamma\left( { N }_{ i }\rightarrow{ \ell  }_{ \alpha  }H \right)  -\Gamma\left( { N }_{ i }\rightarrow \overline { { \ell  }_{ \alpha  } } { H }^{ * } \right)  }{ \Gamma \left( { N }_{ i }\rightarrow{ \ell  }_{ \alpha  }H \right) +\Gamma\left( { N }_{ i }\rightarrow \overline { { \ell  }_{ \alpha  } } { H }^{ * } \right) }\simeq {\varepsilon}'_{ i }+{\varepsilon}_{ i }.
\end{equation}
The ${\varepsilon}'$-type CP asymmetry is obtained from the vertex contribution,
\begin{equation}
{ { \varepsilon  } }_{ i }^{ \prime  }=\frac { {\rm Im}{ \left( { y }^{ \dagger  }y \right)  }_{ ij }^{ 2 } }{ { \left( { y }^{ \dagger  }y \right)  }_{ ii }{ \left( { y }^{ \dagger  }y \right)  }_{ jj } } \left( \frac { { \Gamma  }_{ { N }_{ j } } }{ { m }_{ { N }_{ j } } }  \right) f\left( \frac { { m }_{ { N }_{ j } }^{ 2 } }{ { m }_{ { N }_{ i } }^{ 2 } }  \right),
\end{equation}
where ${ \Gamma  }_{ { N }_{ i } }$ is the tree-level decay width and $f(x)$ is the loop function, which are given as,
\begin{equation}
{ \Gamma  }_{ { N }_{ i } }=\frac { { \left( { y }^{ \dagger  }y \right)  }_{ jj } }{ 8\pi  } { m }_{ { N }_{ j } },\quad f\left( x \right) =\sqrt { x } \left[ 1-\left( 1+x \right) \ln { \left( \frac { 1+x }{ x }  \right)  }  \right].
\end{equation}
In the degenerate heavy Majorna neutrino mass limit ($m_{N_{2}} \approx m_{N_{3}}$), the CP asymmetry ${ \varepsilon  }_{ 2 }^{ \prime  }$
cancels with ${ \varepsilon  }_{ 3 }^{ \prime  }$ and no net CP asymmetry can be obtained from the vertex contribution.

The ${\varepsilon}$-type CP asymmetry arises from the self-energy contribution,
\begin{equation}
{ { \varepsilon  } }_{ i }=\frac { {\rm Im}{ \left( { y }^{ \dagger  }y \right)  }_{ ij }^{ 2 } }{ { \left( { y }^{ \dagger  }y \right)  }_{ ii }{ \left( { y }^{ \dagger  }y \right)  }_{ jj } } \left( \frac { { \Gamma  }_{ { N }_{ j } } }{ { m }_{ { N }_{ j } } }  \right) \frac { \left( { m }_{ { N }_{ i } }^{ 2 }-{ m }_{ { N }_{ j } }^{ 2 } \right) { m }_{ { N }_{ i } }{ m }_{ { N }_{ j } } }{ { \left( { m }_{ { N }_{ i } }^{ 2 }-{ m }_{ { N }_{ j } }^{ 2 } \right)  }^{ 2 }+{ m }_{ { N }_{ i }^{2} }{ \Gamma  }_{ { N }_{ j } }^{ 2 } }.
\end{equation}
In the limit ($m_{N_{2}} \approx m_{N_{3}}$), ${ \varepsilon  }_{ i }$ dominates over ${ \varepsilon  }_{ i }^{ \prime  }$. Furthermore, ${ \varepsilon  }_{ 2 }$ and ${ \varepsilon  }_{ 3 }$ have the same sign. In the limit $m_{N_{2}} \approx m_{N_{3}}$, the total CP asymmetry involving the $N_{2}$ contribution is given by,
\begin{equation}
{ { \epsilon  } }_{ 2}\simeq\frac { {\rm Im}{ \left( { y }^{ \dagger  }y \right)  }_{ 23 }^{ 2 } }{ { \left( { y }^{ \dagger  }y \right)  }_{ 22 }{ \left( { y }^{ \dagger  }y \right)  }_{ 33 } } \left( \frac { { \Gamma  }_{ { N }_{ 3 } } }{ { m }_{ { N }_{ 3} } }  \right) \frac { \left( { m }_{ { N }_{ 2 } }^{ 2 }-{ m }_{ { N }_{ 3 } }^{ 2 } \right) { m }_{ { N }_{ 2 } }{ m }_{ { N }_{ 3 } } }{ { \left( { m }_{ { N }_{ 2 } }^{ 2 }-{ m }_{ { N }_{ 3 } }^{ 2 } \right)  }^{ 2 }+{ m }_{ { N }_{ 2 }}^{2}{ \Gamma  }_{ { N }_{ 3} }^{ 2 } }. 
\end{equation}
In order to generate an $O(1)$ lepton asymmetry, it is necessary to satisfy the following two conditions~\cite{Pilaftsis:1997jf},
\begin{equation}
m_{N_{3}}-m_{N_{2}}\approx \frac{1}{2}{ \Gamma  }_{ { N }_{ 3,2 } },\quad
\frac { {\rm Im}{ \left( { y }^{ \dagger  }y \right)  }_{ 23 }^{ 2 } }{ { \left( { y }^{ \dagger  }y \right)  }_{ 22 }{ \left( { y }^{ \dagger  }y \right)  }_{ 33 } } \approx 1.
\end{equation}
In general, the mass difference is larger than the tree-level decay width ($\Delta m_{N_{32}}>{ \Gamma  }_{ { N }_{ 3,2 } }$), such that,
\begin{eqnarray}
{ { \epsilon  } }_{ 2}&\simeq&\frac { {\rm Im}{ \left( { y }^{ \dagger  }y \right)  }_{ 23 }^{ 2 } }{ { \left( { y }^{ \dagger  }y \right)  }_{ 22 }{ \left( { y }^{ \dagger  }y \right)  }_{ 33 } } \frac { \left( { m }_{ { N }_{ 2 } }^{ 2 }-{ m }_{ { N }_{ 3 } }^{ 2 } \right) { m }_{ { N }_{ 2 } }{ m }_{ { N }_{ 3 } } }{ { \left( { m }_{ { N }_{ 2 } }^{ 2 }-{ m }_{ { N }_{ 3 } }^{ 2 } \right)  }^{ 2 }+{ m }_{ { N }_{ 2 }}^{2}{ \Gamma  }_{ { N }_{ 3} }^{ 2 } }\left( \frac { { \Gamma  }_{ { N }_{ 3,2 } } }{ { m }_{ { N }_{ 3} } }  \right)\nonumber \\ 
&\simeq&-\frac { {\rm Im}{ \left( { y }^{ \dagger  }y \right)  }_{ 23 }^{ 2 } }{ { \left( { y }^{ \dagger  }y \right)  }_{ 22 }{ \left( { y }^{ \dagger  }y \right)  }_{ 33 } }\frac { \Delta m_{N_{32}}{ m }_{ { N }_{ 2 } }^{2}{ m }_{ { N }_{ 3 } } }{ \Delta m_{N_{32}}^{2}{ m }_{ { N }_{ 2 } }^{2}+{ m }_{ { N }_{ 2 }}^{2}{ \Gamma  }_{ { N }_{ 3,2} }^{ 2 } }\left( \frac { { \Gamma  }_{ { N }_{ 3,2 } } }{ { m }_{ { N }_{ 3} } }  \right)\nonumber \\
&\simeq&-\frac { {\rm Im}{ \left( { y }^{ \dagger  }y \right)  }_{ 23 }^{ 2 } }{ { \left( { y }^{ \dagger  }y \right)  }_{ 22 }{ \left( { y }^{ \dagger  }y \right)  }_{ 33 } }\frac{{ \Gamma  }_{ { N }_{ 3,2 }}}{\Delta m_{N_{32}}}.
\end{eqnarray}
When we assume $\frac { {\rm Im}{ \left( { y }^{ \dagger  }y \right)  }_{ 23 }^{ 2 } }{ { \left( { y }^{ \dagger  }y \right)  }_{ 22 }{ \left( { y }^{ \dagger  }y \right)  }_{ 33 } } \approx 10^{-3}$, $m_{N_{2,3}}=1$ TeV and $\Delta m_{N_{32}}\approx 1$ MeV, the CP asymmetry factor becomes ${ \epsilon  }_{ 2}\approx10^{-9}$. 

In thermal leptogensis, the right-handed neutrino yeild ${ Y }_{ N_{2} }$ and the lepton asymmetry ${ Y }_{ \Delta L }$ satisfy the following two Boltzmann equations,
\begin{eqnarray}
\frac { d{ Y }_{ { N }_{ 2 } } }{ dT } &=&\left({ D }_{ 2 }+S\right)\left( { Y }_{ { N }_{ 2 } }-Y_{ { N }_{ 2 } }^{ eq } \right), \\
 \frac { d{ Y }_{ \Delta L } }{ dT } &=&{ { -\epsilon  }_{ 2 }D }_{ 2 }\left( { Y }_{ { N }_{ 2 } }
 -Y_{ { N }_{ 2 } }^{ eq } \right) +{ W }_{ ID }{ Y }_{ \Delta L }.
\end{eqnarray}
The scattering term $S$ describes $\Delta L = 1$ scattering effects, but we neglect this contribution for simplicity. The decay and washout terms are expressed as,
\begin{equation}
{ D }_{ 2 }\left( T \right) =\frac { { \Gamma  }_{ { N }_{ 2 } } }{ H\left( T \right)  } \frac { 1 }{ T } \frac { { K }_{ 1 }\left( {m_{N_{2}} }/{ T } \right)  }{ { K }_{ 2 }\left( { m_{N_{2} }}/{ T } \right)  } ,\ \ \ \  { W }_{ ID }\left( T \right) =\frac { 1 }{ 2 } { D }_{ 2 }\left( T \right) \frac { Y_{ { N }_{ 2 } }^{ eq }\left( { m_{N_{2} }}/{ T } \right)  }{ Y_{ \ell }^{ eq } }.
\end{equation}
The equilibrium yields of Majorana neutrinos and leptons are given by,
\begin{equation}
Y_{ { N }_{ 2 } }^{ eq }\left( T \right) =\frac { 45m_{N_{2} }^{ 2 } }{ 2{ \pi  }^{ 4 }{ T }^{ 2 } } \frac { { K }_{ 2 }\left( { m_{N_{2} }}/{ T } \right)  }{ { h }_{ \rm eff } },\ \ \ \  Y_{ \ell }^{ eq }=\frac { 15 }{ 4{ \pi  }^{ 2 }{ h }_{ \rm eff } }. 
\end{equation}
The analytical solution for the lepton asymmetry $Y_{ \Delta L}$ is given by the following formula~\cite{Fong:2013wr,Buchmuller:2005eh,Buchmuller:2004nz},
\begin{eqnarray}
{ Y }_{ \Delta L }\left( T \right) &=&{ Y }_{ \Delta L }\left( { T }_{ RE } \right) { e }^{ \int _{ { T }_{ RE } }^{ T }{ d } T' { W }_{ ID }\left( T'  \right)  }-\int _{ { T }_{ RE } }^{ T }{ d } T' { \epsilon  }_{ 2 }{ \frac { d{ Y }_{ { N }_{ 2 } } }{ dT' } { e }^{ \int _{ { T'}  }^{ T }{ d } T''{ W }_{ ID }\left( T''  \right)  } }  \\
&\simeq& -{ \epsilon  }_{ 2 }\int _{ { T }_{ RE } }^{ T }{ d } T' { \frac { d{ Y }_{ { N }_{ 2 } } }{ dT'  } { e }^{ \int _{ { T }'  }^{ T }{ d } T''{ W }_{ ID }\left( T''  \right)  } }.\label{eq:73}
\end{eqnarray}
If we assume that there is no preexisting lepton asymmetry, ${ Y }_{ \Delta L }\left( { T }_{ RE } \right)=0$, and neglect the washout term, Eq (\ref{eq:73}) reduces to the form,
\begin{equation}
{ Y }_{ \Delta L }\left( { T }_{ 0 } \right) \simeq -{ \epsilon  }_{ 2 }\int _{ { T }_{ RE } }^{ { T }_{ 0 } }{ d } T' \frac { d{ Y }_{ { N }_{ 2 } } }{ dT'  } ={ \epsilon  }_{ 2 }{ Y }_{ { N }_{ 2 } }\left( { T }_{ RE } \right).
\end{equation}
In general, the heavy Majorana neutrinos come into thermal equilibrium. The initial yield is given by the thermal equilibrium yield ${ Y }_{ { N }_{ 2 } }\left( { T }_{ RE } \right)={ Y }_{ { N }_{ 2 } }^{eq}\simeq0.004$ and the lepton asymmetry is approximately given by ${ Y }_{ \Delta L }\left( { T }_{ 0 } \right) \simeq 0.004\ { \epsilon  }_{ 2 }$. If we include washout effects, the final lepton asymmetry can be obtained as~\cite{Buchmuller:2004nz},
\begin{eqnarray}
{ Y }_{ \Delta L }\left( { T }_{ 0 } \right) &\simeq& -\frac { 27 }{ 16 }{ \epsilon  }_{ 2 }{ \left( \frac { { \Gamma  }_{ { N }_{ 2 } } }{ H\left( { m }_{ { N }_{ 2 } } \right)  }  \right)  }^{ 2 }{ Y }_{ { N }_{ 2 } }^{eq}.
\end{eqnarray}

So far we have been describing the thermal leptogenesis scenario without the singlet scalar. Now we discuss how this scenario will be modified by the singlet scalar, both when it is in and out of thermal equilibrium.

\subsection{Leptogenesis via the singlet scalar in thermal equilibrium}\label{sec:6a}
In this subsection, we discuss a new leptogenesis scenario where the singlet scalar is in thermal equilibrium and the Majorana neutrinos are generated by thermal freeze-in via the singlet scalar. In this scenario, the Higgs portal coupling has to be relatively large, $\lambda > 10^{-6}$, and the Yukawa coupling needs to be small, $\kappa_{2}<10^{-6}$. The relevant Boltzmann equations are given by the following formulas,
\begin{eqnarray}
\frac { d{ Y }_{ { N }_{ 2 } } }{ dT } &=&{ D }_{ 2 }\left( { Y }_{ { N }_{ 2 } }-Y_{ { N }_{ 2 } }^{ eq } \right)-2{ D }_{ s }Y_{s}^{ eq }, \\
 \frac { d{ Y }_{ \Delta L } }{ dT } &=&{ { -\epsilon  }_{ 2 }D }_{ 2 }\left( { Y }_{ { N }_{ 2 } }
 -Y_{ { N }_{ 2 } }^{ eq } \right) +{ W }_{ ID }{ Y }_{ \Delta L }.
\end{eqnarray}
The relevant terms are given by,
\begin{eqnarray}
{ D }_{ 2 } { Y }_{ { N }_{ 2 } }
&=&-\sqrt { \frac { 45 }{ { 4\pi  }^{ 3 }{ G }_{ N } }  } \frac { 1 }{ \sqrt { { g }_{\rm eff } }  } \frac { 1 }{ { T }^{ 3 } } \frac { { K }_{ 1 }\left( { { m }_{ N_{2} } }/{ T } \right)  }{ { K }_{ 2 }\left( { { m }_{ N_{2} } }/{ T } \right)  } \Gamma_{{ N }_{ 2 }}{ Y }_{ { N }_{ 2 } } \nonumber  \\
&=& -\frac { 3\sqrt { 5 }  }{ 2{ \pi  }^{ { 3}/{ 2 }  } } \frac { { m }_{ \rm pl } }{ \sqrt { { g }_{ \rm eff } }  } \frac { 1 }{ { T }^{ 3 } } \frac { { K }_{ 1 }\left( { { m }_{ N_{2} } }/{ T } \right)  }{ { K }_{ 2 }\left( { { m }_{ N_{2} } }/{ T } \right)  } \Gamma_{{ N }_{ 2 }}{ Y }_{ { N }_{ 2 } },  \\
{ D }_{ 2} { Y }_{ { N }_{ 2 } }^{eq} &=&-\sqrt { \frac { 45 }{ { 4\pi  }^{ 3 }{ G }_{ N } }  } \frac { 1 }{ \sqrt { { g }_{ \rm eff } }  } \frac { 1 }{ { T }^{ 3 } } \frac { { K }_{ 1 }\left( { { m }_{ N_{2} } }/{ T } \right)  }{ { K }_{ 2 }\left( { { m }_{ N_{2} } }/{ T } \right)  } \Gamma_{{ N }_{ 2 }}{ Y }_{ { N }_{ 2 } }^{ eq }
\nonumber \\
&=& -\frac { 135\sqrt { 5 }  }{ 4{ \pi  }^{ { 11 }/{ 2 }  } } \frac { { m }_{ \rm pl } }{ { h }_{ \rm eff }\sqrt { { g }_{ \rm eff } }  } \frac { { m }_{ N_{2} }^{ 2 }{ K }_{ 1 }\left( { { m }_{ N_{2} } }/{ T } \right)  }{ { T }^{ 5 } }  \Gamma_{{ N }_{ 2 }}, \\
{ D }_{ s } { Y }_{ s }^{eq} &=&-\sqrt { \frac { 45 }{ { 4\pi  }^{ 3 }{ G }_{ N } }  } \frac { 1 }{ \sqrt { { g }_{ \rm eff } }  } \frac { 1 }{ { T }^{ 3 } } \frac { { K }_{ 1 }\left( { { m }_{ s } }/{ T } \right)  }{ { K }_{ 2 }\left( { { m }_{ s } }/{ T } \right)  }\Gamma \left( s\rightarrow { N }_{ 2}{ N }_{ 2 } \right){ Y }_{ s }^{ eq }
\nonumber \\
&=& -\frac { 135\sqrt { 5 }  }{ 8{ \pi  }^{ { 11 }/{ 2 }  } } \frac { { m }_{\rm pl } }{ { h }_{ \rm eff }\sqrt { { g }_{ \rm eff } }  } \frac { { m }_{ s }^{ 2 }{ K }_{ 1 }\left( { { m }_{ s } }/{ T } \right)  }{ { T }^{ 5 } }\Gamma \left( s\rightarrow { N }_{ 2}{ N }_{ 2 } \right).
\end{eqnarray}

We can write down the analytical solution for the lepton asymmetry $Y_{ \Delta L}$ as,
\begin{equation}
{ Y }_{ \Delta L }\left( T \right) =-{ \epsilon  }_{ 2 }\int _{ { T }_{ RE } }^{ T }  \left( \frac { d{ Y }_{ { N }_{ 2 } } }{ dT'  }+2{ D }_{ s }Y_{s}^{ eq } \right){ e }^{ \int _{ { T }'  }^{ T }{ d } T''{ W }_{ ID }\left( T''  \right)  } dT'.
\label{eq:81}
\end{equation}
If we assume that the Majorana neutrinos and the initial lepton asymmetry are zero, ${ Y }_{ { N }_{ 1,2,3 } }={ Y }_{ \Delta L }\left( { T }_{ RE } \right)=0$, and neglect the washout term in Eq.(\ref{eq:81}), then we find,
\begin{eqnarray}
{ Y }_{ \Delta L }\left( { T }_{ 0 } \right) &\simeq& { \epsilon  }_{ 2 }{ Y }_{ { N }_{ 2 } }\left( { T }_{ RE } \right) -{ \epsilon  }_{ 2 }\int _{ { T }_{ RE } }^{ { T }_{ 0 } }2{ D }_{ s }Y_{s}^{ eq } \ dT \\
 &\simeq&-{ \epsilon  }_{ 2 }\int _{ { T }_{ RE } }^{ { T }_{ 0 } }2{ D }_{ s }Y_{s}^{ eq } \ dT.\label{eq:83}
 \end{eqnarray}
Finally, we analytically integrate Eq.(\ref{eq:83}) from $T_{0}=0$ to $T_{RE}=\infty$ and obtain the lepton asymmetry as,
\begin{eqnarray}
 { Y }_{ \Delta L }\left( { T }_{ 0 } \right)&\simeq& \frac { 135\sqrt { 5 }  }{ 4{ \pi  }^{ { 11 }/{ 2 } } } \frac { { m }_{ \rm pl } }{ { h }_{\rm  eff }\sqrt { g_{ \rm eff } }  }{ \epsilon  }_{ 2 } \int _{ { T }_{ RE } }^{ { T }_{ 0 } }{ \frac { { m }_{ s }^{ 2 }{ K }_{ 1 }\left( { { m }_{ s } }/{ T } \right)  }{ { T }^{ 5 } }  }\Gamma \left( s\rightarrow { N }_{ 2}{ N }_{ 2 } \right) dT \nonumber \\
&\approx&\frac { 135\sqrt { 5 }  }{ 4{ \pi  }^{ { 11 }/{ 2 } } } \frac { { m }_{ \rm pl } }{ { h }_{\rm eff }\sqrt { g_{ \rm eff } }  } { \epsilon  }_{ 2 } \int _{ \infty }^{ 0  }{ \frac { { m }_{ s }^{ 2 }{ K }_{ 1 }\left( { { m }_{ s } }/{ T } \right)  }{ { T }^{ 5 } }  }\Gamma \left( s\rightarrow { N }_{ 2 }{ N }_{ 2 } \right) dT \nonumber \\ 
&\approx&-1.59\times10^{-12} \left(\frac{{ \epsilon  }_{ 2 }}{10^{-9}}\right)\left(\frac{\kappa_{2}}{10^{-7}}\right)^{2}\left(\frac{{\rm TeV}}{m_{s}}\right).
\end{eqnarray}
The $(B + L)$-violating interactions of sphalerons come into thermal equilibrium at temperatures above the electroweak phase transition $T>T_{c}\approx 200\ \rm{GeV}$, and the lepton asymmetry can be converted into a baryon asymmetry as follows~\cite{Harvey:1990qw,Khlebnikov:1988sr,Bochkarev:1987wf},
 \begin{equation}
{ Y }_{ { \Delta  }B }\left( T \right) =\frac { 28 }{ 79 } Y_{ \Delta B-L }\left( T \right) =-\frac { 28 }{ 51 } Y_{ \Delta L }\left( T \right).
 \end{equation}
\begin{figure*}[t]
	\begin{tabular}{cc}
	\begin{minipage}{0.5\hsize}
		\centering
		\subfigure[$\ s\rightarrow N_{2}N_{2}\quad{ \epsilon  }_{ 2 }=10^{-9}$]{
		\includegraphics[width=78mm]{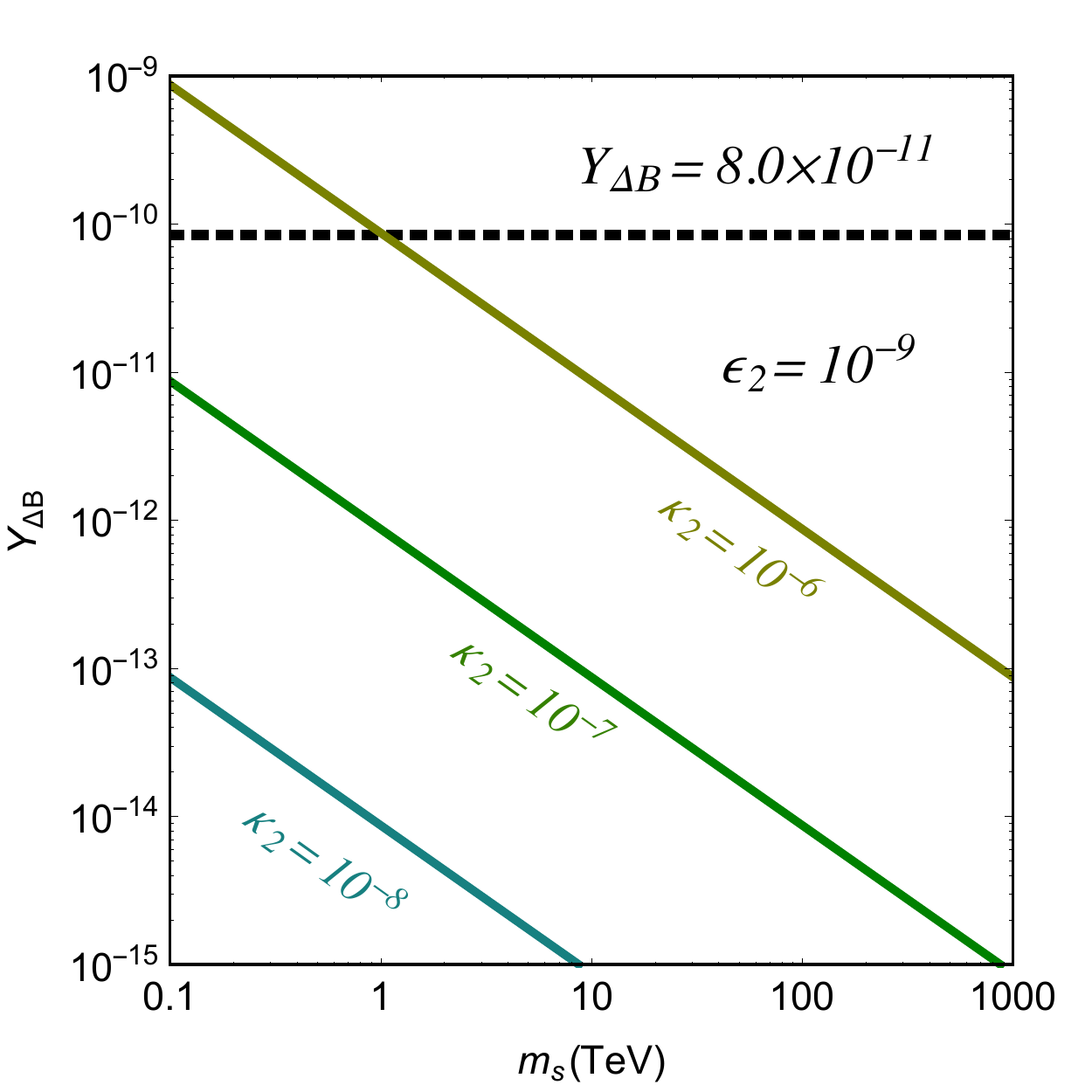}
		\label{Fig:}}
	\end{minipage}
   \begin{minipage}{0.5\hsize}
		\centering
		\subfigure[$\ s\rightarrow N_{2}N_{2}\quad\kappa_{2}=10^{-7}$]{
		\includegraphics[width=78mm]{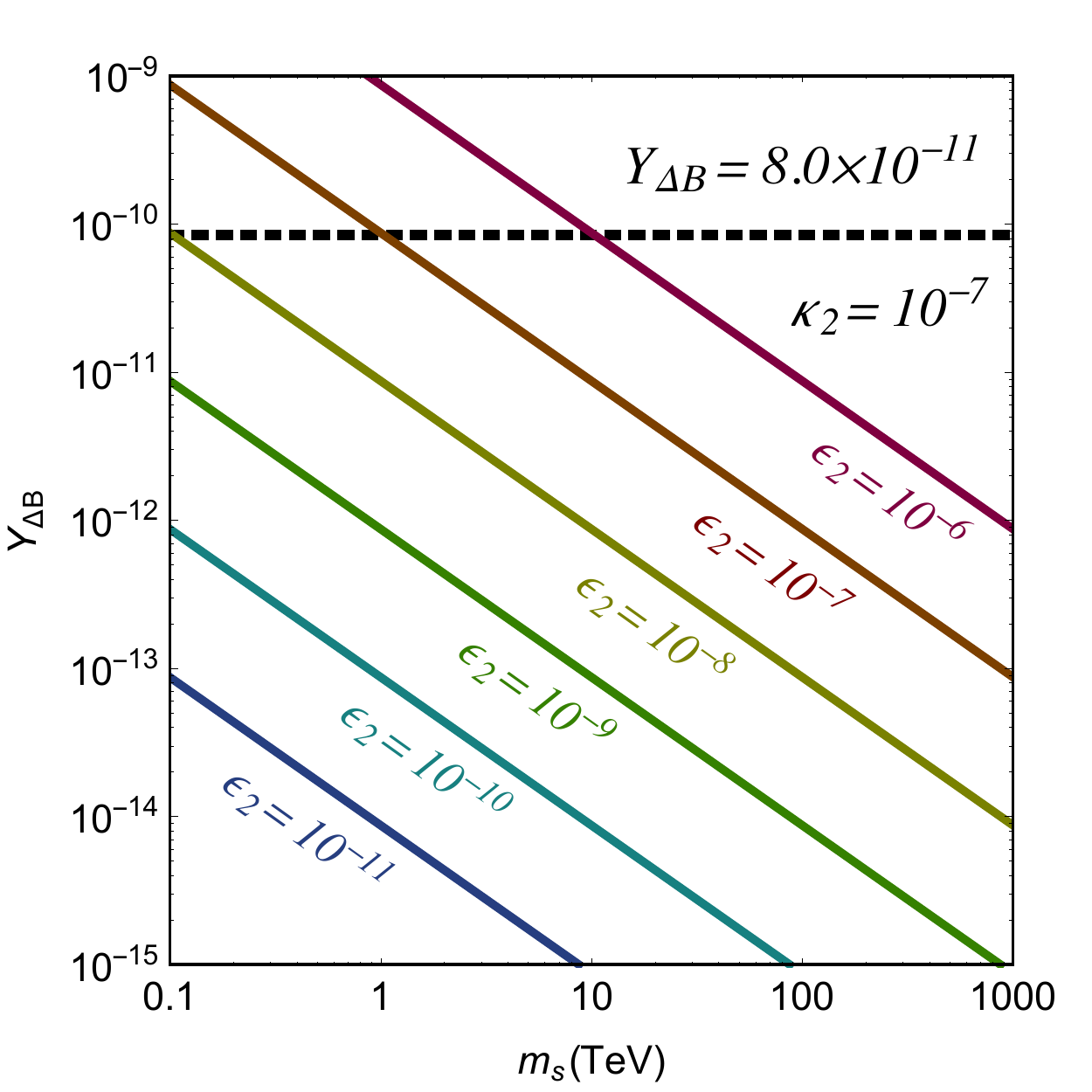}
		\label{Fig:}}
	\end{minipage}	
\end{tabular}
\caption{These figures show the dependence of the baryon asymmetry yield $Y_{ \Delta B}$ on the CP asymmetry factor ${ \epsilon  }_{ 2 }$ and Yukawa coupling $\kappa_{2}$. The Yukawa coupling should satisfy $\kappa_{2}<10^{-6}$, otherwise the sterile neutrino is in  thermal equilibrium.}\label{Fig:7}
\end{figure*}  
Therefore, the final baryon asymmetry is given by,
\begin{eqnarray}
{ Y }_{ { \Delta  }B }\left( T_{0} \right) &\approx& 0.87\times10^{-12} \left(\frac{{ \epsilon  }_{ 2 }}{10^{-9}}\right)\left(\frac{\kappa_{2}}{10^{-7}}\right)^{2}\left(\frac{{\rm TeV}}{m_{s}}\right)\nonumber \\
&\approx& 0.87\times10^{2} \left(\frac{{ \epsilon  }_{ 2 }}{10^{-9}}\right)\left(\frac{m_{N_{2}}}{{\rm TeV}}\right)\left(\frac{{\rm TeV}}{m_{s}}\right)\left(\frac{{\rm TeV}}{ \left< S \right>}\right)^{2}.\label{eq:86}
\end{eqnarray}

From BBN results, the baryon asymmetry is determined to be,
\begin{equation}
{ Y }_{ \Delta B }^{ BBN }=\left( 8.10\pm 0.85 \right) \times { 10 }^{ -11 }.
\end{equation}
From CMB measurements, the baryon asymmetry is determined to be,
\begin{equation}
{ Y }_{ \Delta B }^{ CMB }=\left( 8.79\pm 0.44 \right) \times { 10 }^{ -11 }.
\end{equation}
FIG.\ref{Fig:7} shows the dependence of the baryon asymmetry on the CP asymmetry factor ${ \epsilon  }_{ 2 }$ and the Yukawa coupling $\kappa_{2}$.  The constraint on the Yukawa coupling $\kappa_{2}<10^{-6}$ and
Eq.(\ref{eq:86}) lead to the following constraint on $m_{s}$,
\begin{equation}
\left(\frac{m_{s}}{{\rm TeV}}\right)< 1.09\left(\frac{{ \epsilon  }_{ 2 }}{10^{-9}}\right).
\end{equation}
Therefore, the mass of the singlet scalar cannot be larger than 1 TeV if we are to produce the observed amount of baryon asymmetry. In section~\ref{sec:3a}, we discussed the thermal freeze-in production of keV-MeV sterile neutrino DM and concluded that the singlet scalar should not be heavier than the TeV scale. The mass of the singlet scalar required to achieve leptogenesis is more severely restricted than the mass required for successful dark matter scenarios~\footnote{From Eq.(\ref{eq:21}) and Eq.(\ref{eq:86}) we see that in order to satisfy observational constraints on the DM density and baryon asymmetry in this scenario, the mass of the sterile neutrino must be around 100 MeV.}.

\subsection{Leptogenesis via the singlet scalar out of thermal equilibrium}\label{sec:6b}
When the Higgs portal coupling is small, $\lambda \ll 10^{-6}$, and $s$, $N_2$ and $N_{3}$ do not exist in the early Universe, they do not come into thermal equilibrium. The singlet scalar is then produced by the thermal freeze-in mechanism and decays efficiently into the Majorana neutrinos $N_{2,3}$ which generate a net lepton asymmetry. In this scenario, to determine the yields we have to solve the following Boltzmann equations,
\begin{eqnarray}
\frac { d{ Y }_{ s } }{ dT } &=&\frac { d{ Y }_{ s }^{ A } }{ dT } +{ D }_{ s }Y_{s},\label{eq:90}\\
\frac { d{ Y }_{ { N }_{ 2 } } }{ dT } &=&{ D }_{ 2 }\left( { Y }_{ { N }_{ 2 } }-Y_{ { N }_{ 2 } }^{ eq } \right)-2{ D }_{ s }Y_{s},\label{eq:91}\\
 \frac { d{ Y }_{ \Delta L } }{ dT } &=&{ { -\epsilon  }_{ 2 }D }_{ 2 }\left( { Y }_{ { N }_{ 2 } }
 -Y_{ { N }_{ 2 } }^{ eq } \right) +{ W }_{ ID }{ Y }_{ \Delta L }.\label{eq:92}
\end{eqnarray}
The relevant term in the annihilation process can be expressed as,
\begin{equation}
 \frac { d{ Y }_{ s }^{ A } }{ dT }= -\frac { 135\sqrt { 5 }  }{ 64{ \pi  }^{ { 17 }/{ 2 } } } \frac { { m }_{ \rm pl } }{ { h }_{ \rm eff }\sqrt { { g }_{ \rm eff } }  } \frac { { \lambda  }^{ 2 }{ m }_{ s } }{ { T }^{ 3 } } { K }_{ 1 }\left( { 2{ m }_{ s } }/{ T } \right).
\end{equation}
The singlet scalar decays into the Majorana neutrinos, and the decay term is given by,
\begin{eqnarray}
{ D }_{ s }Y_{s} &=&\sqrt { \frac { 45 }{ { 4\pi  }^{ 3 }{ G }_{ N } }  } \frac { 1 }{ \sqrt { { g }_{ \rm eff } }  } \frac { 1 }{ { T }^{ 3 } } \frac { { K }_{ 1 }\left( { { m }_{ s } }/{ T } \right)  }{ { K }_{ 2 }\left( { { m }_{ s } }/{ T } \right)  } \Gamma \left( s\rightarrow { N }_{ 2 }{ N }_{ 2 } \right) { Y }_{ s } \nonumber  \\
&=&\frac { 3\sqrt { 5 }  }{ 2{ \pi  }^{ { 3}/{ 2 }  } } \frac { { m }_{ \rm pl } }{ \sqrt { { g }_{ \rm eff } }  }\frac { 1 }{ { T }^{ 3 } } \frac { { K }_{ 1 }\left( { { m }_{ s } }/{ T } \right)  }{ { K }_{ 2 }\left( { { m }_{ s } }/{ T } \right)  } \Gamma \left( s\rightarrow { N }_{ 2 }{ N }_{ 2 } \right) { Y }_{ s }.
\end{eqnarray}
We then integrate Eq.(\ref{eq:90}) to estimate the yield of Majorana neutrinos $Y_{N_{2}}$,
\begin{equation}
\int _{ { T }_{ RE } }^{ { T }_{ 0 } }{ \frac { d{ Y }_{ s } }{ dT } dT }
=\int _{ { T }_{ RE } }^{ { T }_{ 0 } }{ \frac { d{ Y }_{ s }^{ A } }{ dT } dT+ } \int _{ { T }_{ RE } }^{ { T }_{ 0 } }{ { D }_{ s }{ Y }_{ s }dT }.
\end{equation}
Taking the initial yield $Y_{s}\left( { T }_{ RE } \right)$ and the final yield  $Y_{s}\left( { T }_{ 0 } \right)$ to be zero, the following equation is obtained,
\begin{equation}
\int _{ { T }_{ RE } }^{ { T }_{ 0 } }{ { D }_{ s }{ Y }_{ s }dT } =-\int _{ { T }_{ RE } }^{ { T }_{ 0 } }{ \frac { d{ Y }_{ s }^{ A } }{ dT } dT }.
\end{equation}

\begin{figure*}[t]
	\begin{tabular}{cc}
	\begin{minipage}{0.5\hsize}
		\centering
		\subfigure[$\ h\rightarrow s\rightarrow N_{2}N_{2}\quad{ \epsilon  }_{ 2 }=10^{-9}$]{
		\includegraphics[width=78mm]{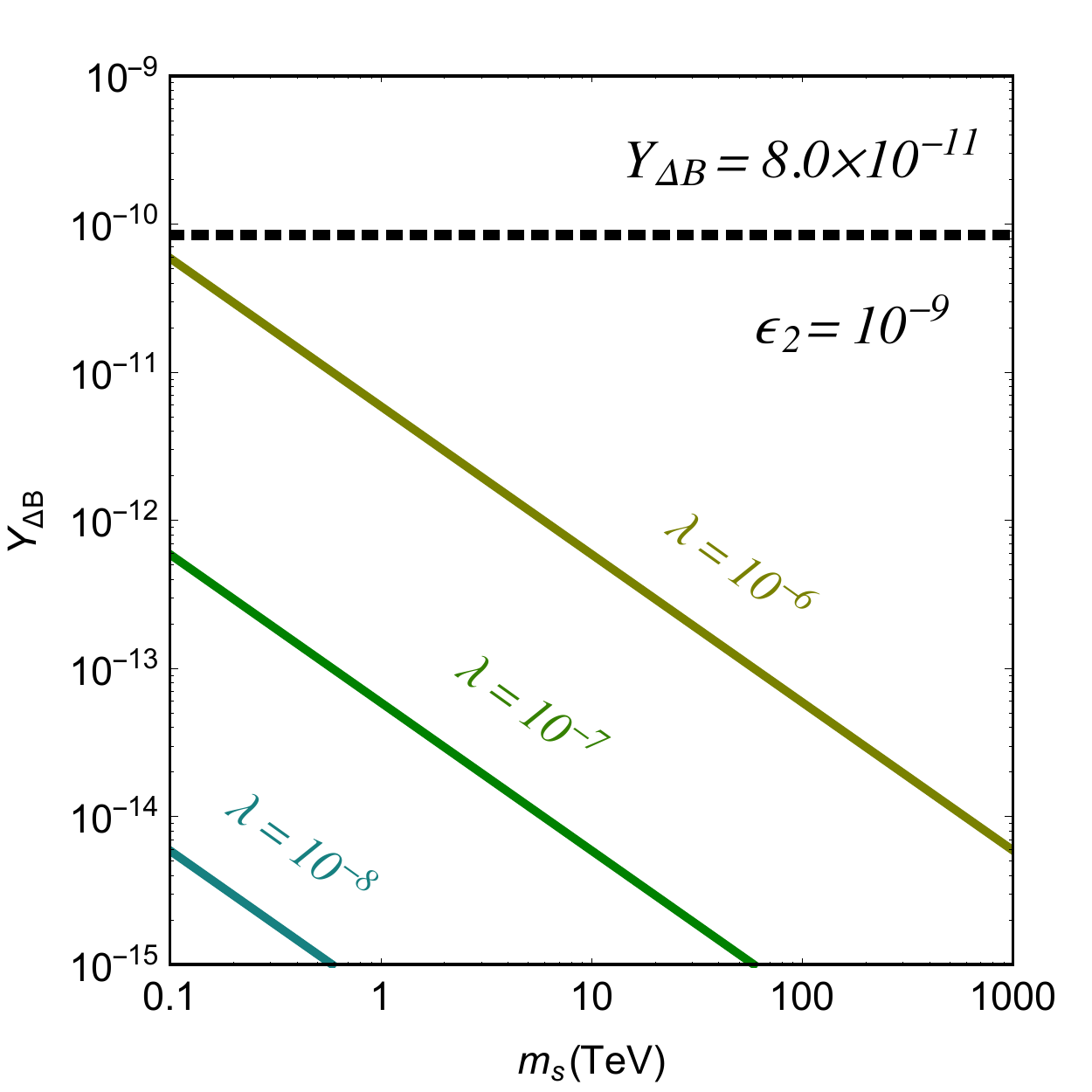}
		\label{Fig:}}
	\end{minipage}
   \begin{minipage}{0.5\hsize}
		\centering
		\subfigure[$\ h\rightarrow s\rightarrow N_{2}N_{2}\quad\kappa_{2}=10^{-7}$]{
		\includegraphics[width=78mm]{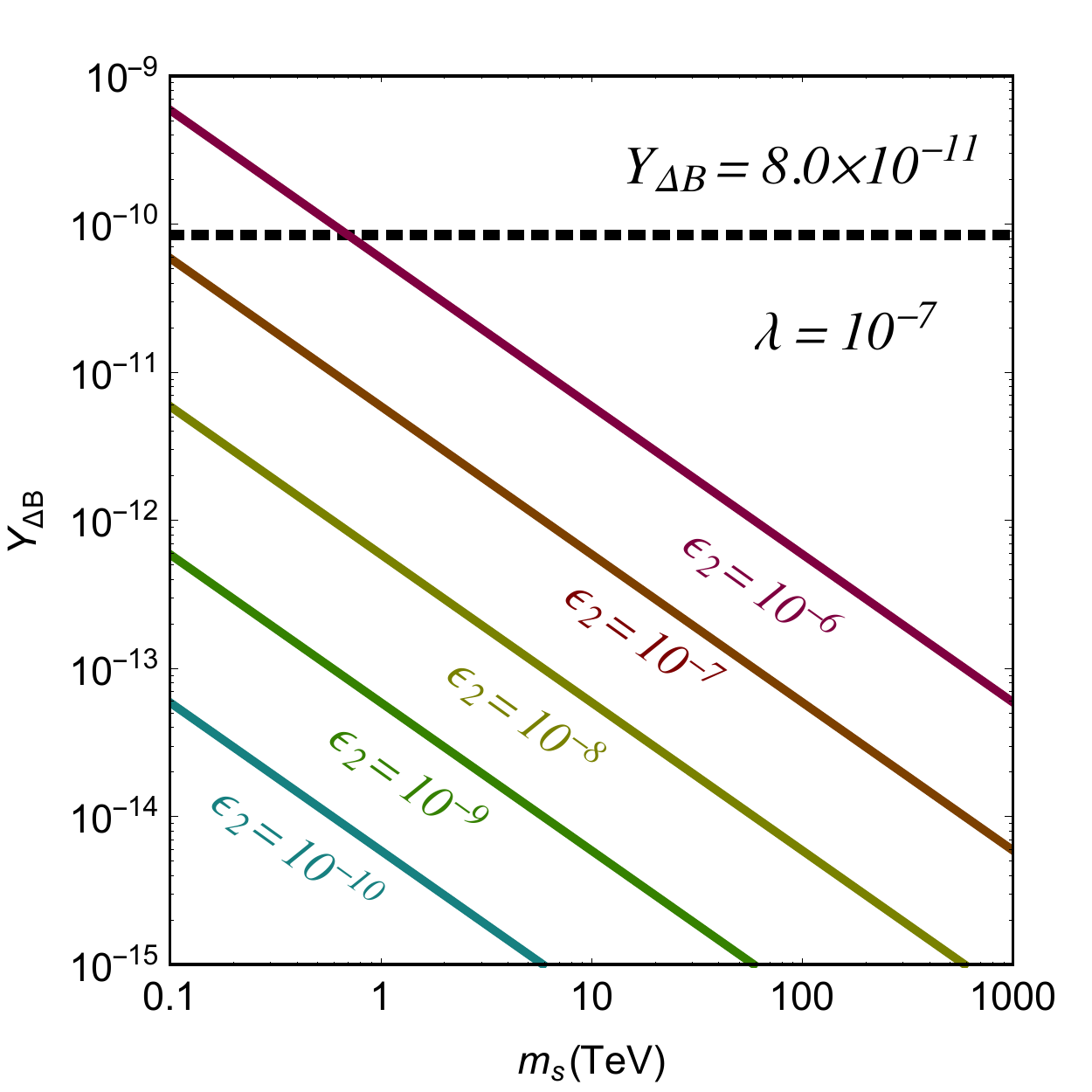}
		\label{Fig:}}
	\end{minipage}	
\end{tabular}
\caption{These figures show the dependence of the baryon asymmetry yield $Y_{ \Delta B}$, on the CP asymmetry factor ${ \epsilon  }_{ 2 }$ and the Higgs portal coupling ${\lambda }$. The Higgs portal coupling has to be small, ${\lambda }<10^{-6}$, in order to prevent the singlet scalar from entering into thermal equilibrium.}
\label{Fig:8}
\end{figure*}  

The lepton asymmetry $Y_{ \Delta L}$ at the temperature of the universe today is obtained as,
\begin{eqnarray}
{ Y }_{ \Delta L }\left( { T }_{ 0 } \right) &\simeq&-{ \epsilon  }_{ 2 }\int _{ { T }_{ RE } }^{ { T }_{ 0 } }2{ D }_{ s }Y_{s} \ dT \nonumber \\
&\simeq& -\frac { 135\sqrt { 5 }  }{ 32{ \pi  }^{ { 17 }/{ 2 } } } \frac { { m }_{\rm pl } }{ { h }_{ \rm eff }\sqrt { { g }_{ \rm eff } }  } { \epsilon  }_{ 2 }\int _{ { T }_{ RE } }^{ { T }_{ 0 } }\frac { { \lambda  }^{ 2 }{ m }_{ s } }{ { T }^{ 3 } } { K }_{ 1 }\left( { 2{ m }_{ s } }/{ T } \right)dT \nonumber \\
&\approx& -\frac { 135\sqrt { 5 }  }{ 32{ \pi  }^{ { 17 }/{ 2 } } } \frac { { m }_{ \rm pl } }{ { h }_{\rm eff }\sqrt { { g }_{ \rm eff } }  }{ \epsilon  }_{ 2 }\int _{ \infty }^{ 0  }\frac { { \lambda  }^{ 2 }{ m }_{ s } }{ { T }^{ 3 } } { K }_{ 1 }\left( { 2{ m }_{ s } }/{ T } \right)dT \nonumber \\
&\approx&1.07\times10^{-13} \left(\frac{{ \epsilon  }_{ 2 }}{10^{-9}}\right)\left(\frac{\lambda }{10^{-7}}\right)^{2}\left(\frac{{\rm TeV}}{m_{s}}\right).
\end{eqnarray}
The final baryon asymmetry is expressed as,
\begin{equation}
{ Y }_{ { \Delta  }B }\approx 0.59\times10^{-13}  \left(\frac{{ \epsilon  }_{ 2 }}{10^{-9}}\right)\left(\frac{\lambda }{10^{-7}}\right)^{2}\left(\frac{{\rm TeV}}{m_{s}}\right).\label{eq:101}
\end{equation}
In this scenario, a TeV-scale singlet scalar can generate the observed amount of baryon asymmetry. 
FIG.\ref{Fig:8} shows the dependence of baryon asymmetry on the CP asymmetry factor ${ \epsilon  }_{ 2 }$ and the Higgs portal coupling $\lambda$. The constraint on the Higgs portal coupling $\lambda<10^{-6}$ and
Eq.(\ref{eq:101}) lead to the following constraint on $m_{s}$,
\begin{equation}
\left(\frac{m_{s}}{{\rm TeV}}\right)< 0.13\left(\frac{{ \epsilon  }_{ 2 }}{10^{-9}}\right).
\end{equation}
In this leptogenesis scenario, the singlet scalar cannot be heavier than 1 TeV in order to produce the observed baryon asymmetry. In section~\ref{sec:3b}, we considered the production of keV-MeV sterile neutrino DM from the non-thermal decay of the scalar singlet, and we showed that the singlet scalar needs to be lighter than the TeV scale. This leptogenesis scenario constrains the mass of the singlet scalar more severely than the scenario that explains the sterile neutrino DM~\footnote{Note that one singlet scalar cannot be used to explain both the dark matter sector and leptogenesis if we consider thermal freeze-in production of the sterile neutrinos.  This is because the production of heavy Majorana neutrinos via non-thermal decay of the singlet scalar dominates.}

\section{Conclusion}\label{sec:7}
In this paper we have considered an extended $\nu$MSM with one additional singlet scalar.
The existence of the singlet scalar and the associated freeze-in production of sterile neutrinos and Majorana neutrinos make it possible to alleviate the usual tensions found when trying to construct dark matter and leptogenesis scenarios within the framework of the $\nu$MSM. 
We have studied two scenarios, making different assumptions about the thermal properties of the singlet scalar in each scenario. If the Higgs portal coupling is relatively large, so that the singlet scalar enters into thermal equilibrium, then the sterile neutrino and the heavy Majorana neutrinos are produced directly via the thermal freeze-in mechanism. If, on the other hand, the Higgs portal coupling is much smaller and the singlet scalar is out of thermal equilibrium, first the singlet scalar is produced by the thermal freeze-in mechanism. Then, the sterile neutrino and heavy Majorana neutrinos are produced via the non-thermal decay of the singlet scalar. In these scenarios, the sterile neutrino DM can evade Lyman-$\alpha$ bounds and X-ray constraints. We found the latter scenario to be more tightly constrained by Lyman-$\alpha$ bounds, with the singlet scalar needing to be lighter than 100 TeV to produce keV-MeV sterile neutrino DM. Thermal freeze-in leptogenesis scenarios severely restrict the singlet scalar mass to be less than 1 TeV in order to generate the observed baryon asymmetry. In summary, in the extended $\nu$MSM with a singlet scalar, a TeV-scale singlet scalar mass is needed in order to generate the observed abundance of dark matter and baryon asymmetry via the thermal freeze-in production mechanism.

\acknowledgments
We would like to thank Kazunori Kohri, Satoshi Iso, Ryuichiro Kitano and Kengo Shimada for valuable advice and useful discussions. 
H.M. would like to thank Alexander Merle and Jonathan White for valuable comments.
This work is supported by JSPS KAKENHI No. 26287039 and   Grant-in-Aid for Scientific research from the Ministry of Education, Science, Sports, and Culture (MEXT), Japan, No. 23104006, and also by World Premier International Research Center Initiative (WPI Initiative), MEXT, Japan.

\bibliography{paper}

\providecommand{\noopsort}[1]{}\providecommand{\singleletter}[1]{#1}%
\begin{thebibliography}{67}%
\makeatletter
\providecommand \@ifxundefined [1]{%
 \@ifx{#1\undefined}
}%
\providecommand \@ifnum [1]{%
 \ifnum #1\expandafter \@firstoftwo
 \else \expandafter \@secondoftwo
 \fi
}%
\providecommand \@ifx [1]{%
 \ifx #1\expandafter \@firstoftwo
 \else \expandafter \@secondoftwo
 \fi
}%
\providecommand \natexlab [1]{#1}%
\providecommand \enquote  [1]{``#1''}%
\providecommand \bibnamefont  [1]{#1}%
\providecommand \bibfnamefont [1]{#1}%
\providecommand \citenamefont [1]{#1}%
\providecommand \href@noop [0]{\@secondoftwo}%
\providecommand \href [0]{\begingroup \@sanitize@url \@href}%
\providecommand \@href[1]{\@@startlink{#1}\@@href}%
\providecommand \@@href[1]{\endgroup#1\@@endlink}%
\providecommand \@sanitize@url [0]{\catcode `\\12\catcode `\$12\catcode
  `\&12\catcode `\#12\catcode `\^12\catcode `\_12\catcode `\%12\relax}%
\providecommand \@@startlink[1]{}%
\providecommand \@@endlink[0]{}%
\providecommand \url  [0]{\begingroup\@sanitize@url \@url }%
\providecommand \@url [1]{\endgroup\@href {#1}{\urlprefix }}%
\providecommand \urlprefix  [0]{URL }%
\providecommand \Eprint [0]{\href }%
\providecommand \doibase [0]{http://dx.doi.org/}%
\providecommand \selectlanguage [0]{\@gobble}%
\providecommand \bibinfo  [0]{\@secondoftwo}%
\providecommand \bibfield  [0]{\@secondoftwo}%
\providecommand \translation [1]{[#1]}%
\providecommand \BibitemOpen [0]{}%
\providecommand \bibitemStop [0]{}%
\providecommand \bibitemNoStop [0]{.\EOS\space}%
\providecommand \EOS [0]{\spacefactor3000\relax}%
\providecommand \BibitemShut  [1]{\csname bibitem#1\endcsname}%
\let\auto@bib@innerbib\@empty
%</preamble>
\bibitem [{\citenamefont {Vecchi}(2013)}]{Vecchi:2013iza}%
  \BibitemOpen
  \bibfield  {author} {\bibinfo {author} {\bibfnamefont {L.}~\bibnamefont
  {Vecchi}},\ }\href@noop {} {\  (\bibinfo {year} {2013})},\ \Eprint
  {http://arxiv.org/abs/1312.5695} {arXiv:1312.5695 [hep-ph]} \BibitemShut
  {NoStop}%
\bibitem [{\citenamefont {Faham}(2014)}]{Faham:2014hza}%
  \BibitemOpen
  \bibfield  {author} {\bibinfo {author} {\bibfnamefont {C.}~\bibnamefont
  {Faham}} (\bibinfo {collaboration} {LUX Collaboration}),\ }\href@noop {} {\
  (\bibinfo {year} {2014})},\ \Eprint {http://arxiv.org/abs/1405.5906}
  {arXiv:1405.5906 [hep-ex]} \BibitemShut {NoStop}%
\bibitem [{\citenamefont {Akerib}\ \emph {et~al.}(2014)\citenamefont {Akerib}
  \emph {et~al.}}]{Akerib:2013tjd}%
  \BibitemOpen
  \bibfield  {author} {\bibinfo {author} {\bibfnamefont {D.}~\bibnamefont
  {Akerib}} \emph {et~al.} (\bibinfo {collaboration} {LUX Collaboration}),\
  }\href {\doibase 10.1103/PhysRevLett.112.091303} {\bibfield  {journal}
  {\bibinfo  {journal} {Phys.Rev.Lett.}\ }\textbf {\bibinfo {volume} {112}},\
  \bibinfo {pages} {091303} (\bibinfo {year} {2014})},\ \Eprint
  {http://arxiv.org/abs/1310.8214} {arXiv:1310.8214 [astro-ph.CO]} \BibitemShut
  {NoStop}%
\bibitem [{\citenamefont {Lavina}(2013)}]{Lavina:2013zxa}%
  \BibitemOpen
  \bibfield  {author} {\bibinfo {author} {\bibfnamefont {L.~S.}\ \bibnamefont
  {Lavina}} (\bibinfo {collaboration} {Collaboration XENON100}),\ }\href@noop
  {} {\  (\bibinfo {year} {2013})},\ \Eprint {http://arxiv.org/abs/1305.0224}
  {arXiv:1305.0224 [hep-ex]} \BibitemShut {NoStop}%
\bibitem [{\citenamefont {Aprile}\ \emph {et~al.}(2013)\citenamefont {Aprile}
  \emph {et~al.}}]{Aprile:2013doa}%
  \BibitemOpen
  \bibfield  {author} {\bibinfo {author} {\bibfnamefont {E.}~\bibnamefont
  {Aprile}} \emph {et~al.} (\bibinfo {collaboration} {XENON100
  Collaboration}),\ }\href {\doibase 10.1103/PhysRevLett.111.021301} {\bibfield
   {journal} {\bibinfo  {journal} {Phys.Rev.Lett.}\ }\textbf {\bibinfo {volume}
  {111}},\ \bibinfo {pages} {021301} (\bibinfo {year} {2013})},\ \Eprint
  {http://arxiv.org/abs/1301.6620} {arXiv:1301.6620 [astro-ph.CO]} \BibitemShut
  {NoStop}%
\bibitem [{\citenamefont {{Aprile}}\ \emph {et~al.}(2012)\citenamefont
  {{Aprile}}, \citenamefont {{Alfonsi}}, \citenamefont {{Arisaka}},
  \citenamefont {{Arneodo}}, \citenamefont {{Balan}}, \citenamefont {{Baudis}},
  \citenamefont {{Bauermeister}}, \citenamefont {{Behrens}}, \citenamefont
  {{Beltrame}}, \citenamefont {{Bokeloh}}, \citenamefont {{Brown}},
  \citenamefont {{Bruno}}, \citenamefont {{Budnik}}, \citenamefont {{Cardoso}},
  \citenamefont {{Chen}}, \citenamefont {{Choi}}, \citenamefont {{Cline}},
  \citenamefont {{Colijn}}, \citenamefont {{Contreras}}, \citenamefont
  {{Cussonneau}}, \citenamefont {{Decowski}}, \citenamefont {{Duchovni}},
  \citenamefont {{Fattori}}, \citenamefont {{Ferella}}, \citenamefont
  {{Fulgione}}, \citenamefont {{Gao}}, \citenamefont {{Garbini}}, \citenamefont
  {{Ghag}}, \citenamefont {{Giboni}}, \citenamefont {{Goetzke}}, \citenamefont
  {{Grignon}}, \citenamefont {{Gross}}, \citenamefont {{Hampel}}, \citenamefont
  {{Kaether}}, \citenamefont {{Kish}}, \citenamefont {{Lamblin}}, \citenamefont
  {{Landsman}}, \citenamefont {{Lang}}, \citenamefont {{Le Calloch}},
  \citenamefont {{Levy}}, \citenamefont {{Lim}}, \citenamefont {{Lin}},
  \citenamefont {{Lindemann}}, \citenamefont {{Lindner}}, \citenamefont
  {{Lopes}}, \citenamefont {{Lung}}, \citenamefont {{Marrod{\'a}n Undagoitia}},
  \citenamefont {{Massoli}}, \citenamefont {{Melgarejo Fernandez}},
  \citenamefont {{Meng}}, \citenamefont {{Molinario}}, \citenamefont {{Nativ}},
  \citenamefont {{Ni}}, \citenamefont {{Oberlack}}, \citenamefont {{Orrigo}},
  \citenamefont {{Pantic}}, \citenamefont {{Persiani}}, \citenamefont
  {{Plante}}, \citenamefont {{Priel}}, \citenamefont {{Rizzo}}, \citenamefont
  {{Rosendahl}}, \citenamefont {{dos Santos}}, \citenamefont {{Sartorelli}},
  \citenamefont {{Schreiner}}, \citenamefont {{Schumann}}, \citenamefont
  {{Scotto Lavina}}, \citenamefont {{Scovell}}, \citenamefont {{Selvi}},
  \citenamefont {{Shagin}}, \citenamefont {{Simgen}}, \citenamefont
  {{Teymourian}}, \citenamefont {{Thers}}, \citenamefont {{Vitells}},
  \citenamefont {{Wang}}, \citenamefont {{Weber}},\ and\ \citenamefont
  {{Weinheimer}}}]{2012PhRvL.109r1301A}%
  \BibitemOpen
  \bibfield  {author} {\bibinfo {author} {\bibfnamefont {E.}~\bibnamefont
  {{Aprile}}}, \bibinfo {author} {\bibfnamefont {M.}~\bibnamefont {{Alfonsi}}},
  \bibinfo {author} {\bibfnamefont {K.}~\bibnamefont {{Arisaka}}}, \bibinfo
  {author} {\bibfnamefont {F.}~\bibnamefont {{Arneodo}}}, \bibinfo {author}
  {\bibfnamefont {C.}~\bibnamefont {{Balan}}}, \bibinfo {author} {\bibfnamefont
  {L.}~\bibnamefont {{Baudis}}}, \bibinfo {author} {\bibfnamefont
  {B.}~\bibnamefont {{Bauermeister}}}, \bibinfo {author} {\bibfnamefont
  {A.}~\bibnamefont {{Behrens}}}, \bibinfo {author} {\bibfnamefont
  {P.}~\bibnamefont {{Beltrame}}}, \bibinfo {author} {\bibfnamefont
  {K.}~\bibnamefont {{Bokeloh}}}, \bibinfo {author} {\bibfnamefont
  {E.}~\bibnamefont {{Brown}}}, \bibinfo {author} {\bibfnamefont
  {G.}~\bibnamefont {{Bruno}}}, \bibinfo {author} {\bibfnamefont
  {R.}~\bibnamefont {{Budnik}}}, \bibinfo {author} {\bibfnamefont {J.~M.~R.}\
  \bibnamefont {{Cardoso}}}, \bibinfo {author} {\bibfnamefont {W.-T.}\
  \bibnamefont {{Chen}}}, \bibinfo {author} {\bibfnamefont {B.}~\bibnamefont
  {{Choi}}}, \bibinfo {author} {\bibfnamefont {D.}~\bibnamefont {{Cline}}},
  \bibinfo {author} {\bibfnamefont {A.~P.}\ \bibnamefont {{Colijn}}}, \bibinfo
  {author} {\bibfnamefont {H.}~\bibnamefont {{Contreras}}}, \bibinfo {author}
  {\bibfnamefont {J.~P.}\ \bibnamefont {{Cussonneau}}}, \bibinfo {author}
  {\bibfnamefont {M.~P.}\ \bibnamefont {{Decowski}}}, \bibinfo {author}
  {\bibfnamefont {E.}~\bibnamefont {{Duchovni}}}, \bibinfo {author}
  {\bibfnamefont {S.}~\bibnamefont {{Fattori}}}, \bibinfo {author}
  {\bibfnamefont {A.~D.}\ \bibnamefont {{Ferella}}}, \bibinfo {author}
  {\bibfnamefont {W.}~\bibnamefont {{Fulgione}}}, \bibinfo {author}
  {\bibfnamefont {F.}~\bibnamefont {{Gao}}}, \bibinfo {author} {\bibfnamefont
  {M.}~\bibnamefont {{Garbini}}}, \bibinfo {author} {\bibfnamefont
  {C.}~\bibnamefont {{Ghag}}}, \bibinfo {author} {\bibfnamefont {K.-L.}\
  \bibnamefont {{Giboni}}}, \bibinfo {author} {\bibfnamefont {L.~W.}\
  \bibnamefont {{Goetzke}}}, \bibinfo {author} {\bibfnamefont {C.}~\bibnamefont
  {{Grignon}}}, \bibinfo {author} {\bibfnamefont {E.}~\bibnamefont {{Gross}}},
  \bibinfo {author} {\bibfnamefont {W.}~\bibnamefont {{Hampel}}}, \bibinfo
  {author} {\bibfnamefont {F.}~\bibnamefont {{Kaether}}}, \bibinfo {author}
  {\bibfnamefont {A.}~\bibnamefont {{Kish}}}, \bibinfo {author} {\bibfnamefont
  {J.}~\bibnamefont {{Lamblin}}}, \bibinfo {author} {\bibfnamefont
  {H.}~\bibnamefont {{Landsman}}}, \bibinfo {author} {\bibfnamefont {R.~F.}\
  \bibnamefont {{Lang}}}, \bibinfo {author} {\bibfnamefont {M.}~\bibnamefont
  {{Le Calloch}}}, \bibinfo {author} {\bibfnamefont {C.}~\bibnamefont
  {{Levy}}}, \bibinfo {author} {\bibfnamefont {K.~E.}\ \bibnamefont {{Lim}}},
  \bibinfo {author} {\bibfnamefont {Q.}~\bibnamefont {{Lin}}}, \bibinfo
  {author} {\bibfnamefont {S.}~\bibnamefont {{Lindemann}}}, \bibinfo {author}
  {\bibfnamefont {M.}~\bibnamefont {{Lindner}}}, \bibinfo {author}
  {\bibfnamefont {J.~A.~M.}\ \bibnamefont {{Lopes}}}, \bibinfo {author}
  {\bibfnamefont {K.}~\bibnamefont {{Lung}}}, \bibinfo {author} {\bibfnamefont
  {T.}~\bibnamefont {{Marrod{\'a}n Undagoitia}}}, \bibinfo {author}
  {\bibfnamefont {F.~V.}\ \bibnamefont {{Massoli}}}, \bibinfo {author}
  {\bibfnamefont {A.~J.}\ \bibnamefont {{Melgarejo Fernandez}}}, \bibinfo
  {author} {\bibfnamefont {Y.}~\bibnamefont {{Meng}}}, \bibinfo {author}
  {\bibfnamefont {A.}~\bibnamefont {{Molinario}}}, \bibinfo {author}
  {\bibfnamefont {E.}~\bibnamefont {{Nativ}}}, \bibinfo {author} {\bibfnamefont
  {K.}~\bibnamefont {{Ni}}}, \bibinfo {author} {\bibfnamefont {U.}~\bibnamefont
  {{Oberlack}}}, \bibinfo {author} {\bibfnamefont {S.~E.~A.}\ \bibnamefont
  {{Orrigo}}}, \bibinfo {author} {\bibfnamefont {E.}~\bibnamefont {{Pantic}}},
  \bibinfo {author} {\bibfnamefont {R.}~\bibnamefont {{Persiani}}}, \bibinfo
  {author} {\bibfnamefont {G.}~\bibnamefont {{Plante}}}, \bibinfo {author}
  {\bibfnamefont {N.}~\bibnamefont {{Priel}}}, \bibinfo {author} {\bibfnamefont
  {A.}~\bibnamefont {{Rizzo}}}, \bibinfo {author} {\bibfnamefont
  {S.}~\bibnamefont {{Rosendahl}}}, \bibinfo {author} {\bibfnamefont
  {J.~M.~F.}\ \bibnamefont {{dos Santos}}}, \bibinfo {author} {\bibfnamefont
  {G.}~\bibnamefont {{Sartorelli}}}, \bibinfo {author} {\bibfnamefont
  {J.}~\bibnamefont {{Schreiner}}}, \bibinfo {author} {\bibfnamefont
  {M.}~\bibnamefont {{Schumann}}}, \bibinfo {author} {\bibfnamefont
  {L.}~\bibnamefont {{Scotto Lavina}}}, \bibinfo {author} {\bibfnamefont
  {P.~R.}\ \bibnamefont {{Scovell}}}, \bibinfo {author} {\bibfnamefont
  {M.}~\bibnamefont {{Selvi}}}, \bibinfo {author} {\bibfnamefont
  {P.}~\bibnamefont {{Shagin}}}, \bibinfo {author} {\bibfnamefont
  {H.}~\bibnamefont {{Simgen}}}, \bibinfo {author} {\bibfnamefont
  {A.}~\bibnamefont {{Teymourian}}}, \bibinfo {author} {\bibfnamefont
  {D.}~\bibnamefont {{Thers}}}, \bibinfo {author} {\bibfnamefont
  {O.}~\bibnamefont {{Vitells}}}, \bibinfo {author} {\bibfnamefont
  {H.}~\bibnamefont {{Wang}}}, \bibinfo {author} {\bibfnamefont
  {M.}~\bibnamefont {{Weber}}}, \ and\ \bibinfo {author} {\bibfnamefont
  {C.}~\bibnamefont {{Weinheimer}}},\ }\href {\doibase
  10.1103/PhysRevLett.109.181301} {\bibfield  {journal} {\bibinfo  {journal}
  {Physical Review Letters}\ }\textbf {\bibinfo {volume} {109}},\ \bibinfo
  {eid} {181301} (\bibinfo {year} {2012})},\ \Eprint
  {http://arxiv.org/abs/1207.5988} {arXiv:1207.5988 [astro-ph.CO]} \BibitemShut
  {NoStop}%
\bibitem [{\citenamefont {Hall}\ \emph {et~al.}(2010)\citenamefont {Hall},
  \citenamefont {Jedamzik}, \citenamefont {March-Russell},\ and\ \citenamefont
  {West}}]{Hall:2009bx}%
  \BibitemOpen
  \bibfield  {author} {\bibinfo {author} {\bibfnamefont {L.~J.}\ \bibnamefont
  {Hall}}, \bibinfo {author} {\bibfnamefont {K.}~\bibnamefont {Jedamzik}},
  \bibinfo {author} {\bibfnamefont {J.}~\bibnamefont {March-Russell}}, \ and\
  \bibinfo {author} {\bibfnamefont {S.~M.}\ \bibnamefont {West}},\ }\href
  {\doibase 10.1007/JHEP03(2010)080} {\bibfield  {journal} {\bibinfo  {journal}
  {JHEP}\ }\textbf {\bibinfo {volume} {1003}},\ \bibinfo {pages} {080}
  (\bibinfo {year} {2010})},\ \Eprint {http://arxiv.org/abs/0911.1120}
  {arXiv:0911.1120 [hep-ph]} \BibitemShut {NoStop}%
\bibitem [{\citenamefont {Blennow}, \citenamefont {Fernandez-Martinez},\ and\
  \citenamefont {Zaldivar}(2014)}]{Blennow:2013jba}%
  \BibitemOpen
  \bibfield  {author} {\bibinfo {author} {\bibfnamefont {M.}~\bibnamefont
  {Blennow}}, \bibinfo {author} {\bibfnamefont {E.}~\bibnamefont
  {Fernandez-Martinez}}, \ and\ \bibinfo {author} {\bibfnamefont
  {B.}~\bibnamefont {Zaldivar}},\ }\href {\doibase
  10.1088/1475-7516/2014/01/003} {\bibfield  {journal} {\bibinfo  {journal}
  {JCAP}\ }\textbf {\bibinfo {volume} {1401}},\ \bibinfo {pages} {003}
  (\bibinfo {year} {2014})},\ \Eprint {http://arxiv.org/abs/1309.7348}
  {arXiv:1309.7348 [hep-ph]} \BibitemShut {NoStop}%
%%CITATION = ARXIV:1309.7348;%%
\bibitem [{\citenamefont {Yaguna}(2012)}]{Yaguna:2011ei}%
  \BibitemOpen
  \bibfield  {author} {\bibinfo {author} {\bibfnamefont {C.~E.}\ \bibnamefont
  {Yaguna}},\ }\href {\doibase 10.1088/1475-7516/2012/02/006} {\bibfield
  {journal} {\bibinfo  {journal} {JCAP}\ }\textbf {\bibinfo {volume} {1202}},\
  \bibinfo {pages} {006} (\bibinfo {year} {2012})},\ \Eprint
  {http://arxiv.org/abs/1111.6831} {arXiv:1111.6831 [hep-ph]} \BibitemShut
  {NoStop}%
%%CITATION = ARXIV:1111.6831;%%
\bibitem [{\citenamefont {Bhupal~Dev}, \citenamefont {Mazumdar},\ and\
  \citenamefont {Qutub}(2014)}]{Dev:2013yza}%
  \BibitemOpen
  \bibfield  {author} {\bibinfo {author} {\bibfnamefont {P.}~\bibnamefont
  {Bhupal~Dev}}, \bibinfo {author} {\bibfnamefont {A.}~\bibnamefont
  {Mazumdar}}, \ and\ \bibinfo {author} {\bibfnamefont {S.}~\bibnamefont
  {Qutub}},\ }\href {\doibase 10.3389/fphy.2014.00026} {\bibfield  {journal}
  {\bibinfo  {journal} {Front.Phys.}\ }\textbf {\bibinfo {volume} {2}},\
  \bibinfo {pages} {26} (\bibinfo {year} {2014})},\ \Eprint
  {http://arxiv.org/abs/1311.5297} {arXiv:1311.5297 [hep-ph]} \BibitemShut
  {NoStop}%
%%CITATION = ARXIV:1311.5297;%%
\bibitem [{\citenamefont {Asaka}, \citenamefont {Blanchet},\ and\ \citenamefont
  {Shaposhnikov}(2005)}]{Asaka:2005an}%
  \BibitemOpen
  \bibfield  {author} {\bibinfo {author} {\bibfnamefont {T.}~\bibnamefont
  {Asaka}}, \bibinfo {author} {\bibfnamefont {S.}~\bibnamefont {Blanchet}}, \
  and\ \bibinfo {author} {\bibfnamefont {M.}~\bibnamefont {Shaposhnikov}},\
  }\href {\doibase 10.1016/j.physletb.2005.09.070} {\bibfield  {journal}
  {\bibinfo  {journal} {Phys.Lett.}\ }\textbf {\bibinfo {volume} {B631}},\
  \bibinfo {pages} {151} (\bibinfo {year} {2005})},\ \Eprint
  {http://arxiv.org/abs/hep-ph/0503065} {arXiv:hep-ph/0503065 [hep-ph]}
  \BibitemShut {NoStop}%
\bibitem [{\citenamefont {Asaka}\ and\ \citenamefont
  {Shaposhnikov}(2005)}]{Asaka:2005pn}%
  \BibitemOpen
  \bibfield  {author} {\bibinfo {author} {\bibfnamefont {T.}~\bibnamefont
  {Asaka}}\ and\ \bibinfo {author} {\bibfnamefont {M.}~\bibnamefont
  {Shaposhnikov}},\ }\href {\doibase 10.1016/j.physletb.2005.06.020} {\bibfield
   {journal} {\bibinfo  {journal} {Phys.Lett.}\ }\textbf {\bibinfo {volume}
  {B620}},\ \bibinfo {pages} {17} (\bibinfo {year} {2005})},\ \Eprint
  {http://arxiv.org/abs/hep-ph/0505013} {arXiv:hep-ph/0505013 [hep-ph]}
  \BibitemShut {NoStop}%
\bibitem [{\citenamefont {Canetti}\ \emph {et~al.}(2013)\citenamefont
  {Canetti}, \citenamefont {Drewes}, \citenamefont {Frossard},\ and\
  \citenamefont {Shaposhnikov}}]{Canetti:2012kh}%
  \BibitemOpen
  \bibfield  {author} {\bibinfo {author} {\bibfnamefont {L.}~\bibnamefont
  {Canetti}}, \bibinfo {author} {\bibfnamefont {M.}~\bibnamefont {Drewes}},
  \bibinfo {author} {\bibfnamefont {T.}~\bibnamefont {Frossard}}, \ and\
  \bibinfo {author} {\bibfnamefont {M.}~\bibnamefont {Shaposhnikov}},\ }\href
  {\doibase 10.1103/PhysRevD.87.093006} {\bibfield  {journal} {\bibinfo
  {journal} {Phys.Rev.}\ }\textbf {\bibinfo {volume} {D87}},\ \bibinfo {pages}
  {093006} (\bibinfo {year} {2013})},\ \Eprint {http://arxiv.org/abs/1208.4607}
  {arXiv:1208.4607 [hep-ph]} \BibitemShut {NoStop}%
\bibitem [{\citenamefont {Dodelson}\ and\ \citenamefont
  {Widrow}(1994)}]{Dodelson:1993je}%
  \BibitemOpen
  \bibfield  {author} {\bibinfo {author} {\bibfnamefont {S.}~\bibnamefont
  {Dodelson}}\ and\ \bibinfo {author} {\bibfnamefont {L.~M.}\ \bibnamefont
  {Widrow}},\ }\href {\doibase 10.1103/PhysRevLett.72.17} {\bibfield  {journal}
  {\bibinfo  {journal} {Phys.Rev.Lett.}\ }\textbf {\bibinfo {volume} {72}},\
  \bibinfo {pages} {17} (\bibinfo {year} {1994})},\ \Eprint
  {http://arxiv.org/abs/hep-ph/9303287} {arXiv:hep-ph/9303287 [hep-ph]}
  \BibitemShut {NoStop}%
\bibitem [{\citenamefont {Akhmedov}, \citenamefont {Rubakov},\ and\
  \citenamefont {Smirnov}(1998)}]{Akhmedov:1998qx}%
  \BibitemOpen
  \bibfield  {author} {\bibinfo {author} {\bibfnamefont {E.~K.}\ \bibnamefont
  {Akhmedov}}, \bibinfo {author} {\bibfnamefont {V.}~\bibnamefont {Rubakov}}, \
  and\ \bibinfo {author} {\bibfnamefont {A.~Y.}\ \bibnamefont {Smirnov}},\
  }\href {\doibase 10.1103/PhysRevLett.81.1359} {\bibfield  {journal} {\bibinfo
   {journal} {Phys.Rev.Lett.}\ }\textbf {\bibinfo {volume} {81}},\ \bibinfo
  {pages} {1359} (\bibinfo {year} {1998})},\ \Eprint
  {http://arxiv.org/abs/hep-ph/9803255} {arXiv:hep-ph/9803255 [hep-ph]}
  \BibitemShut {NoStop}%
%%CITATION = HEP-PH/9803255;%%
\bibitem [{\citenamefont {Canetti}, \citenamefont {Drewes},\ and\ \citenamefont
  {Shaposhnikov}(2013)}]{Canetti:2012vf}%
  \BibitemOpen
  \bibfield  {author} {\bibinfo {author} {\bibfnamefont {L.}~\bibnamefont
  {Canetti}}, \bibinfo {author} {\bibfnamefont {M.}~\bibnamefont {Drewes}}, \
  and\ \bibinfo {author} {\bibfnamefont {M.}~\bibnamefont {Shaposhnikov}},\
  }\href {\doibase 10.1103/PhysRevLett.110.061801} {\bibfield  {journal}
  {\bibinfo  {journal} {Phys.Rev.Lett.}\ }\textbf {\bibinfo {volume} {110}},\
  \bibinfo {pages} {061801} (\bibinfo {year} {2013})},\ \Eprint
  {http://arxiv.org/abs/1204.3902} {arXiv:1204.3902 [hep-ph]} \BibitemShut
  {NoStop}%
\bibitem [{\citenamefont {Bezrukov}\ and\ \citenamefont
  {Shaposhnikov}(2008)}]{Bezrukov:2007ep}%
  \BibitemOpen
  \bibfield  {author} {\bibinfo {author} {\bibfnamefont {F.~L.}\ \bibnamefont
  {Bezrukov}}\ and\ \bibinfo {author} {\bibfnamefont {M.}~\bibnamefont
  {Shaposhnikov}},\ }\href {\doibase 10.1016/j.physletb.2007.11.072} {\bibfield
   {journal} {\bibinfo  {journal} {Phys.Lett.}\ }\textbf {\bibinfo {volume}
  {B659}},\ \bibinfo {pages} {703} (\bibinfo {year} {2008})},\ \Eprint
  {http://arxiv.org/abs/0710.3755} {arXiv:0710.3755 [hep-th]} \BibitemShut
  {NoStop}%
\bibitem [{\citenamefont {Bezrukov}\ \emph {et~al.}(2011)\citenamefont
  {Bezrukov}, \citenamefont {Magnin}, \citenamefont {Shaposhnikov},\ and\
  \citenamefont {Sibiryakov}}]{Bezrukov:2010jz}%
  \BibitemOpen
  \bibfield  {author} {\bibinfo {author} {\bibfnamefont {F.}~\bibnamefont
  {Bezrukov}}, \bibinfo {author} {\bibfnamefont {A.}~\bibnamefont {Magnin}},
  \bibinfo {author} {\bibfnamefont {M.}~\bibnamefont {Shaposhnikov}}, \ and\
  \bibinfo {author} {\bibfnamefont {S.}~\bibnamefont {Sibiryakov}},\ }\href
  {\doibase 10.1007/JHEP01(2011)016} {\bibfield  {journal} {\bibinfo  {journal}
  {JHEP}\ }\textbf {\bibinfo {volume} {1101}},\ \bibinfo {pages} {016}
  (\bibinfo {year} {2011})},\ \Eprint {http://arxiv.org/abs/1008.5157}
  {arXiv:1008.5157 [hep-ph]} \BibitemShut {NoStop}%
%%CITATION = ARXIV:1008.5157;%%
\bibitem [{\citenamefont {Abazajian}\ and\ \citenamefont
  {Koushiappas}(2006)}]{Abazajian:2006yn}%
  \BibitemOpen
  \bibfield  {author} {\bibinfo {author} {\bibfnamefont {K.}~\bibnamefont
  {Abazajian}}\ and\ \bibinfo {author} {\bibfnamefont {S.~M.}\ \bibnamefont
  {Koushiappas}},\ }\href {\doibase 10.1103/PhysRevD.74.023527} {\bibfield
  {journal} {\bibinfo  {journal} {Phys.Rev.}\ }\textbf {\bibinfo {volume}
  {D74}},\ \bibinfo {pages} {023527} (\bibinfo {year} {2006})},\ \Eprint
  {http://arxiv.org/abs/astro-ph/0605271} {arXiv:astro-ph/0605271 [astro-ph]}
  \BibitemShut {NoStop}%
%%CITATION = ASTRO-PH/0605271;%%
\bibitem [{\citenamefont {Shaposhnikov}\ and\ \citenamefont
  {Tkachev}(2006)}]{Shaposhnikov:2006xi}%
  \BibitemOpen
  \bibfield  {author} {\bibinfo {author} {\bibfnamefont {M.}~\bibnamefont
  {Shaposhnikov}}\ and\ \bibinfo {author} {\bibfnamefont {I.}~\bibnamefont
  {Tkachev}},\ }\href {\doibase 10.1016/j.physletb.2006.06.063} {\bibfield
  {journal} {\bibinfo  {journal} {Phys.Lett.}\ }\textbf {\bibinfo {volume}
  {B639}},\ \bibinfo {pages} {414} (\bibinfo {year} {2006})},\ \Eprint
  {http://arxiv.org/abs/hep-ph/0604236} {arXiv:hep-ph/0604236 [hep-ph]}
  \BibitemShut {NoStop}%
\bibitem [{\citenamefont {Kusenko}(2006)}]{Kusenko:2006rh}%
  \BibitemOpen
  \bibfield  {author} {\bibinfo {author} {\bibfnamefont {A.}~\bibnamefont
  {Kusenko}},\ }\href {\doibase 10.1103/PhysRevLett.97.241301} {\bibfield
  {journal} {\bibinfo  {journal} {Phys.Rev.Lett.}\ }\textbf {\bibinfo {volume}
  {97}},\ \bibinfo {pages} {241301} (\bibinfo {year} {2006})},\ \Eprint
  {http://arxiv.org/abs/hep-ph/0609081} {arXiv:hep-ph/0609081 [hep-ph]}
  \BibitemShut {NoStop}%
%%CITATION = HEP-PH/0609081;%%
\bibitem [{\citenamefont {Petraki}\ and\ \citenamefont
  {Kusenko}(2008)}]{Petraki:2007gq}%
  \BibitemOpen
  \bibfield  {author} {\bibinfo {author} {\bibfnamefont {K.}~\bibnamefont
  {Petraki}}\ and\ \bibinfo {author} {\bibfnamefont {A.}~\bibnamefont
  {Kusenko}},\ }\href {\doibase 10.1103/PhysRevD.77.065014} {\bibfield
  {journal} {\bibinfo  {journal} {Phys.Rev.}\ }\textbf {\bibinfo {volume}
  {D77}},\ \bibinfo {pages} {065014} (\bibinfo {year} {2008})},\ \Eprint
  {http://arxiv.org/abs/0711.4646} {arXiv:0711.4646 [hep-ph]} \BibitemShut
  {NoStop}%
%%CITATION = ARXIV:0711.4646;%%
\bibitem [{\citenamefont {Merle}, \citenamefont {Niro},\ and\ \citenamefont
  {Schmidt}(2014)}]{Merle:2013wta}%
  \BibitemOpen
  \bibfield  {author} {\bibinfo {author} {\bibfnamefont {A.}~\bibnamefont
  {Merle}}, \bibinfo {author} {\bibfnamefont {V.}~\bibnamefont {Niro}}, \ and\
  \bibinfo {author} {\bibfnamefont {D.}~\bibnamefont {Schmidt}},\ }\href
  {\doibase 10.1088/1475-7516/2014/03/028} {\bibfield  {journal} {\bibinfo
  {journal} {JCAP}\ }\textbf {\bibinfo {volume} {1403}},\ \bibinfo {pages}
  {028} (\bibinfo {year} {2014})},\ \Eprint {http://arxiv.org/abs/1306.3996}
  {arXiv:1306.3996 [hep-ph]} \BibitemShut {NoStop}%
%%CITATION = ARXIV:1306.3996;%%
\bibitem [{\citenamefont {Adulpravitchai}\ and\ \citenamefont
  {Schmidt}(2014)}]{Adulpravitchai:2014xna}%
  \BibitemOpen
  \bibfield  {author} {\bibinfo {author} {\bibfnamefont {A.}~\bibnamefont
  {Adulpravitchai}}\ and\ \bibinfo {author} {\bibfnamefont {M.~A.}\
  \bibnamefont {Schmidt}},\ }\href@noop {} {\  (\bibinfo {year} {2014})},\
  \Eprint {http://arxiv.org/abs/1409.4330} {arXiv:1409.4330 [hep-ph]}
  \BibitemShut {NoStop}%
%%CITATION = ARXIV:1409.4330;%%
\bibitem [{\citenamefont {Roland}, \citenamefont {Shakya},\ and\ \citenamefont
  {Wells}(2014)}]{Roland:2014vba}%
  \BibitemOpen
  \bibfield  {author} {\bibinfo {author} {\bibfnamefont {S.~B.}\ \bibnamefont
  {Roland}}, \bibinfo {author} {\bibfnamefont {B.}~\bibnamefont {Shakya}}, \
  and\ \bibinfo {author} {\bibfnamefont {J.~D.}\ \bibnamefont {Wells}},\
  }\href@noop {} {\  (\bibinfo {year} {2014})},\ \Eprint
  {http://arxiv.org/abs/1412.4791} {arXiv:1412.4791 [hep-ph]} \BibitemShut
  {NoStop}%
%%CITATION = ARXIV:1412.4791;%%
\bibitem [{\citenamefont {Merle}\ and\ \citenamefont
  {Schneider}(2014)}]{Merle:2014xpa}%
  \BibitemOpen
  \bibfield  {author} {\bibinfo {author} {\bibfnamefont {A.}~\bibnamefont
  {Merle}}\ and\ \bibinfo {author} {\bibfnamefont {A.}~\bibnamefont
  {Schneider}},\ }\href@noop {} {\  (\bibinfo {year} {2014})},\ \Eprint
  {http://arxiv.org/abs/1409.6311} {arXiv:1409.6311 [hep-ph]} \BibitemShut
  {NoStop}%
%%CITATION = ARXIV:1409.6311;%%
\bibitem [{\citenamefont {Merle}\ and\ \citenamefont
  {Totzauer}(2015)}]{Merle:2015oja}%
  \BibitemOpen
  \bibfield  {author} {\bibinfo {author} {\bibfnamefont {A.}~\bibnamefont
  {Merle}}\ and\ \bibinfo {author} {\bibfnamefont {M.}~\bibnamefont
  {Totzauer}},\ }\href@noop {} {\  (\bibinfo {year} {2015})},\ \Eprint
  {http://arxiv.org/abs/1502.01011} {arXiv:1502.01011 [hep-ph]} \BibitemShut
  {NoStop}%
%%CITATION = ARXIV:1502.01011;%%
\bibitem [{\citenamefont {Chen}\ and\ \citenamefont
  {Tang}(2012)}]{Chen:2012faa}%
  \BibitemOpen
  \bibfield  {author} {\bibinfo {author} {\bibfnamefont {C.-S.}\ \bibnamefont
  {Chen}}\ and\ \bibinfo {author} {\bibfnamefont {Y.}~\bibnamefont {Tang}},\
  }\href {\doibase 10.1007/JHEP04(2012)019} {\bibfield  {journal} {\bibinfo
  {journal} {JHEP}\ }\textbf {\bibinfo {volume} {1204}},\ \bibinfo {pages}
  {019} (\bibinfo {year} {2012})},\ \Eprint {http://arxiv.org/abs/1202.5717}
  {arXiv:1202.5717 [hep-ph]} \BibitemShut {NoStop}%
%%CITATION = ARXIV:1202.5717;%%
\bibitem [{\citenamefont {Lebedev}(2012)}]{Lebedev:2012zw}%
  \BibitemOpen
  \bibfield  {author} {\bibinfo {author} {\bibfnamefont {O.}~\bibnamefont
  {Lebedev}},\ }\href {\doibase 10.1140/epjc/s10052-012-2058-2} {\bibfield
  {journal} {\bibinfo  {journal} {Eur.Phys.J.}\ }\textbf {\bibinfo {volume}
  {C72}},\ \bibinfo {pages} {2058} (\bibinfo {year} {2012})},\ \Eprint
  {http://arxiv.org/abs/1203.0156} {arXiv:1203.0156 [hep-ph]} \BibitemShut
  {NoStop}%
%%CITATION = ARXIV:1203.0156;%%
\bibitem [{\citenamefont {Giudice}\ and\ \citenamefont
  {Lee}(2011)}]{Giudice:2010ka}%
  \BibitemOpen
  \bibfield  {author} {\bibinfo {author} {\bibfnamefont {G.~F.}\ \bibnamefont
  {Giudice}}\ and\ \bibinfo {author} {\bibfnamefont {H.~M.}\ \bibnamefont
  {Lee}},\ }\href {\doibase 10.1016/j.physletb.2010.10.035} {\bibfield
  {journal} {\bibinfo  {journal} {Phys.Lett.}\ }\textbf {\bibinfo {volume}
  {B694}},\ \bibinfo {pages} {294} (\bibinfo {year} {2011})},\ \Eprint
  {http://arxiv.org/abs/1010.1417} {arXiv:1010.1417 [hep-ph]} \BibitemShut
  {NoStop}%
%%CITATION = ARXIV:1010.1417;%%
\bibitem [{\citenamefont {Kang}(2014)}]{Kang:2014cia}%
  \BibitemOpen
  \bibfield  {author} {\bibinfo {author} {\bibfnamefont {Z.}~\bibnamefont
  {Kang}},\ }\href@noop {} {\  (\bibinfo {year} {2014})},\ \Eprint
  {http://arxiv.org/abs/1411.2773} {arXiv:1411.2773 [hep-ph]} \BibitemShut
  {NoStop}%
%%CITATION = ARXIV:1411.2773;%%
\bibitem [{\citenamefont {Zeldovich}, \citenamefont {Kobzarev},\ and\
  \citenamefont {Okun}(1974)}]{Zeldovich:1974uw}%
  \BibitemOpen
  \bibfield  {author} {\bibinfo {author} {\bibfnamefont {Y.}~\bibnamefont
  {Zeldovich}}, \bibinfo {author} {\bibfnamefont {I.~Y.}\ \bibnamefont
  {Kobzarev}}, \ and\ \bibinfo {author} {\bibfnamefont {L.}~\bibnamefont
  {Okun}},\ }\href@noop {} {\bibfield  {journal} {\bibinfo  {journal}
  {Zh.Eksp.Teor.Fiz.}\ }\textbf {\bibinfo {volume} {67}},\ \bibinfo {pages} {3}
  (\bibinfo {year} {1974})}\BibitemShut {NoStop}%
%%CITATION = ZETFA,67,3;%%
\bibitem [{\citenamefont {Vilenkin}(1985)}]{Vilenkin:1984ib}%
  \BibitemOpen
  \bibfield  {author} {\bibinfo {author} {\bibfnamefont {A.}~\bibnamefont
  {Vilenkin}},\ }\href {\doibase 10.1016/0370-1573(85)90033-X} {\bibfield
  {journal} {\bibinfo  {journal} {Phys.Rept.}\ }\textbf {\bibinfo {volume}
  {121}},\ \bibinfo {pages} {263} (\bibinfo {year} {1985})}\BibitemShut
  {NoStop}%
%%CITATION = PRPLC,121,263;%%
\bibitem [{\citenamefont {Kibble}(1976)}]{Kibble:1976sj}%
  \BibitemOpen
  \bibfield  {author} {\bibinfo {author} {\bibfnamefont {T.}~\bibnamefont
  {Kibble}},\ }\href {\doibase 10.1088/0305-4470/9/8/029} {\bibfield  {journal}
  {\bibinfo  {journal} {J.Phys.}\ }\textbf {\bibinfo {volume} {A9}},\ \bibinfo
  {pages} {1387} (\bibinfo {year} {1976})}\BibitemShut {NoStop}%
%%CITATION = JPAGA,A9,1387;%%
\bibitem [{\citenamefont {McDonald}(1994)}]{McDonald:1993ex}%
  \BibitemOpen
  \bibfield  {author} {\bibinfo {author} {\bibfnamefont {J.}~\bibnamefont
  {McDonald}},\ }\href {\doibase 10.1103/PhysRevD.50.3637} {\bibfield
  {journal} {\bibinfo  {journal} {Phys.Rev.}\ }\textbf {\bibinfo {volume}
  {D50}},\ \bibinfo {pages} {3637} (\bibinfo {year} {1994})},\ \Eprint
  {http://arxiv.org/abs/hep-ph/0702143} {arXiv:hep-ph/0702143 [HEP-PH]}
  \BibitemShut {NoStop}%
%%CITATION = HEP-PH/0702143;%%
\bibitem [{\citenamefont {Vilja}(1994)}]{Vilja:1993uw}%
  \BibitemOpen
  \bibfield  {author} {\bibinfo {author} {\bibfnamefont {I.}~\bibnamefont
  {Vilja}},\ }\href {\doibase 10.1016/0370-2693(94)90407-3} {\bibfield
  {journal} {\bibinfo  {journal} {Phys.Lett.}\ }\textbf {\bibinfo {volume}
  {B324}},\ \bibinfo {pages} {197} (\bibinfo {year} {1994})},\ \Eprint
  {http://arxiv.org/abs/hep-ph/9312283} {arXiv:hep-ph/9312283 [hep-ph]}
  \BibitemShut {NoStop}%
%%CITATION = HEP-PH/9312283;%%
\bibitem [{\citenamefont {Enqvist}, \citenamefont {Kainulainen},\ and\
  \citenamefont {Vilja}(1993)}]{Enqvist:1992va}%
  \BibitemOpen
  \bibfield  {author} {\bibinfo {author} {\bibfnamefont {K.}~\bibnamefont
  {Enqvist}}, \bibinfo {author} {\bibfnamefont {K.}~\bibnamefont
  {Kainulainen}}, \ and\ \bibinfo {author} {\bibfnamefont {I.}~\bibnamefont
  {Vilja}},\ }\href {\doibase 10.1016/0550-3213(93)90369-Z} {\bibfield
  {journal} {\bibinfo  {journal} {Nucl.Phys.}\ }\textbf {\bibinfo {volume}
  {B403}},\ \bibinfo {pages} {749} (\bibinfo {year} {1993})}\BibitemShut
  {NoStop}%
%%CITATION = NUPHA,B403,749;%%
\bibitem [{\citenamefont {Ade}\ \emph {et~al.}(2014)\citenamefont {Ade} \emph
  {et~al.}}]{Ade:2013zuv}%
  \BibitemOpen
  \bibfield  {author} {\bibinfo {author} {\bibfnamefont {P.}~\bibnamefont
  {Ade}} \emph {et~al.} (\bibinfo {collaboration} {Planck Collaboration}),\
  }\href {\doibase 10.1051/0004-6361/201321591} {\bibfield  {journal} {\bibinfo
   {journal} {Astron.Astrophys.}\ }\textbf {\bibinfo {volume} {571}},\ \bibinfo
  {pages} {A16} (\bibinfo {year} {2014})},\ \Eprint
  {http://arxiv.org/abs/1303.5076} {arXiv:1303.5076 [astro-ph.CO]} \BibitemShut
  {NoStop}%
%%CITATION = ARXIV:1303.5076;%%
\bibitem [{\citenamefont {Abazajian}(2006)}]{Abazajian:2005gj}%
  \BibitemOpen
  \bibfield  {author} {\bibinfo {author} {\bibfnamefont {K.}~\bibnamefont
  {Abazajian}},\ }\href {\doibase 10.1103/PhysRevD.73.063506} {\bibfield
  {journal} {\bibinfo  {journal} {Phys.Rev.}\ }\textbf {\bibinfo {volume}
  {D73}},\ \bibinfo {pages} {063506} (\bibinfo {year} {2006})},\ \Eprint
  {http://arxiv.org/abs/astro-ph/0511630} {arXiv:astro-ph/0511630 [astro-ph]}
  \BibitemShut {NoStop}%
%%CITATION = ASTRO-PH/0511630;%%
\bibitem [{\citenamefont {{Fuller}}, \citenamefont {{Kishimoto}},\ and\
  \citenamefont {{Kusenko}}(2011)}]{2011arXiv1110.6479F}%
  \BibitemOpen
  \bibfield  {author} {\bibinfo {author} {\bibfnamefont {G.~M.}\ \bibnamefont
  {{Fuller}}}, \bibinfo {author} {\bibfnamefont {C.~T.}\ \bibnamefont
  {{Kishimoto}}}, \ and\ \bibinfo {author} {\bibfnamefont {A.}~\bibnamefont
  {{Kusenko}}},\ }\href@noop {} {\bibfield  {journal} {\bibinfo  {journal}
  {ArXiv e-prints}\ } (\bibinfo {year} {2011})},\ \Eprint
  {http://arxiv.org/abs/1110.6479} {arXiv:1110.6479 [astro-ph.CO]} \BibitemShut
  {NoStop}%
\bibitem [{\citenamefont {Abazajian}, \citenamefont {Fuller},\ and\
  \citenamefont {Patel}(2001)}]{Abazajian:2001nj}%
  \BibitemOpen
  \bibfield  {author} {\bibinfo {author} {\bibfnamefont {K.}~\bibnamefont
  {Abazajian}}, \bibinfo {author} {\bibfnamefont {G.~M.}\ \bibnamefont
  {Fuller}}, \ and\ \bibinfo {author} {\bibfnamefont {M.}~\bibnamefont
  {Patel}},\ }\href {\doibase 10.1103/PhysRevD.64.023501} {\bibfield  {journal}
  {\bibinfo  {journal} {Phys.Rev.}\ }\textbf {\bibinfo {volume} {D64}},\
  \bibinfo {pages} {023501} (\bibinfo {year} {2001})},\ \Eprint
  {http://arxiv.org/abs/astro-ph/0101524} {arXiv:astro-ph/0101524 [astro-ph]}
  \BibitemShut {NoStop}%
%%CITATION = ASTRO-PH/0101524;%%
\bibitem [{\citenamefont {Abazajian}, \citenamefont {Fuller},\ and\
  \citenamefont {Tucker}(2001)}]{Abazajian:2001vt}%
  \BibitemOpen
  \bibfield  {author} {\bibinfo {author} {\bibfnamefont {K.}~\bibnamefont
  {Abazajian}}, \bibinfo {author} {\bibfnamefont {G.~M.}\ \bibnamefont
  {Fuller}}, \ and\ \bibinfo {author} {\bibfnamefont {W.~H.}\ \bibnamefont
  {Tucker}},\ }\href {\doibase 10.1086/323867} {\bibfield  {journal} {\bibinfo
  {journal} {Astrophys.J.}\ }\textbf {\bibinfo {volume} {562}},\ \bibinfo
  {pages} {593} (\bibinfo {year} {2001})},\ \Eprint
  {http://arxiv.org/abs/astro-ph/0106002} {arXiv:astro-ph/0106002 [astro-ph]}
  \BibitemShut {NoStop}%
%%CITATION = ASTRO-PH/0106002;%%
\bibitem [{\citenamefont {Bulbul}\ \emph {et~al.}(2014)\citenamefont {Bulbul},
  \citenamefont {Markevitch}, \citenamefont {Foster}, \citenamefont {Smith},
  \citenamefont {Loewenstein} \emph {et~al.}}]{Bulbul:2014sua}%
  \BibitemOpen
  \bibfield  {author} {\bibinfo {author} {\bibfnamefont {E.}~\bibnamefont
  {Bulbul}}, \bibinfo {author} {\bibfnamefont {M.}~\bibnamefont {Markevitch}},
  \bibinfo {author} {\bibfnamefont {A.}~\bibnamefont {Foster}}, \bibinfo
  {author} {\bibfnamefont {R.~K.}\ \bibnamefont {Smith}}, \bibinfo {author}
  {\bibfnamefont {M.}~\bibnamefont {Loewenstein}},  \emph {et~al.},\ }\href
  {\doibase 10.1088/0004-637X/789/1/13} {\bibfield  {journal} {\bibinfo
  {journal} {Astrophys.J.}\ }\textbf {\bibinfo {volume} {789}},\ \bibinfo
  {pages} {13} (\bibinfo {year} {2014})},\ \Eprint
  {http://arxiv.org/abs/1402.2301} {arXiv:1402.2301 [astro-ph.CO]} \BibitemShut
  {NoStop}%
%%CITATION = ARXIV:1402.2301;%%
\bibitem [{\citenamefont {Boyarsky}\ \emph {et~al.}(2014)\citenamefont
  {Boyarsky}, \citenamefont {Ruchayskiy}, \citenamefont {Iakubovskyi},\ and\
  \citenamefont {Franse}}]{Boyarsky:2014jta}%
  \BibitemOpen
  \bibfield  {author} {\bibinfo {author} {\bibfnamefont {A.}~\bibnamefont
  {Boyarsky}}, \bibinfo {author} {\bibfnamefont {O.}~\bibnamefont
  {Ruchayskiy}}, \bibinfo {author} {\bibfnamefont {D.}~\bibnamefont
  {Iakubovskyi}}, \ and\ \bibinfo {author} {\bibfnamefont {J.}~\bibnamefont
  {Franse}},\ }\href {\doibase 10.1103/PhysRevLett.113.251301} {\bibfield
  {journal} {\bibinfo  {journal} {Phys.Rev.Lett.}\ }\textbf {\bibinfo {volume}
  {113}},\ \bibinfo {pages} {251301} (\bibinfo {year} {2014})},\ \Eprint
  {http://arxiv.org/abs/1402.4119} {arXiv:1402.4119 [astro-ph.CO]} \BibitemShut
  {NoStop}%
%%CITATION = ARXIV:1402.4119;%%
\bibitem [{\citenamefont {Lumb}\ \emph {et~al.}(2002)\citenamefont {Lumb},
  \citenamefont {Warwick}, \citenamefont {Page},\ and\ \citenamefont
  {De~Luca}}]{Lumb:2002sw}%
  \BibitemOpen
  \bibfield  {author} {\bibinfo {author} {\bibfnamefont {D.}~\bibnamefont
  {Lumb}}, \bibinfo {author} {\bibfnamefont {R.}~\bibnamefont {Warwick}},
  \bibinfo {author} {\bibfnamefont {M.}~\bibnamefont {Page}}, \ and\ \bibinfo
  {author} {\bibfnamefont {A.}~\bibnamefont {De~Luca}},\ }\href {\doibase
  10.1051/0004-6361:20020531} {\bibfield  {journal} {\bibinfo  {journal}
  {Astron.Astrophys.}\ }\textbf {\bibinfo {volume} {389}},\ \bibinfo {pages}
  {93} (\bibinfo {year} {2002})},\ \Eprint
  {http://arxiv.org/abs/astro-ph/0204147} {arXiv:astro-ph/0204147 [astro-ph]}
  \BibitemShut {NoStop}%
%%CITATION = ASTRO-PH/0204147;%%
\bibitem [{\citenamefont {Read}\ and\ \citenamefont
  {Ponman}(2003)}]{Read:2003hw}%
  \BibitemOpen
  \bibfield  {author} {\bibinfo {author} {\bibfnamefont {A.~M.}\ \bibnamefont
  {Read}}\ and\ \bibinfo {author} {\bibfnamefont {T.~J.}\ \bibnamefont
  {Ponman}},\ }\href@noop {} {\bibfield  {journal} {\bibinfo  {journal}
  {Astron.Astrophys.}\ }\textbf {\bibinfo {volume} {409}},\ \bibinfo {pages}
  {395} (\bibinfo {year} {2003})},\ \Eprint
  {http://arxiv.org/abs/astro-ph/0304147} {arXiv:astro-ph/0304147 [astro-ph]}
  \BibitemShut {NoStop}%
%%CITATION = ASTRO-PH/0304147;%%
\bibitem [{\citenamefont {Gruber}\ \emph {et~al.}(1999)\citenamefont {Gruber},
  \citenamefont {Matteson}, \citenamefont {Peterson},\ and\ \citenamefont
  {Jung}}]{Gruber:1999yr}%
  \BibitemOpen
  \bibfield  {author} {\bibinfo {author} {\bibfnamefont {D.}~\bibnamefont
  {Gruber}}, \bibinfo {author} {\bibfnamefont {J.}~\bibnamefont {Matteson}},
  \bibinfo {author} {\bibfnamefont {L.}~\bibnamefont {Peterson}}, \ and\
  \bibinfo {author} {\bibfnamefont {G.}~\bibnamefont {Jung}},\ }\href {\doibase
  10.1086/307450} {\bibfield  {journal} {\bibinfo  {journal} {Astrophys.J.}\
  }\textbf {\bibinfo {volume} {520}},\ \bibinfo {pages} {124} (\bibinfo {year}
  {1999})},\ \Eprint {http://arxiv.org/abs/astro-ph/9903492}
  {arXiv:astro-ph/9903492 [astro-ph]} \BibitemShut {NoStop}%
%%CITATION = ASTRO-PH/9903492;%%
\bibitem [{\citenamefont {Boyarsky}\ \emph
  {et~al.}(2006{\natexlab{a}})\citenamefont {Boyarsky}, \citenamefont
  {Neronov}, \citenamefont {Ruchayskiy},\ and\ \citenamefont
  {Shaposhnikov}}]{Boyarsky:2005us}%
  \BibitemOpen
  \bibfield  {author} {\bibinfo {author} {\bibfnamefont {A.}~\bibnamefont
  {Boyarsky}}, \bibinfo {author} {\bibfnamefont {A.}~\bibnamefont {Neronov}},
  \bibinfo {author} {\bibfnamefont {O.}~\bibnamefont {Ruchayskiy}}, \ and\
  \bibinfo {author} {\bibfnamefont {M.}~\bibnamefont {Shaposhnikov}},\ }\href
  {\doibase 10.1111/j.1365-2966.2006.10458.x} {\bibfield  {journal} {\bibinfo
  {journal} {Mon.Not.Roy.Astron.Soc.}\ }\textbf {\bibinfo {volume} {370}},\
  \bibinfo {pages} {213} (\bibinfo {year} {2006}{\natexlab{a}})},\ \Eprint
  {http://arxiv.org/abs/astro-ph/0512509} {arXiv:astro-ph/0512509 [astro-ph]}
  \BibitemShut {NoStop}%
%%CITATION = ASTRO-PH/0512509;%%
\bibitem [{\citenamefont {Boyarsky}\ \emph
  {et~al.}(2006{\natexlab{b}})\citenamefont {Boyarsky}, \citenamefont
  {Neronov}, \citenamefont {Ruchayskiy},\ and\ \citenamefont
  {Shaposhnikov}}]{Boyarsky:2006zi}%
  \BibitemOpen
  \bibfield  {author} {\bibinfo {author} {\bibfnamefont {A.}~\bibnamefont
  {Boyarsky}}, \bibinfo {author} {\bibfnamefont {A.}~\bibnamefont {Neronov}},
  \bibinfo {author} {\bibfnamefont {O.}~\bibnamefont {Ruchayskiy}}, \ and\
  \bibinfo {author} {\bibfnamefont {M.}~\bibnamefont {Shaposhnikov}},\ }\href
  {\doibase 10.1103/PhysRevD.74.103506} {\bibfield  {journal} {\bibinfo
  {journal} {Phys.Rev.}\ }\textbf {\bibinfo {volume} {D74}},\ \bibinfo {pages}
  {103506} (\bibinfo {year} {2006}{\natexlab{b}})},\ \Eprint
  {http://arxiv.org/abs/astro-ph/0603368} {arXiv:astro-ph/0603368 [astro-ph]}
  \BibitemShut {NoStop}%
%%CITATION = ASTRO-PH/0603368;%%
\bibitem [{\citenamefont {Weinberg}\ \emph {et~al.}(2013)\citenamefont
  {Weinberg}, \citenamefont {Bullock}, \citenamefont {Governato}, \citenamefont
  {de~Naray},\ and\ \citenamefont {Peter}}]{Weinberg:2013aya}%
  \BibitemOpen
  \bibfield  {author} {\bibinfo {author} {\bibfnamefont {D.~H.}\ \bibnamefont
  {Weinberg}}, \bibinfo {author} {\bibfnamefont {J.~S.}\ \bibnamefont
  {Bullock}}, \bibinfo {author} {\bibfnamefont {F.}~\bibnamefont {Governato}},
  \bibinfo {author} {\bibfnamefont {R.~K.}\ \bibnamefont {de~Naray}}, \ and\
  \bibinfo {author} {\bibfnamefont {A.~H.~G.}\ \bibnamefont {Peter}},\
  }\href@noop {} {\  (\bibinfo {year} {2013})},\ \Eprint
  {http://arxiv.org/abs/1306.0913} {arXiv:1306.0913 [astro-ph.CO]} \BibitemShut
  {NoStop}%
%%CITATION = ARXIV:1306.0913;%%
\bibitem [{\citenamefont {Kamada}\ \emph {et~al.}(2013)\citenamefont {Kamada},
  \citenamefont {Yoshida}, \citenamefont {Kohri},\ and\ \citenamefont
  {Takahashi}}]{Kamada:2013sh}%
  \BibitemOpen
  \bibfield  {author} {\bibinfo {author} {\bibfnamefont {A.}~\bibnamefont
  {Kamada}}, \bibinfo {author} {\bibfnamefont {N.}~\bibnamefont {Yoshida}},
  \bibinfo {author} {\bibfnamefont {K.}~\bibnamefont {Kohri}}, \ and\ \bibinfo
  {author} {\bibfnamefont {T.}~\bibnamefont {Takahashi}},\ }\href {\doibase
  10.1088/1475-7516/2013/03/008} {\bibfield  {journal} {\bibinfo  {journal}
  {JCAP}\ }\textbf {\bibinfo {volume} {1303}},\ \bibinfo {pages} {008}
  (\bibinfo {year} {2013})},\ \Eprint {http://arxiv.org/abs/1301.2744}
  {arXiv:1301.2744 [astro-ph.CO]} \BibitemShut {NoStop}%
%%CITATION = ARXIV:1301.2744;%%
\bibitem [{\citenamefont {Boyanovsky}(2008)}]{Boyanovsky:2008nc}%
  \BibitemOpen
  \bibfield  {author} {\bibinfo {author} {\bibfnamefont {D.}~\bibnamefont
  {Boyanovsky}},\ }\href {\doibase 10.1103/PhysRevD.78.103505} {\bibfield
  {journal} {\bibinfo  {journal} {Phys.Rev.}\ }\textbf {\bibinfo {volume}
  {D78}},\ \bibinfo {pages} {103505} (\bibinfo {year} {2008})},\ \Eprint
  {http://arxiv.org/abs/0807.0646} {arXiv:0807.0646 [astro-ph]} \BibitemShut
  {NoStop}%
%%CITATION = ARXIV:0807.0646;%%
\bibitem [{\citenamefont {Kaplinghat}(2005)}]{Kaplinghat:2005sy}%
  \BibitemOpen
  \bibfield  {author} {\bibinfo {author} {\bibfnamefont {M.}~\bibnamefont
  {Kaplinghat}},\ }\href {\doibase 10.1103/PhysRevD.72.063510} {\bibfield
  {journal} {\bibinfo  {journal} {Phys.Rev.}\ }\textbf {\bibinfo {volume}
  {D72}},\ \bibinfo {pages} {063510} (\bibinfo {year} {2005})},\ \Eprint
  {http://arxiv.org/abs/astro-ph/0507300} {arXiv:astro-ph/0507300 [astro-ph]}
  \BibitemShut {NoStop}%
%%CITATION = ASTRO-PH/0507300;%%
\bibitem [{\citenamefont {Hisano}, \citenamefont {Kohri},\ and\ \citenamefont
  {Nojiri}(2001)}]{Hisano:2000dz}%
  \BibitemOpen
  \bibfield  {author} {\bibinfo {author} {\bibfnamefont {J.}~\bibnamefont
  {Hisano}}, \bibinfo {author} {\bibfnamefont {K.}~\bibnamefont {Kohri}}, \
  and\ \bibinfo {author} {\bibfnamefont {M.~M.}\ \bibnamefont {Nojiri}},\
  }\href {\doibase 10.1016/S0370-2693(01)00395-1} {\bibfield  {journal}
  {\bibinfo  {journal} {Phys.Lett.}\ }\textbf {\bibinfo {volume} {B505}},\
  \bibinfo {pages} {169} (\bibinfo {year} {2001})},\ \Eprint
  {http://arxiv.org/abs/hep-ph/0011216} {arXiv:hep-ph/0011216 [hep-ph]}
  \BibitemShut {NoStop}%
%%CITATION = HEP-PH/0011216;%%
\bibitem [{\citenamefont {Strigari}, \citenamefont {Kaplinghat},\ and\
  \citenamefont {Bullock}(2007)}]{Strigari:2006jf}%
  \BibitemOpen
  \bibfield  {author} {\bibinfo {author} {\bibfnamefont {L.~E.}\ \bibnamefont
  {Strigari}}, \bibinfo {author} {\bibfnamefont {M.}~\bibnamefont
  {Kaplinghat}}, \ and\ \bibinfo {author} {\bibfnamefont {J.~S.}\ \bibnamefont
  {Bullock}},\ }\href {\doibase 10.1103/PhysRevD.75.061303} {\bibfield
  {journal} {\bibinfo  {journal} {Phys.Rev.}\ }\textbf {\bibinfo {volume}
  {D75}},\ \bibinfo {pages} {061303} (\bibinfo {year} {2007})},\ \Eprint
  {http://arxiv.org/abs/astro-ph/0606281} {arXiv:astro-ph/0606281 [astro-ph]}
  \BibitemShut {NoStop}%
%%CITATION = ASTRO-PH/0606281;%%
\bibitem [{\citenamefont {Aoyama}\ \emph {et~al.}(2011)\citenamefont {Aoyama},
  \citenamefont {Ichiki}, \citenamefont {Nitta},\ and\ \citenamefont
  {Sugiyama}}]{Aoyama:2011ba}%
  \BibitemOpen
  \bibfield  {author} {\bibinfo {author} {\bibfnamefont {S.}~\bibnamefont
  {Aoyama}}, \bibinfo {author} {\bibfnamefont {K.}~\bibnamefont {Ichiki}},
  \bibinfo {author} {\bibfnamefont {D.}~\bibnamefont {Nitta}}, \ and\ \bibinfo
  {author} {\bibfnamefont {N.}~\bibnamefont {Sugiyama}},\ }\href {\doibase
  10.1088/1475-7516/2011/09/025} {\bibfield  {journal} {\bibinfo  {journal}
  {JCAP}\ }\textbf {\bibinfo {volume} {1109}},\ \bibinfo {pages} {025}
  (\bibinfo {year} {2011})},\ \Eprint {http://arxiv.org/abs/1106.1984}
  {arXiv:1106.1984 [astro-ph.CO]} \BibitemShut {NoStop}%
%%CITATION = ARXIV:1106.1984;%%
\bibitem [{\citenamefont {Viel}\ \emph {et~al.}(2006)\citenamefont {Viel},
  \citenamefont {Lesgourgues}, \citenamefont {Haehnelt}, \citenamefont
  {Matarrese},\ and\ \citenamefont {Riotto}}]{Viel:2006kd}%
  \BibitemOpen
  \bibfield  {author} {\bibinfo {author} {\bibfnamefont {M.}~\bibnamefont
  {Viel}}, \bibinfo {author} {\bibfnamefont {J.}~\bibnamefont {Lesgourgues}},
  \bibinfo {author} {\bibfnamefont {M.~G.}\ \bibnamefont {Haehnelt}}, \bibinfo
  {author} {\bibfnamefont {S.}~\bibnamefont {Matarrese}}, \ and\ \bibinfo
  {author} {\bibfnamefont {A.}~\bibnamefont {Riotto}},\ }\href {\doibase
  10.1103/PhysRevLett.97.071301} {\bibfield  {journal} {\bibinfo  {journal}
  {Phys.Rev.Lett.}\ }\textbf {\bibinfo {volume} {97}},\ \bibinfo {pages}
  {071301} (\bibinfo {year} {2006})},\ \Eprint
  {http://arxiv.org/abs/astro-ph/0605706} {arXiv:astro-ph/0605706 [astro-ph]}
  \BibitemShut {NoStop}%
%%CITATION = ASTRO-PH/0605706;%%
\bibitem [{\citenamefont {Viel}\ \emph {et~al.}(2005)\citenamefont {Viel},
  \citenamefont {Lesgourgues}, \citenamefont {Haehnelt}, \citenamefont
  {Matarrese},\ and\ \citenamefont {Riotto}}]{Viel:2005qj}%
  \BibitemOpen
  \bibfield  {author} {\bibinfo {author} {\bibfnamefont {M.}~\bibnamefont
  {Viel}}, \bibinfo {author} {\bibfnamefont {J.}~\bibnamefont {Lesgourgues}},
  \bibinfo {author} {\bibfnamefont {M.~G.}\ \bibnamefont {Haehnelt}}, \bibinfo
  {author} {\bibfnamefont {S.}~\bibnamefont {Matarrese}}, \ and\ \bibinfo
  {author} {\bibfnamefont {A.}~\bibnamefont {Riotto}},\ }\href {\doibase
  10.1103/PhysRevD.71.063534} {\bibfield  {journal} {\bibinfo  {journal}
  {Phys.Rev.}\ }\textbf {\bibinfo {volume} {D71}},\ \bibinfo {pages} {063534}
  (\bibinfo {year} {2005})},\ \Eprint {http://arxiv.org/abs/astro-ph/0501562}
  {arXiv:astro-ph/0501562 [astro-ph]} \BibitemShut {NoStop}%
%%CITATION = ASTRO-PH/0501562;%%
\bibitem [{\citenamefont {Davidson}\ and\ \citenamefont
  {Ibarra}(2002)}]{Davidson:2002qv}%
  \BibitemOpen
  \bibfield  {author} {\bibinfo {author} {\bibfnamefont {S.}~\bibnamefont
  {Davidson}}\ and\ \bibinfo {author} {\bibfnamefont {A.}~\bibnamefont
  {Ibarra}},\ }\href {\doibase 10.1016/S0370-2693(02)01735-5} {\bibfield
  {journal} {\bibinfo  {journal} {Phys.Lett.}\ }\textbf {\bibinfo {volume}
  {B535}},\ \bibinfo {pages} {25} (\bibinfo {year} {2002})},\ \Eprint
  {http://arxiv.org/abs/hep-ph/0202239} {arXiv:hep-ph/0202239 [hep-ph]}
  \BibitemShut {NoStop}%
%%CITATION = HEP-PH/0202239;%%
\bibitem [{\citenamefont {Pilaftsis}(1997)}]{Pilaftsis:1997jf}%
  \BibitemOpen
  \bibfield  {author} {\bibinfo {author} {\bibfnamefont {A.}~\bibnamefont
  {Pilaftsis}},\ }\href {\doibase 10.1103/PhysRevD.56.5431} {\bibfield
  {journal} {\bibinfo  {journal} {Phys.Rev.}\ }\textbf {\bibinfo {volume}
  {D56}},\ \bibinfo {pages} {5431} (\bibinfo {year} {1997})},\ \Eprint
  {http://arxiv.org/abs/hep-ph/9707235} {arXiv:hep-ph/9707235 [hep-ph]}
  \BibitemShut {NoStop}%
%%CITATION = HEP-PH/9707235;%%
\bibitem [{\citenamefont {Pilaftsis}\ and\ \citenamefont
  {Underwood}(2004)}]{Pilaftsis:2003gt}%
  \BibitemOpen
  \bibfield  {author} {\bibinfo {author} {\bibfnamefont {A.}~\bibnamefont
  {Pilaftsis}}\ and\ \bibinfo {author} {\bibfnamefont {T.~E.}\ \bibnamefont
  {Underwood}},\ }\href {\doibase 10.1016/j.nuclphysb.2004.05.029} {\bibfield
  {journal} {\bibinfo  {journal} {Nucl.Phys.}\ }\textbf {\bibinfo {volume}
  {B692}},\ \bibinfo {pages} {303} (\bibinfo {year} {2004})},\ \Eprint
  {http://arxiv.org/abs/hep-ph/0309342} {arXiv:hep-ph/0309342 [hep-ph]}
  \BibitemShut {NoStop}%
%%CITATION = HEP-PH/0309342;%%
\bibitem [{\citenamefont {Fong}, \citenamefont {Nardi},\ and\ \citenamefont
  {Riotto}(2012)}]{Fong:2013wr}%
  \BibitemOpen
  \bibfield  {author} {\bibinfo {author} {\bibfnamefont {C.~S.}\ \bibnamefont
  {Fong}}, \bibinfo {author} {\bibfnamefont {E.}~\bibnamefont {Nardi}}, \ and\
  \bibinfo {author} {\bibfnamefont {A.}~\bibnamefont {Riotto}},\ }\href
  {\doibase 10.1155/2012/158303} {\bibfield  {journal} {\bibinfo  {journal}
  {Adv.High Energy Phys.}\ }\textbf {\bibinfo {volume} {2012}},\ \bibinfo
  {pages} {158303} (\bibinfo {year} {2012})},\ \Eprint
  {http://arxiv.org/abs/1301.3062} {arXiv:1301.3062 [hep-ph]} \BibitemShut
  {NoStop}%
%%CITATION = ARXIV:1301.3062;%%
\bibitem [{\citenamefont {Buchmuller}, \citenamefont {Peccei},\ and\
  \citenamefont {Yanagida}(2005)}]{Buchmuller:2005eh}%
  \BibitemOpen
  \bibfield  {author} {\bibinfo {author} {\bibfnamefont {W.}~\bibnamefont
  {Buchmuller}}, \bibinfo {author} {\bibfnamefont {R.}~\bibnamefont {Peccei}},
  \ and\ \bibinfo {author} {\bibfnamefont {T.}~\bibnamefont {Yanagida}},\
  }\href {\doibase 10.1146/annurev.nucl.55.090704.151558} {\bibfield  {journal}
  {\bibinfo  {journal} {Ann.Rev.Nucl.Part.Sci.}\ }\textbf {\bibinfo {volume}
  {55}},\ \bibinfo {pages} {311} (\bibinfo {year} {2005})},\ \Eprint
  {http://arxiv.org/abs/hep-ph/0502169} {arXiv:hep-ph/0502169 [hep-ph]}
  \BibitemShut {NoStop}%
%%CITATION = HEP-PH/0502169;%%
\bibitem [{\citenamefont {Buchmuller}, \citenamefont {Di~Bari},\ and\
  \citenamefont {Plumacher}(2005)}]{Buchmuller:2004nz}%
  \BibitemOpen
  \bibfield  {author} {\bibinfo {author} {\bibfnamefont {W.}~\bibnamefont
  {Buchmuller}}, \bibinfo {author} {\bibfnamefont {P.}~\bibnamefont {Di~Bari}},
  \ and\ \bibinfo {author} {\bibfnamefont {M.}~\bibnamefont {Plumacher}},\
  }\href {\doibase 10.1016/j.aop.2004.02.003} {\bibfield  {journal} {\bibinfo
  {journal} {Annals Phys.}\ }\textbf {\bibinfo {volume} {315}},\ \bibinfo
  {pages} {305} (\bibinfo {year} {2005})},\ \Eprint
  {http://arxiv.org/abs/hep-ph/0401240} {arXiv:hep-ph/0401240 [hep-ph]}
  \BibitemShut {NoStop}%
%%CITATION = HEP-PH/0401240;%%
\bibitem [{\citenamefont {Harvey}\ and\ \citenamefont
  {Turner}(1990)}]{Harvey:1990qw}%
  \BibitemOpen
  \bibfield  {author} {\bibinfo {author} {\bibfnamefont {J.~A.}\ \bibnamefont
  {Harvey}}\ and\ \bibinfo {author} {\bibfnamefont {M.~S.}\ \bibnamefont
  {Turner}},\ }\href {\doibase 10.1103/PhysRevD.42.3344} {\bibfield  {journal}
  {\bibinfo  {journal} {Phys.Rev.}\ }\textbf {\bibinfo {volume} {D42}},\
  \bibinfo {pages} {3344} (\bibinfo {year} {1990})}\BibitemShut {NoStop}%
%%CITATION = PHRVA,D42,3344;%%
\bibitem [{\citenamefont {Khlebnikov}\ and\ \citenamefont
  {Shaposhnikov}(1988)}]{Khlebnikov:1988sr}%
  \BibitemOpen
  \bibfield  {author} {\bibinfo {author} {\bibfnamefont {S.~Y.}\ \bibnamefont
  {Khlebnikov}}\ and\ \bibinfo {author} {\bibfnamefont {M.}~\bibnamefont
  {Shaposhnikov}},\ }\href {\doibase 10.1016/0550-3213(88)90133-2} {\bibfield
  {journal} {\bibinfo  {journal} {Nucl.Phys.}\ }\textbf {\bibinfo {volume}
  {B308}},\ \bibinfo {pages} {885} (\bibinfo {year} {1988})}\BibitemShut
  {NoStop}%
%%CITATION = NUPHA,B308,885;%%
\bibitem [{\citenamefont {Bochkarev}\ and\ \citenamefont
  {Shaposhnikov}(1987)}]{Bochkarev:1987wf}%
  \BibitemOpen
  \bibfield  {author} {\bibinfo {author} {\bibfnamefont {A.}~\bibnamefont
  {Bochkarev}}\ and\ \bibinfo {author} {\bibfnamefont {M.}~\bibnamefont
  {Shaposhnikov}},\ }\href {\doibase 10.1142/S0217732387000537} {\bibfield
  {journal} {\bibinfo  {journal} {Mod.Phys.Lett.}\ }\textbf {\bibinfo {volume}
  {A2}},\ \bibinfo {pages} {417} (\bibinfo {year} {1987})}\BibitemShut
  {NoStop}%
%%CITATION = MPLAE,A2,417;%%
\end{thebibliography}%

\end{document}